\documentclass[aps,pre,groupedaddress,twocolumn,floatfix]{revtex4-2}

\newcommand{\beq}{\begin{equation}}
\newcommand{\eeq}{\end{equation}}
\def\NOTE#1{\textcolor{black}{#1}}

\usepackage[dvipsnames]{xcolor}

\usepackage{amsmath,amssymb,natbib}
\usepackage{graphicx}
\usepackage{diagbox}
\usepackage{lipsum}
\usepackage{bbm}
\definecolor{mypink1}{rgb}{255, 0, 255}
\colorlet{LightRubineRed}{RubineRed!70!}

\begin{document}


\title{L\'evy on-off intermittency}


\author{Adrian van Kan} \email[]{adrian.van.kan@phys.ens.fr}
\author{Alexandros Alexakis} \email[]{alexakis@phys.ens.fr} \author{Marc-Etienne Brachet} \email[]{marc.brachet@gmail.com}
\affiliation{Laboratoire de Physique de l’Ecole normale supérieure, ENS, Université PSL, CNRS, Sorbonne Université, Université de Paris, F-75005 Paris, France}


\date{\today}

\begin{abstract}
We present a new form of intermittency, {\it L\'evy on-off intermittency}, which arises from multiplicative $\alpha$-stable white noise close to an instability threshold. We study this problem in the linear and nonlinear regimes, both theoretically and numerically, for the case of a pitchfork bifurcation with fluctuating growth rate. We compute the stationary distribution analytically and numerically from the associated fractional Fokker-Planck equation in the Stratonovich interpretation. We characterize the system in the parameter space $(\alpha,\beta)$ of the noise, with stability parameter $\alpha\in (0,2)$ and skewness parameter $\beta\in[-1,1]$. Five regimes are identified in this parameter space, in addition to the well-studied Gaussian case $\alpha=2$. Three regimes are located at $1<\alpha<2$, where the noise has finite mean but infinite variance. They are differentiated by $\beta$ and all display a critical transition at the deterministic instability threshold, with on-off intermittency close to onset. Critical exponents are computed from the stationary distribution. Each regime is characterized by a specific form of the density and specific critical exponents, which differ starkly from the Gaussian case. A finite or infinite number of \NOTE{integer-order} moments may converge, depending on parameters. Two more regimes are found at $0<\alpha\leq 1$. There, the mean of the noise diverges, and no critical transition occurs. In one case the origin is always unstable, independently of the distance $\mu$ from the deterministic threshold. In the other case, the origin is conversely always stable, independently of $\mu$. \NOTE{We thus demonstrate} that an instability subject to non-equilibrium, power-law-distributed fluctuations can display substantially different properties than for Gaussian thermal fluctuations, in terms of statistics and critical behavior.

\end{abstract}


\maketitle



\section{Introdution}
\label{sec:intro}
On-off intermittency is a common phenomenon in nonequilibrium physical systems, which is characterized by an aperiodic switching between a large-amplitude ``on" state and a small-amplitude ``off" state. It was originally studied theoretically in the context of low-dimensional deterministic chaos and nonlinear maps \cite{fujisaka1985new,platt1993off,ott1994blowout,heagy1994characterization} and has since then been observed in numerous experimental setups ranging from electronic devices \cite{hammer1994experimental}, spin-wave instabilities \cite{rodelsperger1995off}, liquid crystals \cite{john1999off,vella2003off} and plasmas \cite{feng1998off} to multistable laser fibers \cite{huerta2008experimental}, sediment transport \cite{benavides2020multiplicative}, human balancing motion \cite{cabrera2002off,cabrera2004stick} and blinking quantum dots in semiconductor nanocrystals \cite{margolin2005power,frantsuzov2008universal}. On-off intermittency has also been observed in numerical simulations of turbulence in thin layers \cite{benavides2017critical,van2019condensates} and magneto-hydroydnamic dynamo flows \cite{sweet2001blowout,alexakis2008effect,raynaud2013intermittency}.

From a theoretical perspective, on-off intermittency arises in the presence of multiplicative noise close to an instability threshold. Therefore it is natural to study it using appropriate stochastic models, such as 
\begin{equation}
    \frac{dX}{dt} = (f(t)+\mu)X - \gamma X^3, \label{eq:langevin}
\end{equation}
i.e. a supercritical pitchfork bifurcation \cite{strogatz2018nonlinear} with a fluctuating growth rate, where $\mu$ is the deterministic growth rate, and $f(t)$ is usually zero-mean, Gaussian, white noise, $\langle f(t) \rangle = 0$, $\langle f(t) f(t') \rangle = 2 \delta(t-t')$, in terms of the ensemble average $\langle \cdot \rangle$. In this study, we adopt the Stratonovich interpretation  \cite{stratonovich1966new} of eqn. (\ref{eq:langevin}), unless stated otherwise. We may take $X$ to be non-negative, since sign changes are incompatible with the exact solution of (\ref{eq:langevin}) given in \cite{aumaitre2007noise} and in appendix \ref{sec:app_exact_sol}. For Gaussian noise, the exact stationary probability density function (PDF) is known  to be  $p(x)= N x^{-1+\mu}e^{-\frac{\gamma}{2} x^2}$ with normalisation $N$, for $\mu>0$ \cite{schenzle1979multiplicative}. \NOTE{For $\mu\leq 0$, the distribution approaches $\delta(x)$ at late times, with the cumulative distribution function (CDF) (the integral of the PDF up to $x$) converging to $1$ for all $x>0$. In that case, all moments of the stationary density vanish.} For $\mu>0$ the moments of $X$ scale as $\langle X^n \rangle \propto \mu^{c_n}$ with the critical exponents $c_n=1$ for all $n>0$, see \cite{aumaitre2007noise}. The $c_n$ for Gaussian noise are different from their deterministic ``mean field" values, which are $c_n=n/2$. This is an instance of anomalous scaling, a phenomenon which has received much attention in various areas of physics, in particular in the context of continuous phase transitions at equilibrium and critical phenomena \cite{kadanoff1967static,goldenfeld2018lectures}, as well as in turbulence \cite{eyink1994analogies,goldenfeld2017turbulence}.\\

Here, we introduce \textit{L\'evy on-off intermittency} as the case where $f(t)$ is given by L\'evy white noise, whose PDF is an $\alpha$-stable distribution featuring power-law tails associated with extreme events in terms of noise amplitude \cite{shlesinger1995levy, chechkin2008introduction}. The Gaussian distribution (which is a special case of $\alpha$-stable distributions) is of fundamental importance due to its stability: by the central limit theorem \cite{feller2008introduction}, it constitutes an attractor in the space of PDFs with finite variance. Similarly, by the generalized central limit theorem \cite{gnedenko1954limit,uchaikin2011chance}, non-Gaussian $\alpha$-stable distributions constitute an attractor in the space of PDFs whose variance does not exist. Non-Gaussian fluctuations, which may often be modeled as $\alpha$-stable, are found in incompletely thermalized systems or, in general, in systems driven away from thermal equilibrium: non-equilibrated heat reservoirs can be considered as a source of non-Gaussian noise \cite{shlesinger1995levy,dybiec2004resonant}.

If $X(t)$ solves equation (\ref{eq:langevin}) with $f(t)$ being L\'evy white noise, then $Y=\log X(t)$ is said to perform a {\it L\'evy flight} in a particular anharmonic potential. L\'evy flights were first introduced by Mandelbrot in \cite{mandelbrot1983fractal} and have since found numerous applications, such as anomalous diffusion, for instance in different fluid flows, \cite{shlesinger1987levy,solomon1993observation,metzler2000random,dubkov2008levy}, the statistics of 2-D fluid turbulence \cite{dubrulle1998truncated}, plasma turbulence \cite{del2005nondiffusive}, finance \cite{schinckus2013physicists}, climatology \cite{ditlevsen1999anomalous,ditlevsen1999observation}, animal foraging \cite{viswanathan1996levy,sims2008scaling}, human mobility \cite{rhee2011levy} (although a debate about the applicability in the latter two cases is ongoing \cite{gonzalez2008understanding,edwards2007revisiting}), COVID-19 spreading \cite{gross2020spatio}, human balancing motion \cite{cabrera2004human} and more \cite{metzler2004restaurant, applebaum2004levy}. \textcolor{black}{We stress that, while L\'evy flights are characterized by rare, large jumps that may be called intermittent, the phenomenon of on-off intermittency is distinct from L\'evy flights, in that it specifically arises from multiplicative noise near an instability threshold.} Fluctuations obeying heavy-tailed distributions have also been observed for neuron activity patterns in the human brain \cite{roberts2015heavy}. Moreover, L\'evy walks, a class of random processes similar to L\'evy flights with increments following a heavy-tailed PDF, but with each step taking finite time \cite{shlesinger1986levy,zaburdaev2015levy}, have been proposed as a model of blinking quantum dots in semiconductor nanocrystals \cite{jung2002lineshape,margolin2005nonergodicity}. We highlight that blinking quantum dots and human balancing motion are two examples which exhibit both L\'evy statistics and on-off intermittency. 
Furthermore, very recently, in an idealized model of three-dimensional perturbations in two-dimensional flows, described in the companion paper to this study, \cite{vankan2021levy}, it was found that the perturbation amplitude obeyed equation (\ref{eq:langevin}) with an approximately white noise whose PDF had power-law tails due to the power-law structure of the velocity fields involved. The findings of the companion paper originally motivated the present study and suggested a rationale for numerically observed jump-like growth signals of three-dimensional perturbations in rapidly rotating turbulence \cite{seshasayanan2020onset}. 

A significant body of theoretical literature is devoted to L\'evy flights in potentials, driven by additive L\'evy noise, \cite{jespersen1999levy, chechkin2002stationary,chechkin2003bifurcation, chechkin2004levy,dybiec2007stationary,denisov2008steady,dybiec2010stationary,padash2019first}, as well as to stochastic processes driven by multiplicative L\'evy noise \cite{srokowski2009fractional,srokowski2009multiplicative,la2010dynamics,srokowski2010nonlinear,srokowski2012multiplicative}. For additive noise, it has been shown that L\'evy flights in a quartic or steeper potential possess finite mean and variance, for all parameters of the L\'evy noise \cite{chechkin2003bifurcation}. Many classical problems which are well studied for Gaussian noise have been revisited using L\'evy noise, such as the escape from a potential well \cite{chechkin2005barrier,dybiec2007escape,chechkin2007barrier,koren2007leapover,capala2020levy}, noise-induced transitions and stochastic resonance \cite{zeng2007effects,dybiec2009levy,dybiec2009levy2,xu2013levy,yamapi2019levy}, oscillators under the influence of noise \cite{chechkin2002stationary,sokolov2011harmonic,tanaka2020low}, the Verhulst model \cite{dubkov2008verhulst}, the L\'evy rachet \cite{dybiec2008transport} \textcolor{black}{and Josephson junctions subject to L\'evy noise \cite{guarcello2013role,guarcello2020voltage,guarcello2017anomalous,valenti2014switching,guarcello2016effects,guarcello2019josephson}.} However, despite this impressive body of work, while the impact of colored noise  \cite{ding1995distribution,aumaitre2005low,aumaitre2006effects,aumaitre2007noise,alexakis2012critical,petrelis2012anomalous} and higher dimensions \cite{alexakis2009planar} on on-off intermittency have received attention, the theory of on-off intermittency due to multiplicative L\'evy noise close to an instability threshold has not been studied systematically before, to the best of our knowledge. 

Here, we show theoretically and numerically that for L\'evy white noise, the phenomenology of equation (\ref{eq:langevin}) can differ starkly from the case of  Gaussian white noise. In some cases, the origin never changes stability -- there is no critical point. When there is a critical point, the critical behavior and the properties of on-off intermittency near onset depend non-trivially on the parameters of the L\'evy noise. It is shown that in stationary state a finite or infinite number of integer-order moments may exist, depending on the parameters of the noise.

The remainder of this paper is structured as follows. In section \ref{sec:theory_bg}, we present the theoretical background of this study. In section \ref{sec:lin_theory}, we analyse the linear ($\gamma=0$) regime. In section \ref{sec:nonlinear}, we present analytical results on the nonlinear ($\gamma>0$) statistically stationary state and verify our results against numerical solutions of the stationary fractional Fokker-Planck and Langevin equations. Finally in section \ref{sec:conclusions}, we discuss our results and conclude.
\section{Theoretical background}
\label{sec:theory_bg}
Here, we introduce aspects of the theory of stable PDFs and describe how they are related to L\'evy flights.
\subsection{Properties of $\alpha$-stable probability densities}
\begin{figure}[ht]
    \centering
    \includegraphics[width=8.6cm]{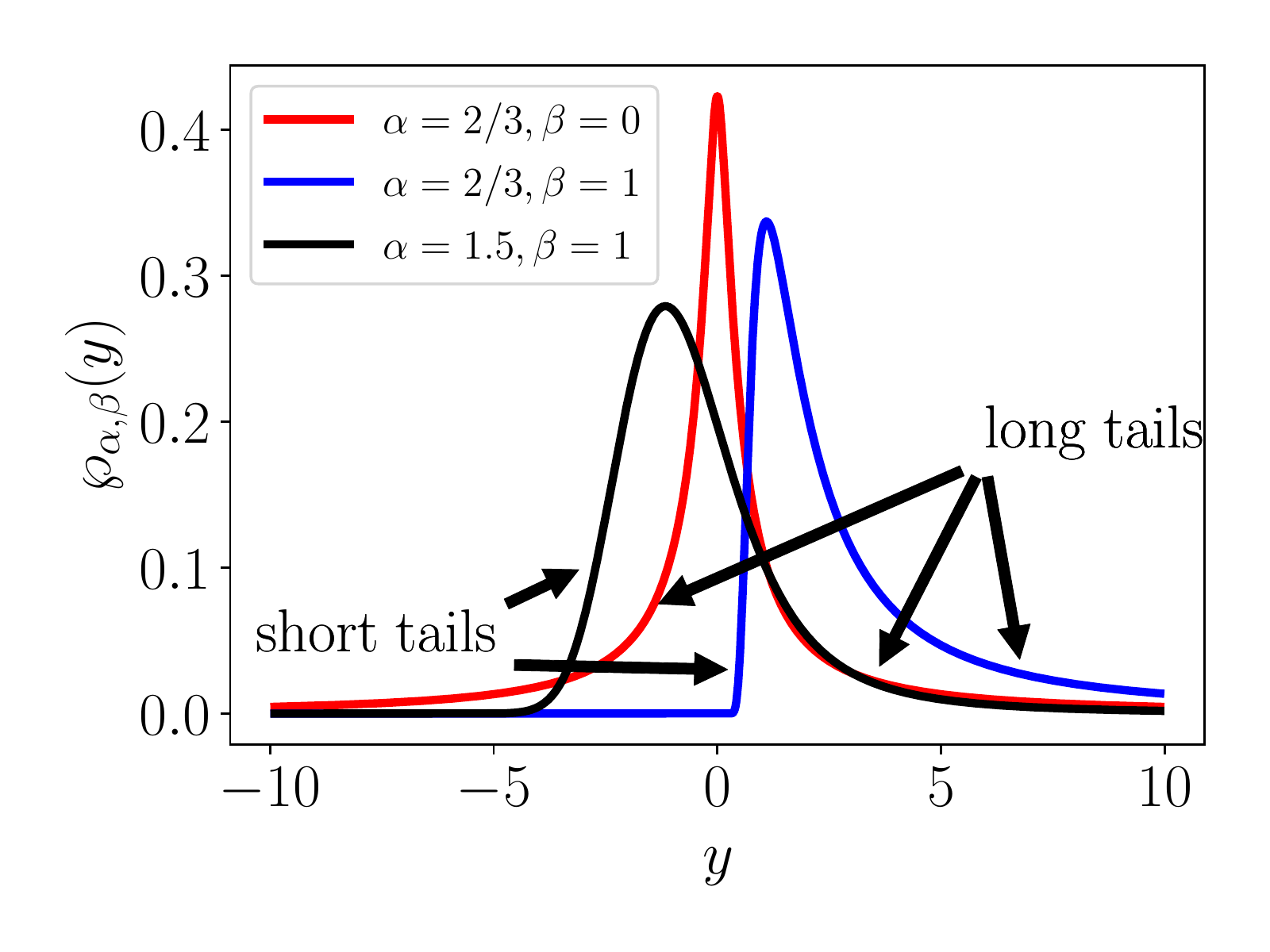}
    \caption{Illustration of long power-law tails and short exponential tails in the stable distributions discussed in the text.}
    \label{fig:ill_stable_dist}
\end{figure}

For parameters $\alpha\in (0,2],\beta\in [-1,1]$, the $\alpha$-stable PDF for a random variable $Y$ is denoted by 
$\wp_{\alpha,\beta}(y)$ and defined by its characteristic function (i.e. Fourier transform),
\begin{equation}
    \varphi_{\alpha,\beta}(k) = \exp\Bigg\lbrace - |k|^\alpha [1-i\beta\mathrm{sgn}(k) \Phi(k)] \Bigg\rbrace, \label{eq:def_stab_cf}
\end{equation}
 with
\begin{equation}
    \Phi(k) = \begin{cases}  \tan\left( \frac{\pi \alpha}{2} \right) \hspace{1.3cm} \alpha \neq 1 \\ - \frac{2}{\pi} \log(|k|) \hspace{1cm} \alpha =1 \end{cases},
\end{equation}
see \cite{uchaikin2011chance}. \textcolor{black}{A standard method for simulating stable random variables is given in \cite{chambers1976method}.} Note that (\ref{eq:def_stab_cf}) is not the most general form possible: there may be a scale parameter in the exponential, which we set equal to one. One refers to $\alpha$ as the stability parameter. For $\alpha=2$, where $\beta$ is irrelevant since $\Phi=0$, one recovers the Gaussian distribution. In the following, we consider $\alpha<2$. The parameter $\beta$, known as the skewness parameter, measures the asymmetry of of the distribution, where $\beta=0$ corresponds to a symmetric PDF, while $|\beta|= 1$ is referred to as maximally skewed. \textcolor{black}{When $\beta\neq0$, the most probable value of $y$, given by the maximum of $\wp_{\alpha,\beta}(y)$, differs from the average valye of $y$, which is equal to zero here when it exists.} We highlight the symmetry relation 
\begin{equation}
    \wp_{\alpha,\beta}(y) = \wp_{\alpha,-\beta}(-y),
    \label{eq:symm}
\end{equation} which follows directly from the definition. Importantly, there are two different possible asymptotic behaviors that a stable distribution can display. When $|\beta| < 1$, there are two long (``heavy") power-law tails, at $y\to \pm \infty$,
\begin{equation}
\wp_{\alpha,\beta}(|y|\to\infty)  
\propto  \lbrace 1+\beta \text{sign}(y)\rbrace  |y|^{-1-\alpha}. \label{eq:long_tails}
\end{equation}
The presence of power-law tails implies that the stable PDF has a finite mean (equal to zero), but a diverging variance for $1<\alpha<2$, while both mean and variance diverge for $\alpha\leq 1$. For $\beta=\pm 1$, the asymptotics given in (\ref{eq:long_tails}) break down on one side. In this case, there is a short exponential tail on the side where the power law breaks down and only a single long power-law tail remains. For $1\leq \alpha<2$, $\wp_{\alpha,\beta=\pm 1}(y)$ 
is supported on $\mathbb{R}$. 
By contrast, for $\alpha<1$ and $\beta=\pm 1$, the probability density is one-sided, with the exponential tail vanishing at the origin, such that $\wp_{\alpha,\beta=1}(y)=0$ at $y\leq 0$ and $\wp_{\alpha,\beta=-1}(y)=0$ at $y\geq 0$, which is consistent with the symmetry (\ref{eq:symm}). Both for $1<\alpha<2$, $\beta=-1$ as $y\to+\infty$, and for $\alpha<1$, $\beta=1$ as $y\to 0^+$, the leading-order asymptotic form of the short tail of the stable PDF can be obtained by Laplace's method and is given by 
\begin{equation}
    \wp_{\alpha,\beta}(y) \sim c_0 y^{\frac{1-\alpha/2}{\alpha-1}} \exp\left(-c_1 y^{\frac{\alpha}{\alpha-1}}\right), \label{eq:short_tails} 
\end{equation}
where $c_0$, $c_1$ are positive, $\alpha$-dependent constants, cf. theorem 4.7.1 in \cite{uchaikin2011chance}. Note that this reduces to a Gaussian when $\alpha=2$. By the symmetry of eq. (\ref{eq:symm}), the same result holds, with $y$ replaced by $-y$, for $1<\alpha<2$, $\beta=+1$ as $y\to-\infty$ and at $\alpha<1$, $\beta=-1$ as $y\to 0^-$.
The different behaviors are illustrated for three cases in figure \ref{fig:ill_stable_dist}. Unfortunately, useful explicit expressions for the stable PDF only exist in a small number of special cases.
\subsection{L\'evy flights and the space-fractional Fokker-Planck equation}
Consider the Langevin equation (\ref{eq:langevin}) with $f(t)$ being white ``L\'evy" noise. \textcolor{black}{More precisely, for a given time step $dt$, we let $f(t)dt = dt^{1/\alpha}F(t),$ where $F(t)$ obeys the alpha-stable PDF $\wp_{\alpha,\beta}(F)$, defined by (\ref{eq:def_stab_cf}), and is drawn independently for any time $t$,} \cite{dubkov2008levy}. 
Since the Langevin equation (\ref{eq:langevin}) involves a multiplicative noise term, one needs to decide on an interpretation thereof. As has been discussed in the literature,  \cite{srokowski2009multiplicative,srokowski2012multiplicative}, like in the Gaussian case, the two standard interpretations are the Stratonovich \cite{stratonovich1966new} interpretation, which preserves the rules of standard calculus, and the non-anticipating It\^o \cite{ito1944stochasticintegral} interpretation. According to the choice of interpretation, the probability density will be governed by a different form of the (space-)fractional Fokker-Planck equation (FFPE), so called since it involves fractional derivatives in the state variable. First consider the Stratonovich interpretation, such that $Y=\log(X)$ obeys the following equation with additive noise,
\begin{equation}
    \frac{dY}{dt} = \mu - \gamma e^{2Y} + f(t), \label{eq:langevin_y}
\end{equation}
which says that $Y(t)$ performs a L\'evy flight in the potential $V(Y) = -\mu Y + \frac{\gamma}{2} e^{2Y}$. The density associated with $Y(t)$, denoted by $p_y(y,t)$, then obeys the FFPE
\begin{equation}
    \partial_t p_y(y,t)= - \partial_y\left[\left(\mu - \gamma e^{2y}\right)p_y(y,t)\right] + \mathcal{D}^{\alpha,\beta}_y p_y(y,t), \label{eq:ffpe_stratonovich}
\end{equation}
\cite{denisov2009generalized}, where the fractional derivative operator
\begin{equation}
    \mathcal{D}_y^{\alpha,\beta}g(y)  = - \frac{ (1+\beta)D_+^\alpha g(y) + (1-\beta)D_-^\alpha g(y)}{2\cos\left(\frac{\alpha \pi}{2}\right)},
\end{equation}
for an arbitrary function $g(y)$, is known as the Riesz-Feller fractional derivative of order $\alpha$ and skewness $\beta$ \cite{Mainardi_thefundamental}. It can be expressed in terms of the left and right Riemann-Liouville fractional derivatives, which for $1 < \alpha < 2$ are given by \cite{chechkin2004levy,samko1993fractional},
\begin{equation}
    (D_+^\alpha g)(y) = \frac{1}{\Gamma(2-\alpha)} \frac{d^2}{d y^2}\int_{-\infty}^y \frac{g(z)dz}{(y-z)^{\alpha-1}} \label{eq:Dplus_riemann_derivative}
\end{equation}
and
\begin{equation}
    (D^\alpha_- g)(y) = \frac{1}{\Gamma(2-\alpha)} \frac{d^2}{d y^2}\int_{y}^\infty \frac{g(z)dz}{(z-y)^{\alpha-1}}.\label{eq:Dminus_riemann_derivative}
\end{equation}
For $0<\alpha<1$, the definitions are similar \cite{samko1993fractional}, 
\begin{equation}
    (D_+^\alpha g)(y) = + \frac{1}{\Gamma(1-\alpha)} \frac{d}{dy} \int_{-\infty}^y \frac{g(z)dz}{(y-z)^\alpha},  \label{eq:Dplus_riemann_derivative_alpha_lt_1}
\end{equation}
and
\begin{equation}
    (D_-^\alpha g)(y) = - \frac{1}{\Gamma(1-\alpha)} \frac{d}{dy} \int_y^\infty  \frac{g(z)dz}{(z-y)^\alpha}. \label{eq:Dminus_riemann_derivative_alpha_lt_1}
\end{equation}
For $\alpha=2$, one has $\mathcal{D}^{\alpha,\beta}_y=\partial_y^2$. The Riemann-Liouville fractional derivatives have a simple Fourier transform, $\mathcal{F}[D_\pm^\alpha f](k)= (\pm i k)^\alpha \mathcal{F}[f](k)$, see chapter 7 of \cite{samko1993fractional}, which is often invoked. However, our analysis will be performed mostly in physical space. 
Once the solution to equation (\ref{eq:ffpe_stratonovich}) is known, then the probability density $p_x(x,t)$ associated with the original variable $X(t)$ is given by
\begin{equation}
    p_x(x,t) = \frac{1}{x} p_y(\log(x),t).
\end{equation}
\text{ }\\
If, instead of the Stratonovich interpretation, one adopts the It\^o interpretation, then the FFPE is given by
\begin{equation}
    \partial_t p_x(x,t) = - \partial_x \left[ (\mu x - \gamma x^3)p_x(x,t) \right] + \mathcal{D}_x^{\alpha,\beta} (x^\alpha p_x(x,t)), \label{eq:ffpe_ito}
\end{equation}
as derived in \cite{denisov2009generalized}. \\

We continue in the Stratonovich interpretation. In the absence of nonlinearity, when $\gamma=0$, one can 
solve in Fourier space for a delta-peaked initial condition, e.g. $X(0)=1$, which leads to the fundamental solution given in \cite{Mainardi_thefundamental},
\begin{equation}
    p_x(x,t)  
    = \frac{\wp_{\alpha,\beta}\left(\frac{\log(x)-\mu t}{t^{1/\alpha}}\right)}{t^{1/\alpha} x} \label{eq:logstable_pdf}
\end{equation}
where $\wp_{\alpha,\beta}(\cdot)$ is the $\alpha$-stable PDF whose Fourier transform is given in eq. (\ref{eq:def_stab_cf}). The corresponding cumulative probability distribution (CDF) is
\begin{equation}
    P(x<\chi) = \mathcal{P}_{\alpha,\beta}\left(\frac{\log(\chi)-\mu t}{t^{1/\alpha}}\right), \label{eq:logstable_cdf}
\end{equation}
in terms of the $\alpha$-stable CDF $\mathcal{P}_{\alpha,\beta}(z) = \int_{-\infty}^z \wp_{\alpha,\beta}(z')dz'$.
Clearly, equation (\ref{eq:logstable_pdf}) holds for the Gaussian case of $\alpha=2$, the familiar log-normal distribution. By analogy with the latter, for $0<\alpha<2$ the PDF in eq. (\ref{eq:logstable_pdf}) is known as the {log-stable PDF} and the associated process as the {\it log-stable process}. For $0<t<\infty$, the moments of the log-stable PDF are only finite for $\beta=-1$. \textcolor{black}{This is because it is the only case where the $\alpha$-stable PDF does not have a heavy tail of the form (\ref{eq:long_tails}) at $+\infty$. When a heavy tail is present ($\beta>-1$), then averaging over $e^{ny}=x^n$ for any $n>0$ does not give a finite result. 
For this reason the associated stochastic process with $\beta=-1$ is also known as the {\it finite-moment log-stable process}. It is well known, in particular in finance, see \cite{carr2003finite} (there, only $1<\alpha< 2$ is considered).} 
\section{Linear theory}
\label{sec:lin_theory}
Here we study the late-time limit of solution  (\ref{eq:logstable_pdf}) corresponding to eq. (\ref{eq:langevin}) with $\gamma=0$, starting from a localized initial condition at $x>0$, to determine the stability of the origin $x=0$. 
This will be helpful later for interpreting the nonlinear 
$(\gamma> 0)$ results.
\subsection{The Gaussian case}
First, for illustration, consider the Gaussian case $\alpha=2$ in (\ref{eq:logstable_pdf}), which gives the log-normal PDF for $X$
\begin{equation}
    p_x(x,t) = \frac{1}{x\sqrt{2\pi t}} \exp\left(\frac{-(\log(x)-\mu t)^2}{2t}\right).
\end{equation}
The probability $P(x<\chi)$ to find the system at $x<\chi$ after time $t$ is given by the CDF in eq. (\ref{eq:logstable_cdf}), which here equals
\begin{equation}
    P(x<\chi) = \frac{1}{2} \left[1+\mathrm{erf}\left(\frac{\log(\chi)-\mu t}{\sqrt{2t}}\right)\right],
\end{equation}
where $\mathrm{erf}(x)$ is the error function. Considering the limit of late times $t\to \infty$ for fixed $\chi$, using that $\mathrm{erf}(x\to\pm \infty) = \pm 1$, one deduces that $P(x<\chi)\to 1$ if $\mu<0$, while $P(x<\chi)\to 0$ if $\mu >0$. This indicates that at $\mu=0$ the origin $x=0$ goes from asymptotically stable to unstable.

 Alternatively, one might attempt to determine the stability of the origin by studying the moments of $X$ as a function of time. For $\alpha=2$, the FFPE in (\ref{eq:ffpe_stratonovich}) reduces to the the ordinary Fokker-Planck equation
\begin{equation}
    \partial_t p_y(y,t) =  - \mu \partial_y p_y(y,t) + \partial_y^2 p_y(y,t).
\end{equation}
Multiplying by $\exp(n y)=x^n$ and integrating over $y$, one arrives, upon integrating by parts, at the relation
\begin{equation}
    \partial_t \langle X^n \rangle = \left( \mu n+ n^2\right) \langle X^n \rangle,
\end{equation}
which implies that 
\begin{equation} 
\langle X^n(t)\rangle = X_0^n e^{\lambda_n(\mu) t}, \label{eq:gaussian_moments_growth}
\end{equation}
for an initial condition $X(0)=X_0$, with the growth rate $\lambda_n(\mu)=(n \mu  + n^2)$. Importantly, the value of $\mu$ where the growth rate of $\langle X^n\rangle $ vanishes, denoted $\mu_c(n)$, depends on $n$ and is given by $\mu_c(n)=-n$. We have shown based on the CDF that the system is stable for $\mu<0$. However, equation (\ref{eq:gaussian_moments_growth}) indicates that for $n$ large enough, $\langle X^n\rangle$ grows exponentially in time even for $\mu<0$. This is due to rare transient excursions to large $y$, which give a non-negligible contribution since $e^{ny}$ is large. Thus, the moments are not the correct indicator for stability in the system (\ref{eq:langevin}) with $\gamma=0$ and one needs to be careful when concluding stability based on them. However, as discussed in \cite{seshasayanan2018growth}, the limit of $\mu_c(n)$ as $n\to 0$ does indicate the correct threshold, namely $\mu = 0$. This is because that limit is related to the growth of $\langle \log(X(t)) \rangle$, which weighs large-$X$ contributions less strongly.

\subsection{The general $\alpha$-stable case: moments}

In the general $\alpha$-stable case the solution is the log-stable  distribution given in (\ref{eq:logstable_pdf}). When $\beta>-1$,  moments $\langle X^n \rangle$ diverge for any $n>0$, as described above. 
Thus no stability criterion can be derived based on the moments.

For the special case $\beta=-1$, the moments $\langle X^n \rangle$ exist and can be calculated. While the moments have been given in the literature before in the It\^o interpretation, see \cite{carr2003finite}, we give a novel (to our knowledge) derivation in the Stratonovich interpretation. For $\beta=-1$ the FFPE reads 
 \begin{equation}
     \mathcal{D}_y^{\alpha,\beta=-1} f(y) = - \sec(\alpha \pi/2) D_-^\alpha f(y), \label{eq:fract_ibp}
 \end{equation}
 with $D_-^\alpha$ given by (\ref{eq:Dplus_riemann_derivative}). Following the steps made in the Gaussian case, we multiply (\ref{eq:ffpe_stratonovich}) by $e^{ny}$ and integrate over $y$. Fractional integration by parts obeys
\begin{equation}
    \int_{-\infty}^\infty f(y)(D_+^\alpha g)(y) dy = \int_{-\infty}^\infty (D_-^\alpha f)(y) g(y) dy,
\end{equation}
for sufficiently well-behaved functions $f$ and $g$ such that the fractional derivatives and integrals exist, \cite{samko1993fractional}. Here, this requires $\beta=-1$. Furthermore, note that
\begin{equation}
    D_+^\alpha (e^{ny}) =  n^\alpha e^{ny}, \label{eq:frac_deriv_exp}
\end{equation}
 which for $0<\alpha<1$ and $1<\alpha<2$ follows directly from the definition of $D_+^\alpha$ in eqs. (\ref{eq:Dplus_riemann_derivative}), (\ref{eq:Dplus_riemann_derivative_alpha_lt_1}) upon changing integration variables to $u=y-z$. One obtains
\begin{equation}
    \partial_t \langle X^n\rangle = \left[n\mu - \sec(\pi\alpha/2) n^\alpha\right] \langle X^n \rangle,
\end{equation}
such that 
\begin{equation}
    \langle (X(t))^n\rangle = X_0^n e^{\lambda_n^S t} \label{eq:moments_fmlsp},
\end{equation}
with
\begin{equation}
    \lambda_n^S = \left[n\mu -\sec(\alpha \pi/2) n^\alpha\right].
\end{equation}
Hence, the value of $\mu$ where the growth rate vanishes depends on $n$,
 \begin{equation}
     \mu_{c}(n)= \sec(\alpha\pi/2) n^{\alpha-1}. \label{eq:mu_crit_alpha}
 \end{equation}
 
For $\alpha=2$, this reduces to the Gaussian result. For completeness, we note that the growth rate in the It\^o interpretation given in \cite{carr2003finite} is similar (see their eq. (8)),

     \begin{equation}
     \lambda_n^{I}=\lambda_n^{S}+n\sec(\alpha \pi/2).\end{equation}
     
     We have verified equation (\ref{eq:moments_fmlsp}) for both $\alpha>1$ and $\alpha<1$ by computing the moments of the exact solution (\ref{eq:logstable_pdf}) numerically (not shown). 
     %
     However, the moments which we just computed for $\beta=-1$ are ill-suited for studying the linear stability problem. This is  because, as in the Gaussian case, the moments are dominated
     by rare large-amplitude events. \textcolor{black}{However, taking the limit $n\to 0$ in $\mu_c(n)$ following \cite{seshasayanan2018growth}, where large amplitudes are weighted less strongly, one predicts the threshold to be at $\mu=0$ for $\alpha>1$ and at $\mu=\infty$ for $\alpha<1$.} In the following section, we consider the CDF of the log-stable process to deduce the asymptotic stability of the origin and show in particular that the $n\to0$ predictions are correct. 

     
\subsection{The general $\alpha$-stable case: the CDF}
\label{sec:LinCDF}

\begin{figure}
    \centering
    \includegraphics[width=8.6cm]{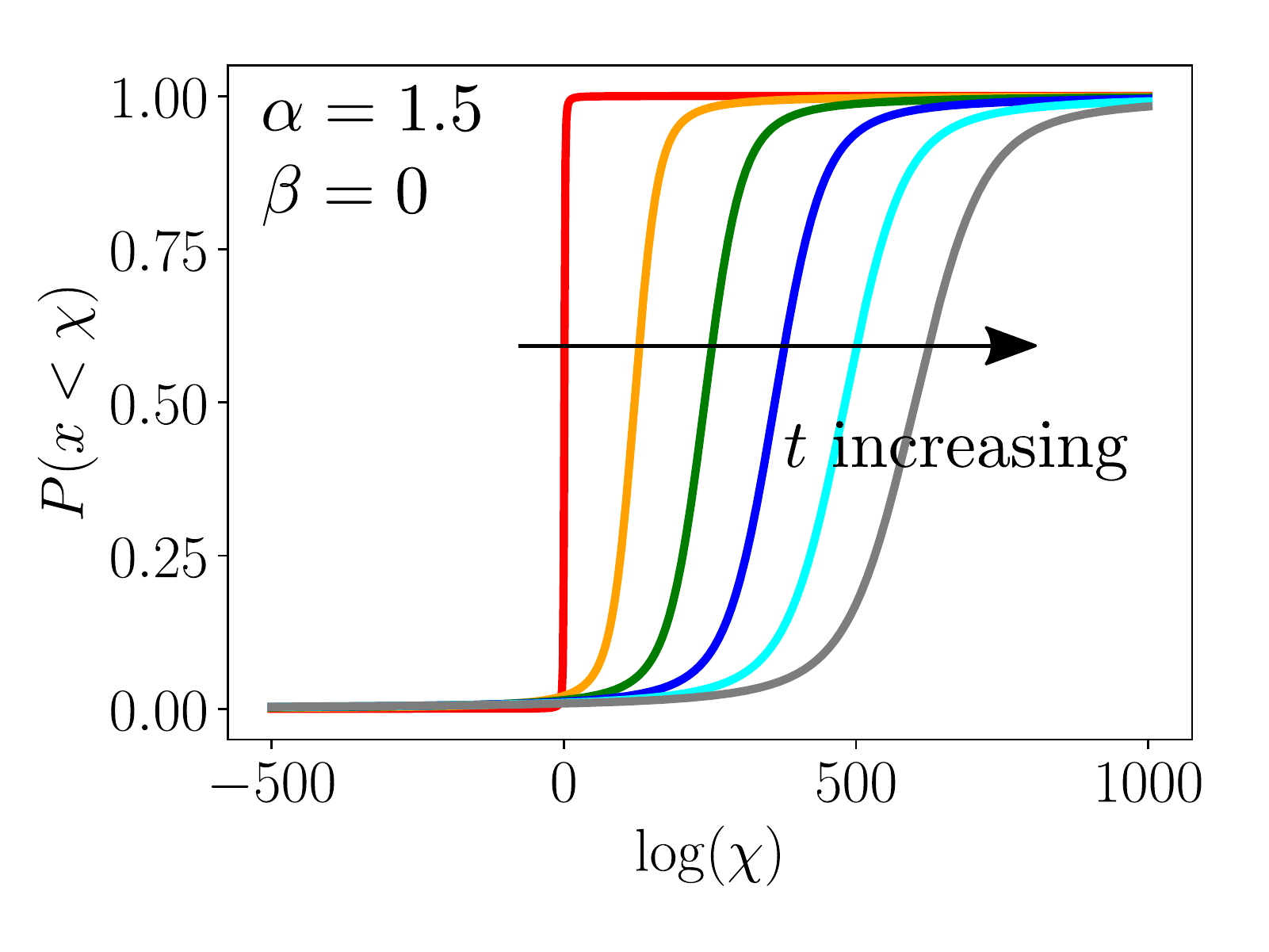}
    \includegraphics[width=8.6cm]{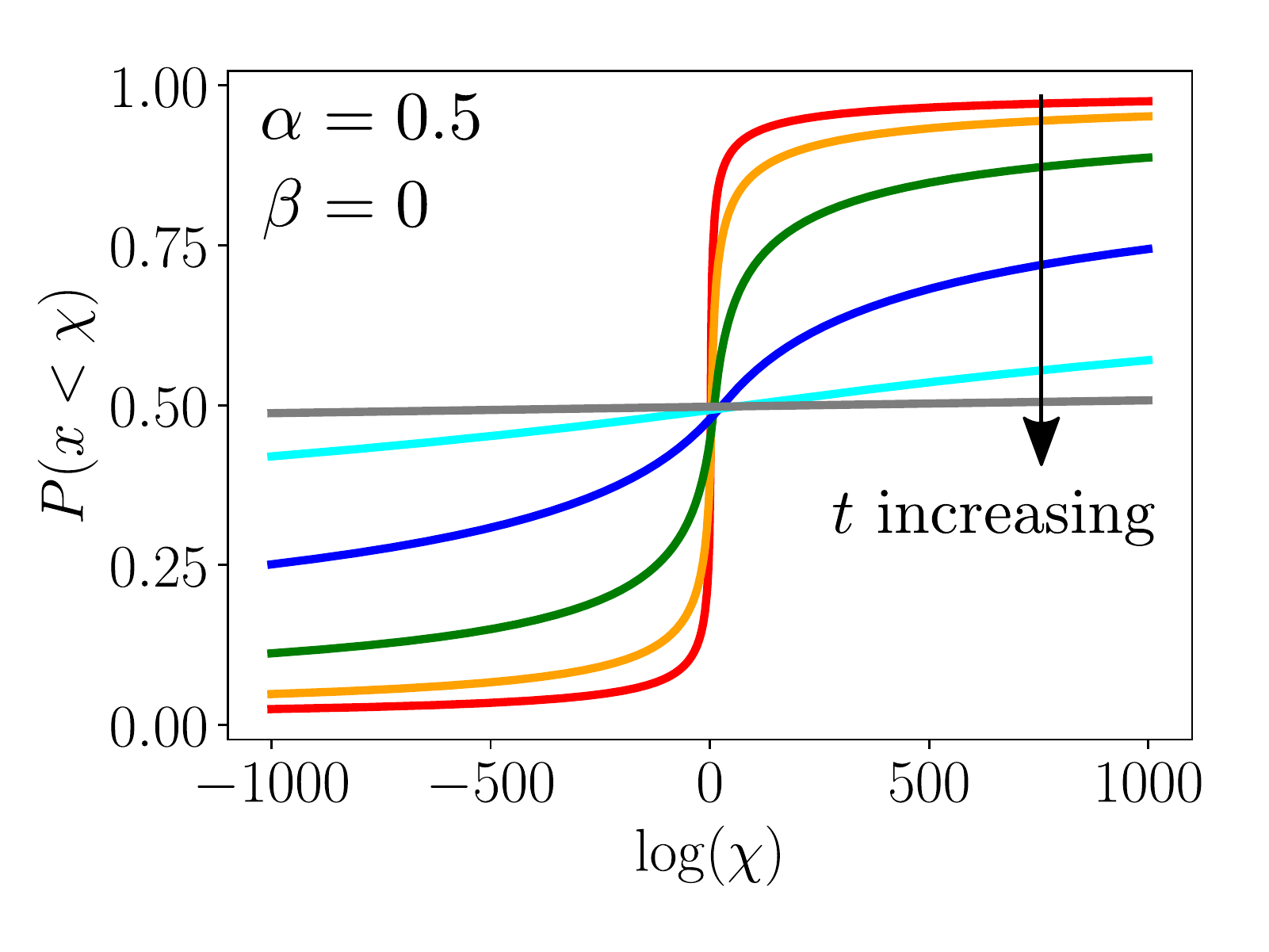}
    \caption{CDF of log-stable law (\ref{eq:logstable_pdf}) for $\beta=0$, $\alpha=1.5$ (top panel) and $\alpha=0.5$ (bottom panel) with $\mu=1$ and time $t$ increasing in the order red, orange, green, blue, cyan, grey. Clearly, the CDF shifts to the right as time increases in the top panel indicating that probability is leaking to $+\infty$, but takes the constant value $0.5$ in the bottom panel, indicating that the probability leaking to both $+\infty$ and $-\infty$.}
    \label{fig:ill_logstable_cdf_late_time}
\end{figure}

Consider the log-stable CDF given in eq. (\ref{eq:logstable_cdf}). Figure \ref{fig:ill_logstable_cdf_late_time} shows the time evolution of the CDF for $\beta=0$ and $\alpha=1.5$ (top panel), $\alpha=0.5$ (bottom panel), both for $\mu=1$. 
One observes that for $\alpha>1$, probability shifts to the right due to the drift, indicating leakage to positive infinity. Conversely, for $\alpha<1$, the CDF approaches a constant value, strictly larger than zero and strictly smaller than one, indicating that probability is leaking to both positive and negative infinity. 

\NOTE{More precisely,
for $1<\alpha<2$ and $\beta<1$, $\mu>0$ one may use eqns. (\ref{eq:long_tails}) and (\ref{eq:logstable_cdf}) to show that at late times ($t\to\infty$), for any given $\chi$, the probability for $x<\chi$, is given by}
\begin{equation}
    P(x<\chi) \propto \frac{(1-\beta)t}{|\mu t - \log(\chi)|^\alpha} \propto (1-\beta)t^{1-\alpha}.
\end{equation}
\NOTE{
Thus $P(x<\chi)$
 decreases as time progresses, in agreement with our conclusion based the top panel of figure \ref{fig:ill_logstable_cdf_late_time}. A similar argument for $\mu<0$ and the same range of $\alpha$ shows that in this case $P(x>\chi)$ decreases in time.}
 
 For $\beta=1$ and the same range of $\alpha$, taking the same limit, for $\mu>0$, $t\to \infty$ and $\chi$ fixed, one finds using (\ref{eq:short_tails}) and (\ref{eq:logstable_cdf}) that 
 \begin{align}
    P(x<\chi) &\propto 
t^\frac{1-\alpha}{2\alpha} 
e^{-c_1 \mu^\frac{\alpha}{\alpha-1} t},
\end{align}
which also decays, this time exponentially fast, as $t$ increases. Similarly for $\mu<0$ and the same range of $\alpha$, one can show that $P(x>\chi)$ decreases in time.
In short, we find that for any $\alpha$ in the range $1<\alpha<2$, the probability leaks to $\log(x)\to\mathrm{sign}(\mu) \infty$ as $t\to \infty$ for the linear ($\gamma=0$) problem. \\

If $0<\alpha<1$, then for any $\mu$ and fixed $\chi$ as $t\to \infty$, the argument of the CDF in (\ref{eq:logstable_cdf}), $(\log(\chi)-\mu t)t^{-1/\alpha}\to 0$, such that
\begin{equation}
P(x<\chi) \to \mathcal{P}_{\alpha,\beta}(0),
\end{equation}
 where the right-hand side is the $\alpha$-stable CDF evaluated at zero, which is a $\mu$-independent constant. For $\beta=0$, the constant is $0.5$ by symmetry, as illustrated in figure \ref{fig:ill_logstable_cdf_late_time}, but in general, it will depend on $\beta$ in a continuous way. In particular, for $\beta=1$, $\mathcal{P}_{\alpha,\beta=1}(0)=0$, since the stable PDF is only supported at positive values in this case. On the other hand, for $\beta=-1$, the constant is $\mathcal{P}_{\alpha,\beta=-1}(0)=1$, since the PDF is only supported at negative values. In short, we find that for any $\alpha$ in the range $0<\alpha<1$ 
 the probability leaks to both $\log(x)\to-\infty$ and $\log(x)\to+\infty$, with the exceptions of $\beta=\pm1$, where probability leaks to $\log(x)\to \beta \infty$. 
 \\
In the marginal case $\alpha=1$, the fact that $(\log(\chi)-\mu t)/t \to -\mu$ for any fixed $\chi$ implies
\begin{equation}
    P(x<\chi) \to \mathcal{P}_{\alpha=1,\beta}(-\mu),
\end{equation}
where the right-hand side is the $\alpha$-stable CDF evaluated at $-\mu$, which is a positive constant for any finite $\mu$ and any $\beta\in[-1,1]$. Hence, at $\alpha=1$, probability leaks to both $\log(x)\to -\infty$ and $\log(x)\to \infty$ for all $\mu$. Only the fraction of the weight escaping in each direction depends on $\mu$.

We note that all of the results obtained above from the exact linear ($\gamma=0$) solution can be understood in terms of a competition between the drift $\mu t$ and the widening of the PDF, which goes as $t^{1/\alpha}$. For $\alpha>1$, the drift is dominant over the widening and probability leaks to $\log(x) \to\mathrm{sign}(\mu) \infty$. On the other hand, for $0<\alpha \leq 1$, the drift no longer dominates and probability spreads out to both $\log(x) \to \pm \infty$, except for one-sided noise. 

In summary, translating the results back to the original variable $x$, we have shown that in the log-stable process, for $1<\alpha<2$, for any $\beta\in [-1,1]$, all the probability leaks to $x\to+\infty$ for $\mu>0$, while for $\mu<0$ all the probability accumulates at the origin $x=0$. On the other hand, for $0<\alpha<1$, the probability leaks both $x\to 0$ and to $x\to \infty$ independently of $\mu$, except for one-sided noise at $\beta=\pm 1$. There, all the probability leaks to $x=0$ for $\beta=-1$ and to $x\to\infty$ for $\beta=1$. At $\alpha=1$, the probability leaks to both $x=0$ and $x=\infty$, independently of $\beta$ and $\mu$. Table \ref{tab:summary_linear_solution} summarizes the late-time behavior of the linear solution.

Finally, we point out that in the only case where the moments exist, at $\beta=-1$, they do not straightforwardly indicate asymptotic stability. For $1<\alpha<2$, and $\mu<0$, the origin is stable. Yet, moments of sufficiently high order will grow. For $0<\alpha\leq 1$, the origin is stable independently of $\mu$, but there also, high-order moments grow. However, taking the moment order $n\to 0$ predicts the correct thresholds $\mu=0$ for $\alpha>1$ and $\mu=\infty$ (no instability at any finite $\mu$) for $\alpha<1$.  
\begin{table}[]
    \centering
    \begin{tabular}{|c|*{3}{c|}}\hline
    \backslashbox[10mm]{$\alpha$}{$\beta$}
     &{$-1$}&{$(-1,1)$}&{$1$}\\\hline
     $(1,2]$ &$\mathrm{sign}(\mu) \infty$& $\mathrm{sign}(\mu) \infty$& $\mathrm{sign}(\mu) \infty$\\\hline
    $1$ &$+\infty \, \& -\infty$ &$+\infty \,\& -\infty$&$+\infty \, \& -\infty$\\\hline
    $(0,1)$ &$-\infty$&$+\infty\, \& -\infty$& $+\infty$\\\hline
    \end{tabular}
    \caption{\textcolor{black}{
Summary of the late-time behavior of the (linear) log-stable process (\ref{eq:logstable_pdf}). For a given combination of $\alpha$ and $\beta$, it is indicated where the weight of the probability will leak to in terms of the variable $Y=\log(X)$.}}
    \label{tab:summary_linear_solution}
\end{table}

\noindent

\begin{figure}
    \centering
    \includegraphics[width=8.6cm]{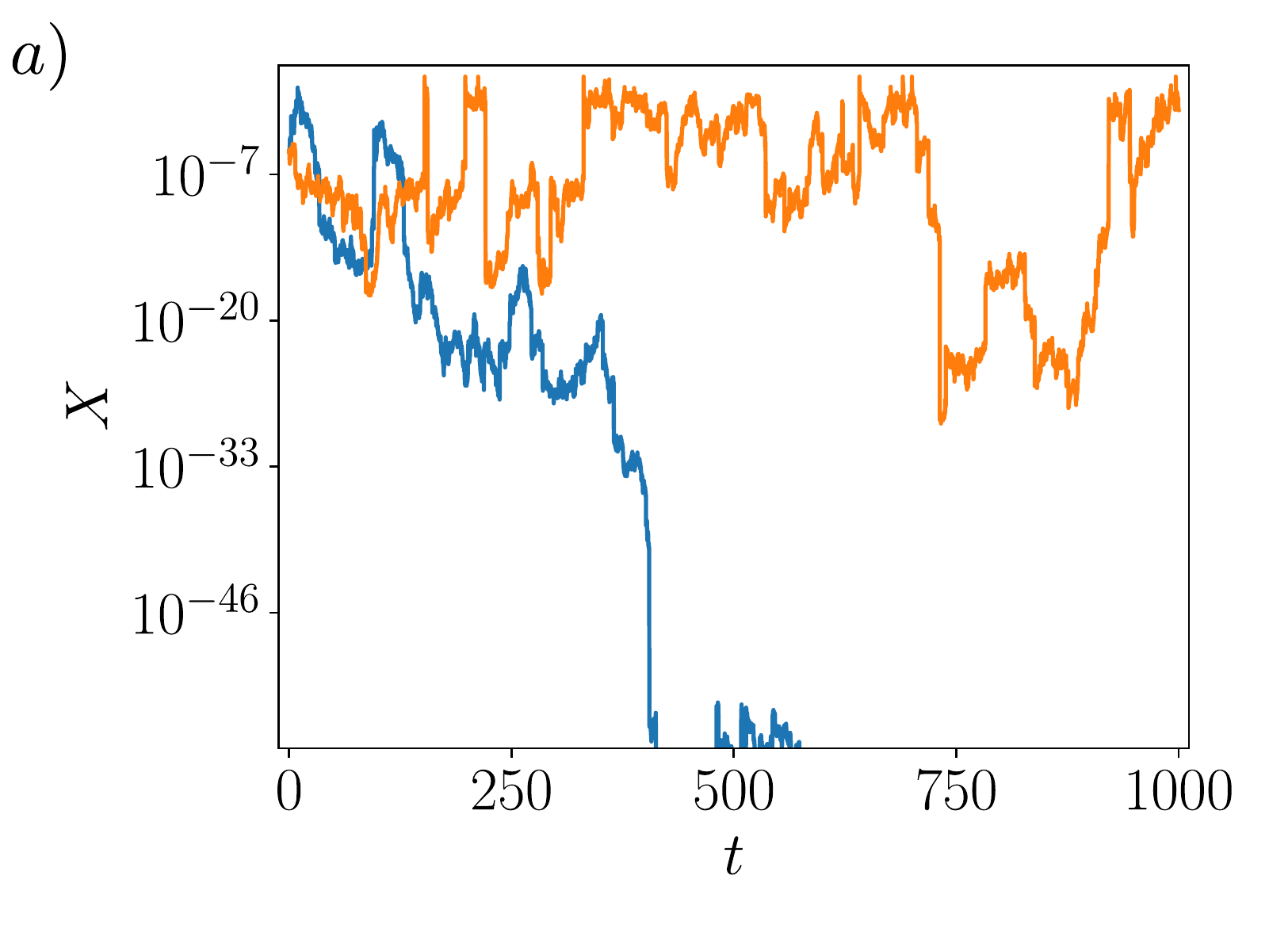}      \includegraphics[width=8.6cm]{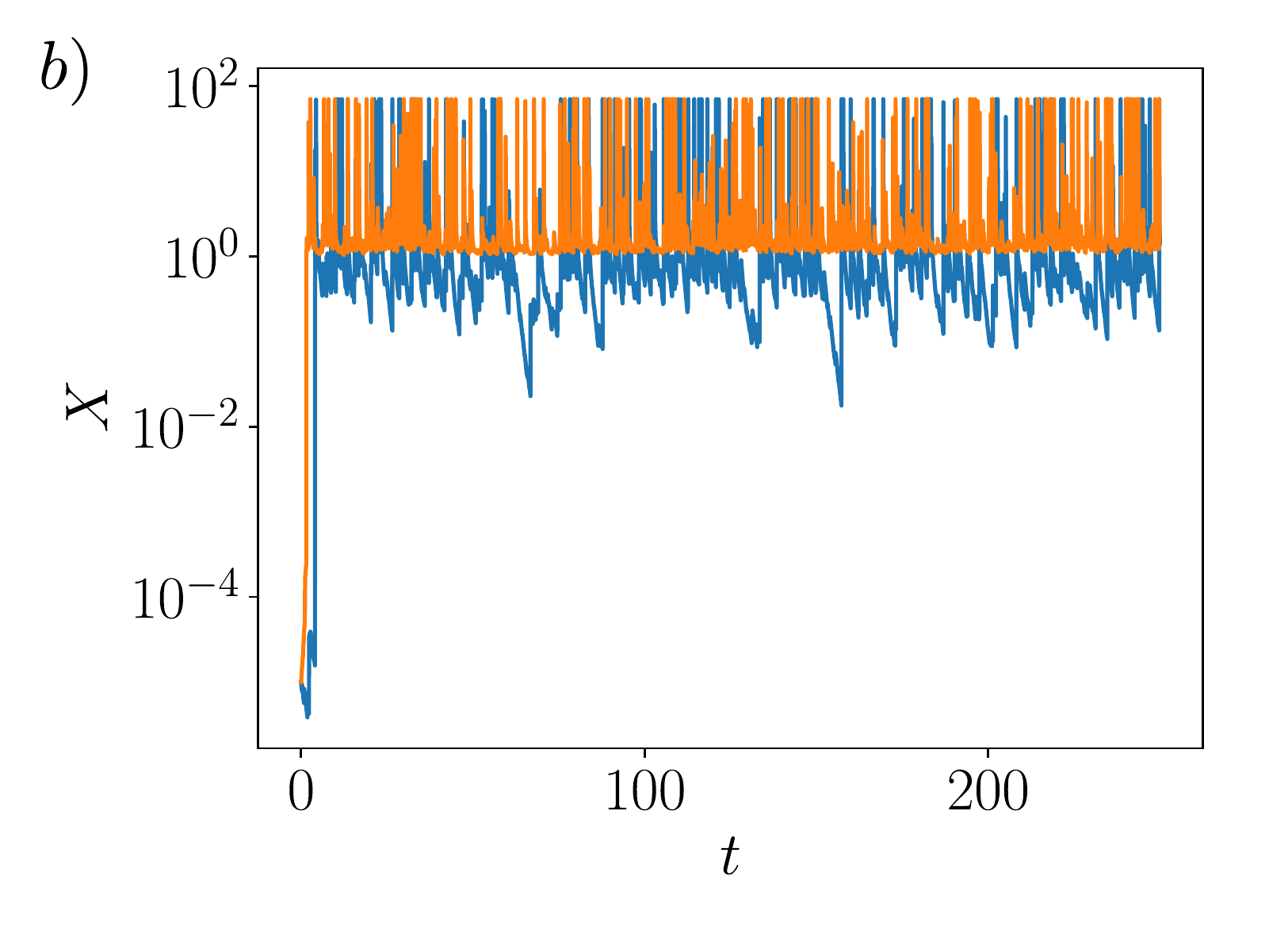}      \includegraphics[width=8.6cm]{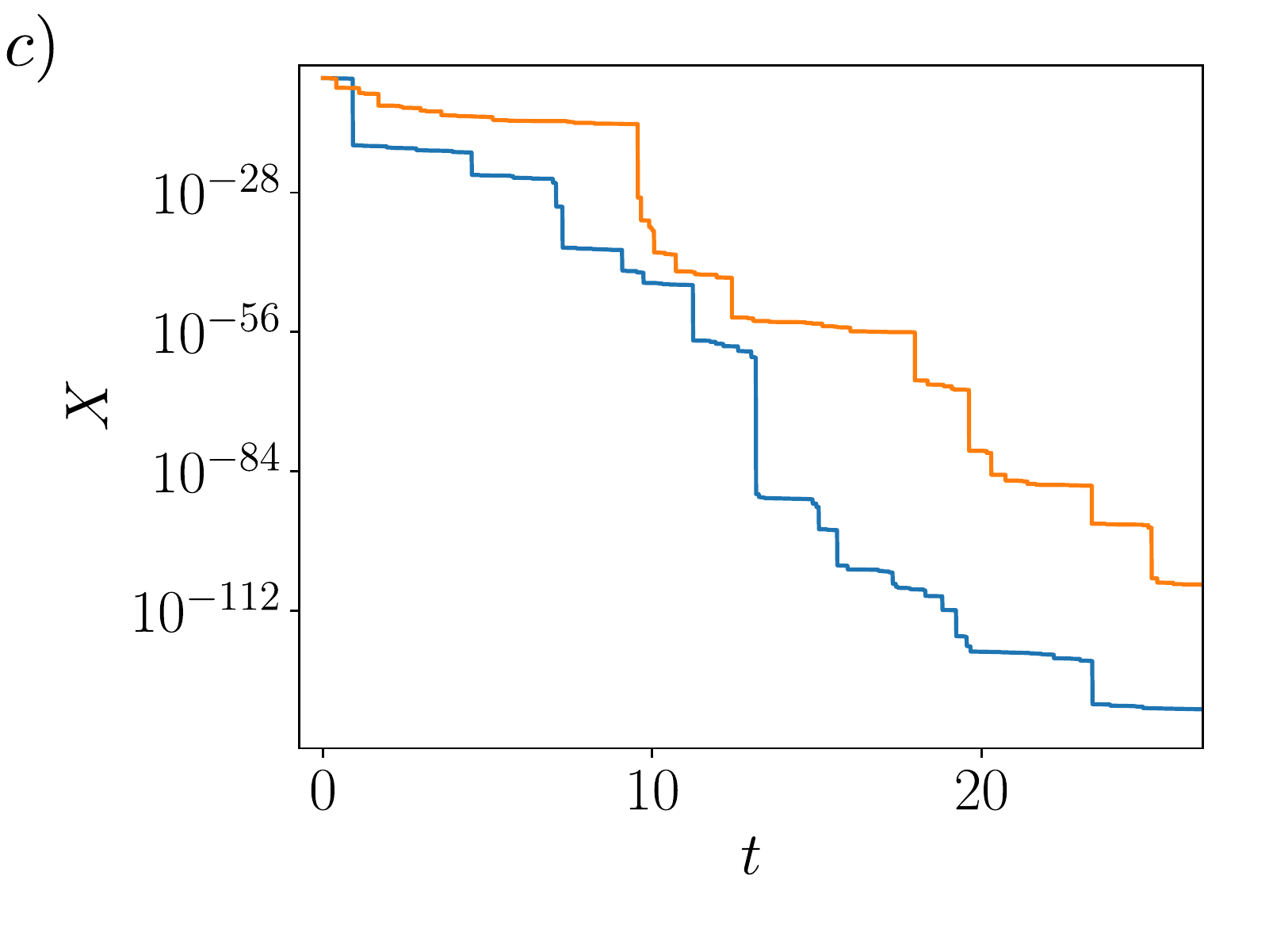}
    \caption{Semi-logarithmic plots of time series $X(t)$ (see eq. (\ref{eq:langevin})). Panel a): $\alpha=1.5$, $\beta=0$, $\gamma=1$. Orange (top): $\mu=0.2$ -- $X(t)$ varies over 20 orders of magnitude, displaying on-off intermittency. Blue (bottom): $\mu=-0.2$ -- $X$ decays to zero. A critical transition occurs between the two, at $\mu=0$. Panel b): $\alpha=0.5$, $\beta=1$, $\gamma=1$. Orange (top): $\mu=1$. Blue (bottom): $\mu=-1$. The origin is unstable for all $\mu$. Panel c): same as panel b) but with $\beta=-1$. Here, the origin is stable for all $\mu$.}
    \label{fig:ill_ts}
\end{figure}
\section{Nonlinear theory}
\label{sec:nonlinear}
In this section, we study the effect of the nonlinear term in equation (\ref{eq:langevin}) with $\gamma>0$
in the development of the instability. 
The nonlinearity will prevent the leakage of probability to $x\to\infty$ that was observed for many cases in the linear regime, thus leading to a stationary distribution that we try to estimate here.

For illustration, typical solutions of the nonlinear Langevin equation (\ref{eq:langevin}) are shown in figure \ref{fig:ill_ts}. The realizations are generated efficiently by integrating eq. (\ref{eq:langevin}) using its exact solution given in appendix \ref{sec:app_exact_sol}. Three different cases are shown: in panel a) $\alpha=1.5$, $\beta=0$, in panel b) $\alpha=0.5,\beta=1$, in panel c) $\alpha=0.5,\beta=-1$. For each case, two typical time series are shown, one for positive $\mu$ and one for negative $\mu$ at fixed $\gamma=1$. In panel a), at negative $\mu$, $X$ decays to zero. At positive $\mu$, there is on-off intermittency: $X$ fluctuates over many orders of magnitude, but does not decay. There is a qualitative change of behavior between $\mu>0$ and $\mu<0$. Typical trajectories at $\beta\neq 0$, $1<\alpha<2$ resemble those in panel a). In panel b), the origin is unstable for both positive and and negative $\mu$. In panel c) the origin is stable for both positive and negative $\mu$. 

\subsection{Exact relation for the second moment}
One important property of equation $(\ref{eq:langevin})$ is that if $\langle f(t) \rangle$ exists (i.e. for $1<\alpha\leq 2$), then for $X>0$ it implies that
\begin{equation}
    \frac{d}{dt}\langle \log(X)\rangle = \mu - \gamma \langle X^2\rangle + \langle f(t) \rangle.\label{eq:evol_mean_log}
\end{equation}
\NOTE{ 
Assuming $\langle f(t)\rangle=0$, then for $\gamma>0,\mu< 0$ the right-hand side is negative, resulting in
\beq \langle \log(X)\rangle \leq \mu t,  \eeq   
which tends to $-\infty$ as $t\to\infty$. However, if $\mu>0$, then a stationary state is reached for which $d\langle \log(X)\rangle/dt=0$ and the second moment satisfies
\begin{equation}
    \langle X^2\rangle = \mu/\gamma.\label{eq:meanx2_id}
    \end{equation}}
\NOTE{
By contrast with the linear regime, for which it was shown above that moments are not a reliable indicator of stability, 
moments in the nonlinear regime are pertinent to the stability of the origin. 
This is due to the fact that the nonlinearity in equation (\ref{eq:langevin}) impedes excursions to large amplitudes, which are the reason why high moments may grow exponentially in the linear case ($\gamma=0$), even when the origin is stable.  Equation (\ref{eq:meanx2_id}) thus already indicates that for $1<\alpha\le2$ 
the system is unstable when $\mu>0$ in agreement with the predictions of section \ref{sec:LinCDF}. \textcolor{black}{Note, however, that 
stability or instability cannot be concluded from (\ref{eq:meanx2_id}) for $0<\alpha\leq1$, since eq. (\ref{eq:meanx2_id}) is not valid there.}}\\ 
\textcolor{black}{From equation (\ref{eq:evol_mean_log}) and the above discussion following from it, it follows that $\mu+\langle f(t)\rangle=\mu$ controls the growth/decay of moments in the nonlinear regime, if $\langle f(t)\rangle$ exists. This is true even if $f(t)$ is asymmetric.}
\subsection{Asymptotics of the PDF at large $x$}
In this section, we study the fractional Fokker-Planck equation under the assumption of stationarity to derive the asymptotics of the stationary density for  $x\to \infty$. Here, we need to consider $\beta>-1$ and $\beta=-1$ separately.
\subsubsection{The case $\beta>-1$}
\label{sec:asymptotics_y_to_infinity_beta_gt_-1}
Let $\beta>-1$, and consider the FFPE in the Stratonovich interpretation, i.e. equation (\ref{eq:ffpe_stratonovich}). 
 For $y\to \infty$, we neglect $D_\alpha^- \ll D_\alpha^+$ and $\mu \ll \gamma \exp(2y)$ \textcolor{black}{to find the following equation for the stationary distribution associated with the process $Y=\log(X)$, denoted by $p_{y,st}(y)$,  }
\begin{equation}
    \gamma e^{2y} p_{y,st}(y) \approx  \frac{(1+\beta)\frac{d}{dy} \int_{-\infty}^y  \frac{p_{y,st}(z)}{(y-z)^{\alpha-1}} dz}{2\cos(\pi\alpha/2)\Gamma(2-\alpha)} .
\end{equation}
Asymptotically, the integral is dominated by $z\ll y$. Hence $(y-z)^{\alpha-1}\approx y^{\alpha-1}$. The remaining integral can be approximated as $\int_{-\infty}^y p_{y,st}(z)dz\approx \int_{-\infty}^\infty p_{y,st}(z)dz=1$. The resulting equation implies the following asymptotic behavior for the stationary density at large $y$,
\begin{equation}
    p_{y,st}(y) \sim \frac{C(1+\beta)}{\gamma} y^{-\alpha}\exp(-2y).
\end{equation}
The prefactor is given by
\begin{equation} C= \sin(\pi\alpha/2)\Gamma(\alpha)/\pi, \label{eq:def_C} \end{equation} which has been simplified using Euler's reflection formula $\Gamma(\alpha)\Gamma(1-\alpha)=\pi/\sin(\pi \alpha )$. \textcolor{black}{In terms of the stationary distribution $p_{x,st}(x)$ associated with the original process $X$, this gives}
\begin{equation}
    p_{x,st}(x) \sim \frac{C(1+\beta)}{\gamma} \log(x)^{-\alpha} x^{-3},
    \label{eq:asymptotics_x_to_infty_beta_gt_m1}
\end{equation}
for $x\to \infty$. While the above derivation is valid for $1<\alpha<2$, one may repeat the same steps for $0<\alpha<1$ with the corresponding fractional derivative from eqns. (\ref{eq:Dplus_riemann_derivative_alpha_lt_1}), and finds the same result. For $\alpha>1$, there is both a finite mean and a finite variance. For $\alpha<1$, the variance in infinite, but the mean is finite. We note that the derivation given here is inspired by a similar argument from \cite{chechkin2004levy}. 

Further, if one chooses the It\^o interpretation, then one may derive the large-$x$ asymptotics in a similar way. One begins by considering the stationary solutions of the associated It\^o FFPE for $p_x(x,t)$, i.e. equation (\ref{eq:ffpe_ito}). Then one takes the limit $x\to \infty$, assuming $D_+^\alpha \gg D_-^\alpha$, and using $\mu x \ll \gamma x^3$ to find
\begin{equation}
    \gamma x^3 p_{x,st}(x) \approx   \frac{(1+\beta)\frac{d}{dx}\int_{-\infty}^x \frac{z^\alpha p_{x,st}(z)dz}{(x-z)^{\alpha-1}}}{2\cos(\pi\alpha/2)\Gamma(2-\alpha)}.
\end{equation}
Now, $\int_{-\infty}^x \frac{z^\alpha p_{x,st}(z)dz}{(x-z)^{\alpha-1}}\approx \frac{1}{x^{\alpha-1}} \int_{-\infty}^\infty z^\alpha p_{x,st}(z) dz$ by a similar reasoning as for the Stratonovich case. The remaining integral cannot be performed explicitly, but it is an $x$-independent constant. Hence, one finds the asymptotic proportionality
\begin{equation}
    p_{x,st}^{\text{It\^o}}(x) \propto (1+\beta) x^{-3-\alpha}, \label{eq:asymptotics_large_x_beta_gt_-1_ito}
\end{equation}
for large $x$. This result is remarkable, since the power law matches exactly the one found for additive noise in a quartic potential, \cite{chechkin2004levy}. In particular, the third moment is finite in the It\^o interpretation for $1<\alpha<2$ (where it diverges in the Stratonovich case), and the variance is finite for $0<\alpha<1$ (where it diverges in the Stratonovich case). The observation that the asymptotic form of the tails of the stationary PDF are altered by a state-dependent L\'evy noise amplitude in the Stratonovich interpretation, but not in the It\^o interpretation, has been made in previous studies with different functional forms of multiplicative noise \cite{srokowski2009fractional,srokowski2009multiplicative}. For the remainder of this paper, we will adopt the Stratonovich interpretation.
\subsubsection{The case $\beta=-1$}
The asymptotics in (\ref{eq:asymptotics_x_to_infty_beta_gt_m1}) and (\ref{eq:asymptotics_large_x_beta_gt_-1_ito}) break down for $\beta=-1$, which is the nonlinear version of the finite-moment log-stable process. For Gaussian noise, $\alpha=2$, the stationary PDF in $y$ is known to be $p_{y,st}(y)=Ne^{\mu y} e^{-\frac{\gamma}{2}e^{2y}}$, which decays faster than $e^{ny}$ at large $y$  for any $n>0$. For $1<\alpha<2$ and $\beta=-1$, the stable PDF has a short tail $+\infty$, decaying faster than a Gaussian (since $\alpha/(\alpha-1)>2$ in eqn. (\ref{eq:short_tails})). This implies that the stationary PDF under such L\'evy noise will decay faster than in the Gaussian case. Hence, the moments of $x$ for any order $n>0$ exist there also. In terms of $p_{y,st}(y)$, one expects a double-exponential function as in the Gaussian case. However, unfortunately, we cannot derive these large-$x$ asymptotics explicitly as we did for $\beta>-1$, since the Riemann-Liouville derivative of such functions is not known in simple terms. Rather, we will rely on numerical solutions to confirm that the PDF of $y$ indeed decays faster than exponentially at $y\to+\infty$. At $0<\alpha<1$, $\beta=-1$, since the origin is stable for all $\mu$ in the linear regime, it will continue to be stable in the nonlinear regime (the nonlinearity in equation (\ref{eq:langevin}) is strictly negative). \NOTE{Thus the stationary PDF is $\delta(x)$ in this case, i.e. the CDF converges to $1$ for all $x>0$ in the long-time limit.}
\subsection{Asymptotics of the PDF at  $x\to 0$}
We now investigate the asymptotic behavior of the stationary density for  $x\to 0$. Here we need to distinguish between the cases $\beta=1$ and $\beta<1$.
\subsubsection{The case $\beta=1$}
Consider $\beta=1$ and $\mu>0$. The FFPE (\ref{eq:ffpe_stratonovich}) in steady state, taking $y\to -\infty$ and neglecting $\gamma e^{2y} \ll \mu$, reads
\begin{equation}
    0 = - \mu \partial_y p_{y,st}(y) - \frac{1}{\cos(\alpha\pi/2)}(D_+^\alpha f)(y).
\end{equation}
Making an exponential ansatz $p_{y,st}(y)\propto e^{Ay}$ and  using the fractional derivative of the exponential given in equation (\ref{eq:frac_deriv_exp}) leads to
\begin{equation}
    0 = - \mu A e^{A y} -  \frac{1}{\cos(\alpha \pi/2)} \mu A^{\alpha} e^{Ay},
\end{equation}
implying
\begin{equation}
    A\equiv A_\alpha(\mu)=\left(- \mu \cos(\alpha\pi/2)\right)^{1/(\alpha-1)} \label{eq:A_diff_from_-1}
\end{equation}
In terms of the original variable $x$, this corresponds to
\begin{equation}
    p_{x,st}(x) \propto x^{-1+A_\alpha(\mu)}, \label{eq:asymptotics_pdf_small_x_beta1}
\end{equation}
which for $\alpha>1$ is integrable when $\mu>0$. The term \textit{integrable} is used here to mean that the integral of a given function over its domain converges. When $\alpha>1$ and $\mu\leq 0$, on the other hand, the solution is non-integrable, which is associated with the absence of a steady-state solution in the space $x>0$. \NOTE{In that case, the stationary density is given by $\delta(x)$, with the CDF converging to $1$ for all $x$ in the long-time limit}. For $\alpha<1$ and $-\infty<\mu<0$, the same result (\ref{eq:asymptotics_pdf_small_x_beta1}) holds with $A>0$, such that the solution is integrable. For $\alpha<1$, $\mu>0$, the solution again ceases to be integrable. From the Langevin equation (\ref{eq:langevin_y}) one deduces that this is due to the fact that in this case $\dot{Y}>0$ for $Y<\frac{1}{2}\ln{(\mu/\gamma)}$, since the noise is strictly positive, and thus the probability to be at $Y\leq \frac{1}{2} \ln(\mu/\gamma)$ vanishes at late times. \NOTE{ The stationary PDF is thus only supported at values of $y$ above the
deterministic saturation point $\frac{1}{2}\ln(\mu/\gamma)$ 
and vanishes for all smaller $y$.} Hence, the exponential ansatz is inappropriate and breaks down. This indicates that for $\alpha<1$, $\mu>0$, we cannot neglect the nonlinear term in equation (\ref{eq:langevin}), since it is the only one that decreases $Y$. 

The above discussion confirms the intuition based on the linear  solution (including an arrest of the leakage of probability to $+\infty$ by the nonlinear term in eq. (\ref{eq:langevin})). For $\beta=1$, $\alpha>1$ a critical transition occurs at $\mu=0$, from all weight of the stationary PDF being at $x=0$ (origin stable) to non-zero weight at $x>0$ (origin unstable). On the other hand, for $0<\alpha<1$, $\beta=1$, the origin is always unstable.
\subsubsection{The case $\beta<1$}
Let $\beta<1$ and $\mu>0$. We follow once again the arguments of \cite{chechkin2004levy}, starting from eq. (\ref{eq:ffpe_stratonovich}). Consider $y\to-\infty$, such that $\mu \gg \gamma e^{2y}$
, and neglect $D_+^\alpha\ll D_-^\alpha$ to find
\begin{equation}
   - \mu p_{y,st}(y) = \frac{(1-\beta)\frac{d}{dy}\int_y^\infty \frac{p_{y,st}(z)dz}{(z-y)^{\alpha-1}}}{2\cos(\pi\alpha/2)\Gamma(2-\alpha)}.
\end{equation}
Using normalisation of the PDF, as for the large-$x$ limit at $\beta>-1$, we find that the stationary PDF is asymptotically given by
\begin{equation}
    p_{y,st}(y) \sim \frac{(1-\beta)C}{\mu} (-y)^{-\alpha},
    \label{eq:asymp_y_to_minus_infty}
\end{equation}
for $y\to-\infty$, where $C$ is given by (\ref{eq:def_C}). In terms of the original variable, this corresponds to
\begin{equation}
    p_{x,st}(x) \sim \frac{C(1-\beta)}{\mu} x^{-1} (\log(1/x))^{-\alpha}
    \label{eq:asymp_x_to_zero_beta_lt_1}
\end{equation}
for $x\to 0$. While the above derivation is for $1<\alpha<2$, the case $0<\alpha<1$ leads to the same result. Clearly, this solution breaks down at negative $\mu$, since the predicted PDF ceases to be positive. In this case, the stationary distribution is $\delta(x)$. \textcolor{black}{For $\alpha<1$, the fact that the solution (\ref{eq:asymp_y_to_minus_infty}) is not integrable at $y=-\infty$ implies that there is no stationary state at $x>0$.} Instead, the stationary density in that case is $\delta(x)$, for all $\mu$. For $1<\alpha<2$ and $\mu>0$, on the other hand, (\ref{eq:asymp_y_to_minus_infty}) gives a consistent, integrable stationary PDF. \\

\begin{table}[h]
    \centering
\begin{tabular}{ |c| c| c| }
 \hline $\beta$ & \quad $p_{x,st}(x\to 0) \quad$ 
                & \quad $p_{x,st}(x\to \infty) \quad$ \\ 
 \hline    $-1$ & $\quad{C(\mu x)^{-1} \log^{-\alpha}(1/x)}\quad$ 
                & \quad exponential decay \quad \\
 \hline  $\quad(-1,1)\quad$ 
                  &  $C(\mu x)^{-1} \log^{-\alpha}(1/x)$
                  &  ${C\gamma^{-1} x^{-3} \log^{-\alpha}(x)} $ \\
 \hline  $1$      &  $ \propto x^{-1+A_\alpha(\mu)}$
                  &  ${C\gamma^{-1} x^{-3} \log^{-\alpha}(x)} $\\  
\hline  
\end{tabular}
    \caption{Summary of the different asymptotic behaviors of $p_{x,st}(x)$ obtained in the previous sections for $\alpha<2$. The domain of validity of the formulas is discussed in the text. The constants $C$ and $A_\alpha(\mu)$ are given in eqns. (\ref{eq:def_C}), (\ref{eq:A_diff_from_-1}).}
    \label{tab:summary_asymptotics}
\end{table}

Table \ref{tab:summary_asymptotics} summarizes the different asymptotic behaviors obtained above. The closest resemblance with the $\alpha=2$ case is seen in the large-$x$ exponential decay at $\beta=-1$ (which is Gaussian for $\alpha=2$), and the small-$x$ $\mu$-dependent power-law at $\beta=1$ (where $A_2(\mu)=\mu$). The asymptotics in the remaining cases are qualitatively different from the Gaussian case. \textcolor{black}{A criterion for a system to be on-off intermittent is whether the stationary density $p_{x,st}(x)$ diverges at $x=0$. In previous studies of on-off intermittency with various types of noise, there is generally a critical value $\mu_c>0$ above which the intermittent behavior disappears \cite{aumaitre2005low,aumaitre2006effects}. The asymptotic results presented above imply that for $1<\alpha\leq 2$, on-off intermittency will cease  when $-1+A_\alpha(\mu)=0$ at $\beta=1$, since the singularity at $x=0$ disappears at this point. However, at $\beta<1$, the $x^{-1} \log^{-\alpha}(1/x)$ behavior at small $x$ remains present for all $\mu>0$. This implies that there is no value of $\mu$ where on-off intermittency ceases to be present in that case, by contrast with all previously known cases of on-off intermittency.}

Summarising, a transition occurs at $\mu=0$ for $1<\alpha<2$, from a stable origin at $\mu<0$ to an unstable origin at $\mu>0$. This is as predicted in the linear theory. \textcolor{black}{When $\alpha<1$, $\beta<1$, the origin is stable for all $\mu$, and for $\alpha<1$, $\beta=1$, the origin is always unstable. These results are also consistent with the linear theory, taking into account saturation by the nonlinearity.}
\label{sec:asymp_x_to_0}




\subsection{PDFs and Moments}
\NOTE{ Here we attempt to deduce the moments 
based on the asymptotic behavior of the PDFs discussed in the previous sections complemented by numerical solutions of the stationary FFPE, using a heuristic approach. The numerical solutions are computed using the finite-difference formulation described in appendix \ref{sec:app_num_ffpe}.}
\begin{figure}[ht]
    \centering
    \includegraphics[width=8.6cm]{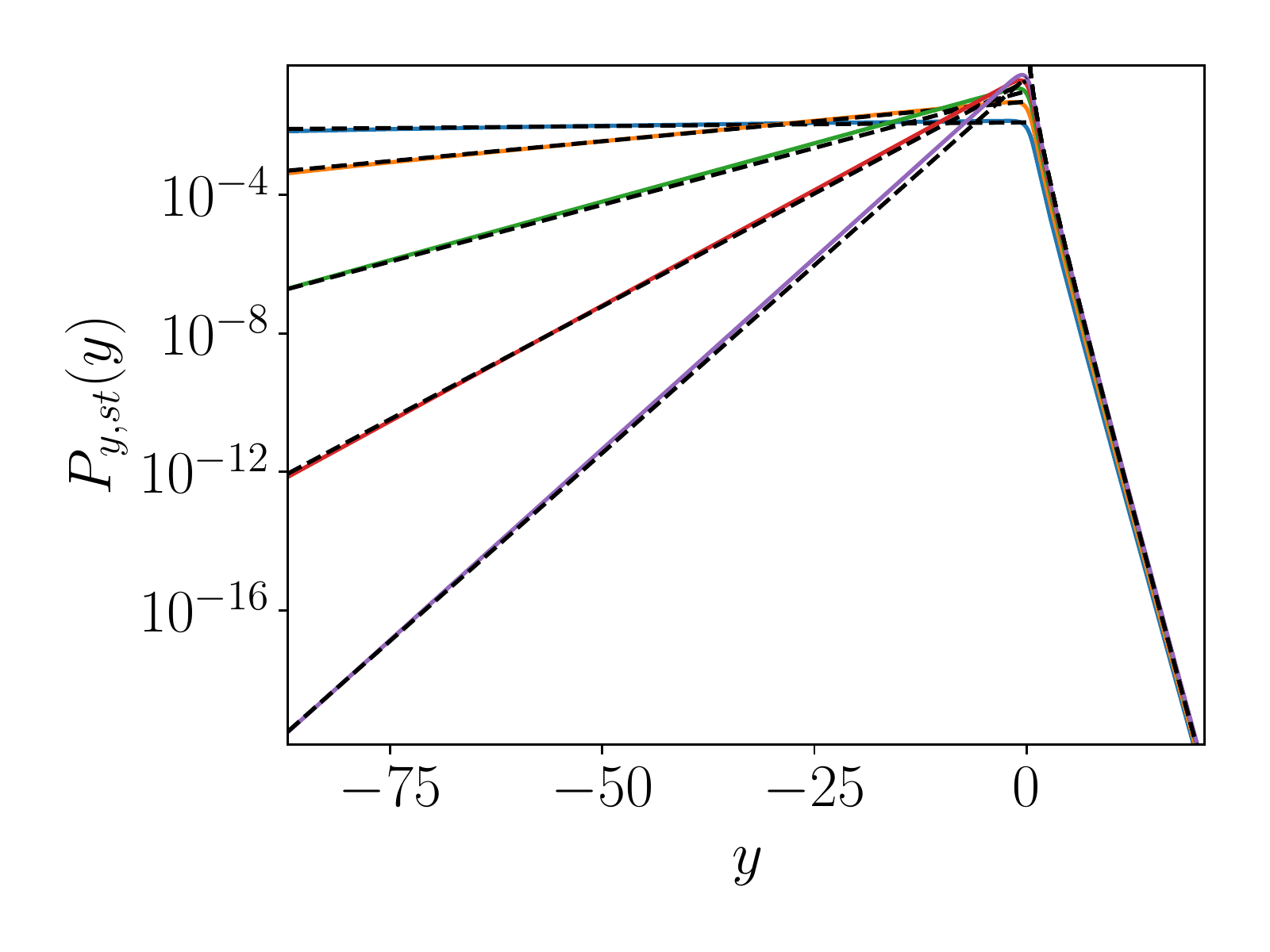}
    \caption{Semi-log plot of numerically obtained stationary PDF for $\alpha=1.5$, $\beta=1.0$, varying $\mu=0.1,0.33,0.55,0.78,1.0$, at $\gamma=1$ fixed. The dashed line on the right is the theoretical prediction (\ref{eq:asymptotics_x_to_infty_beta_gt_m1}) for the the cut-off by non-linearity. The dashed lines on the left show shows the prediction $p_{y,st}(y) \propto \exp(A(\mu)y)$ from (\ref{eq:asymptotics_pdf_small_x_beta1}). }
    \label{fig:pdf_plots}
\end{figure}
\begin{figure}[ht]
    \centering
    \includegraphics[width=8.6cm]{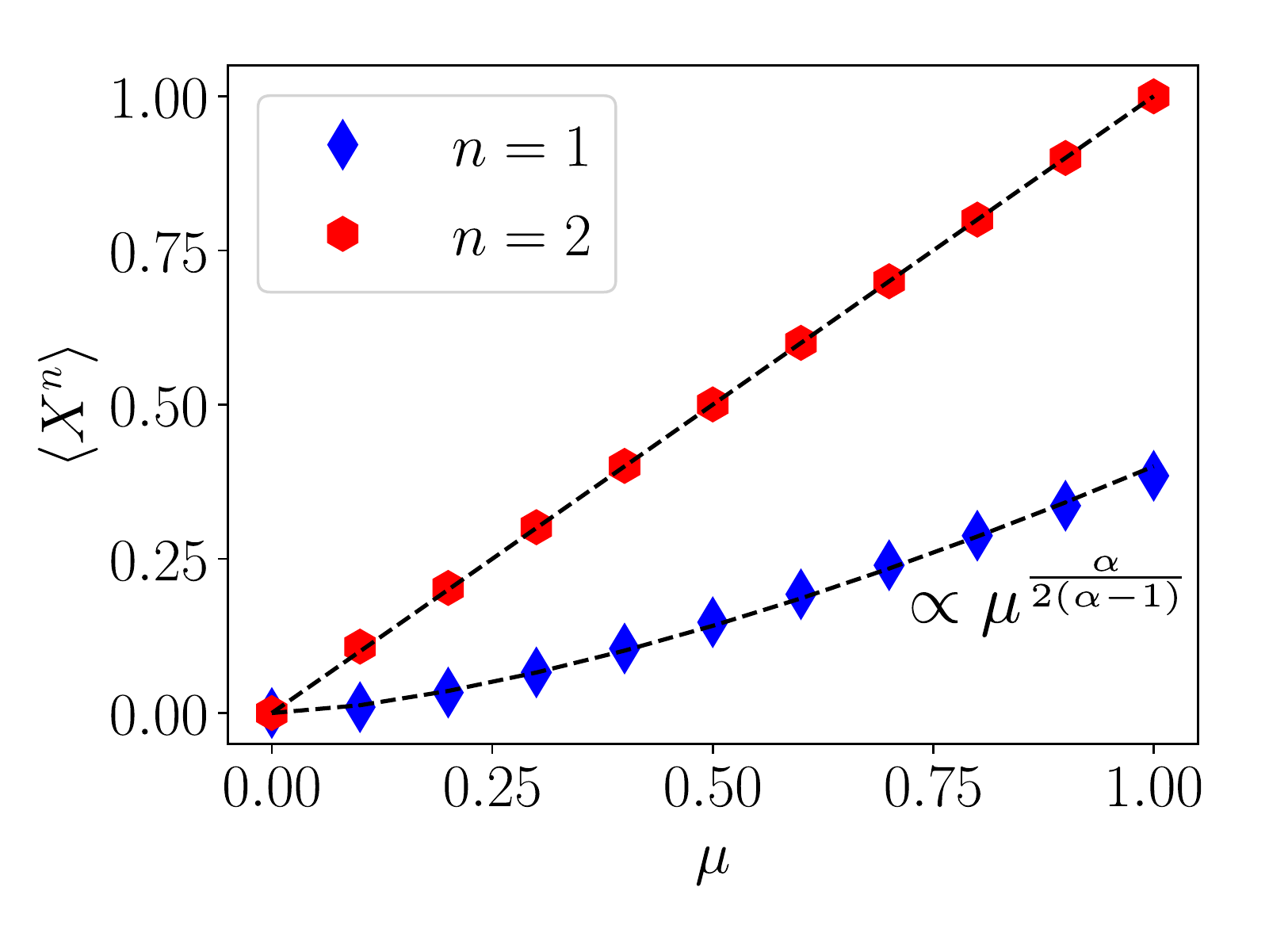}
    \caption{First and second moment of $X$ for $\alpha=1.5$, $\beta=1$ versus $\mu$ at $\gamma=1$. The first moment scales as predicted in (\ref{eq:meanX_vs_mu_beta1}), shown by the curved dashd line, and the second moment is linear, in agreement with (\ref{eq:meanx2_id}).}
    \label{fig:moments_beta_1}
\end{figure}
\subsubsection{The case $1<\alpha<2$, $\beta=1$}
\NOTE{We begin by showing the results from the numerical solution
of the FFPE.
Figure \ref{fig:pdf_plots} indicates an agreement with the theoretical results of the previous section, for 
$x\ll1$ (i.e. $y\to-\infty$) and $x\gg1$ (i.e. $y\to\infty$). In order to calculate the 
scaling with $\mu$  of the different moments we can model
the PDF as}
\begin{equation}
    p_{x,st}(x) \approx \frac{1}{N} \begin{cases} x^{-1+A_\alpha(\mu)} \hspace{0.75cm}: x< x_* \\ B x^{-3} \log^{-\alpha}(x): x\geq x_*\end{cases},
\end{equation}
\NOTE{where $x^*$ and $N$ are model parameters,
$A_\alpha(\mu)$ is as given in equation (\ref{eq:A_diff_from_-1}) and $B=x_*^{2+A_\alpha(\mu)}\log^\alpha(x_*)$ for continuity.} To determine the two unknowns $N$ and $x_*$, we impose normalisation of the PDF and the second moment identity (\ref{eq:meanx2_id}). At small $\mu$, the dominant part of the weight is at negative $y$, i.e. at small $x$, as visible in figure \ref{fig:pdf_plots}. This implies
 \begin{align}
     N 
     &\sim \frac{x_*^{A_\alpha(\mu)} }{A_\alpha(\mu)},
 \end{align}
 as $\mu\to 0^+$.
In addition,
\begin{equation}
     \langle X^2 \rangle = \frac{\left(\frac{1}{2+A_\alpha(\mu)} + \frac{1}{\alpha-1} \log^{1-\alpha}(x_*) \right)x_*^{2+A_\alpha(\mu)}}{N}
 \end{equation}
  and moments of order higher than two diverge. By equation (\ref{eq:meanx2_id}) we have $\langle X^2 \rangle =\mu/\gamma$. For $0<\mu\ll1$, this implies, 
 \begin{equation}
     x_* \approx \sqrt{2\mu/(A_\alpha(\mu)\gamma)},
 \end{equation}
 such that
 \begin{equation}
    \langle X \rangle \approx A_\alpha(\mu) x_* \propto \mu^{\frac{\alpha}{2(\alpha-1)}}
    \label{eq:meanX_vs_mu_beta1},
 \end{equation}
 where $A_\alpha(\mu)$ was inserted from equation (\ref{eq:A_diff_from_-1}). Figure \ref{fig:moments_beta_1} shows that this agrees with the numerical solution of the stationary FFPE for the examplary case $\alpha=1.5$, $\beta=1$. \NOTE{We note that (\ref{eq:meanX_vs_mu_beta1}) can simply be extended to $\langle X^n\rangle\propto \mu^{c_n}$ with $0<n\leq 2$, with $c_n= \frac{2-n(2-\alpha)}{2(\alpha-1)}$ which varies continuously with $n$. In particular $c_n \approx \frac{1}{\alpha-1}$ for small $n$, $c_1= \frac{\alpha}{2(\alpha-1)}$ and, by construction, $c_2 = 1$.}

 \subsubsection{The case $1<\alpha<2$, $\beta=-1$}
\begin{figure}[ht]
     \centering
     \includegraphics[width=8.6cm]{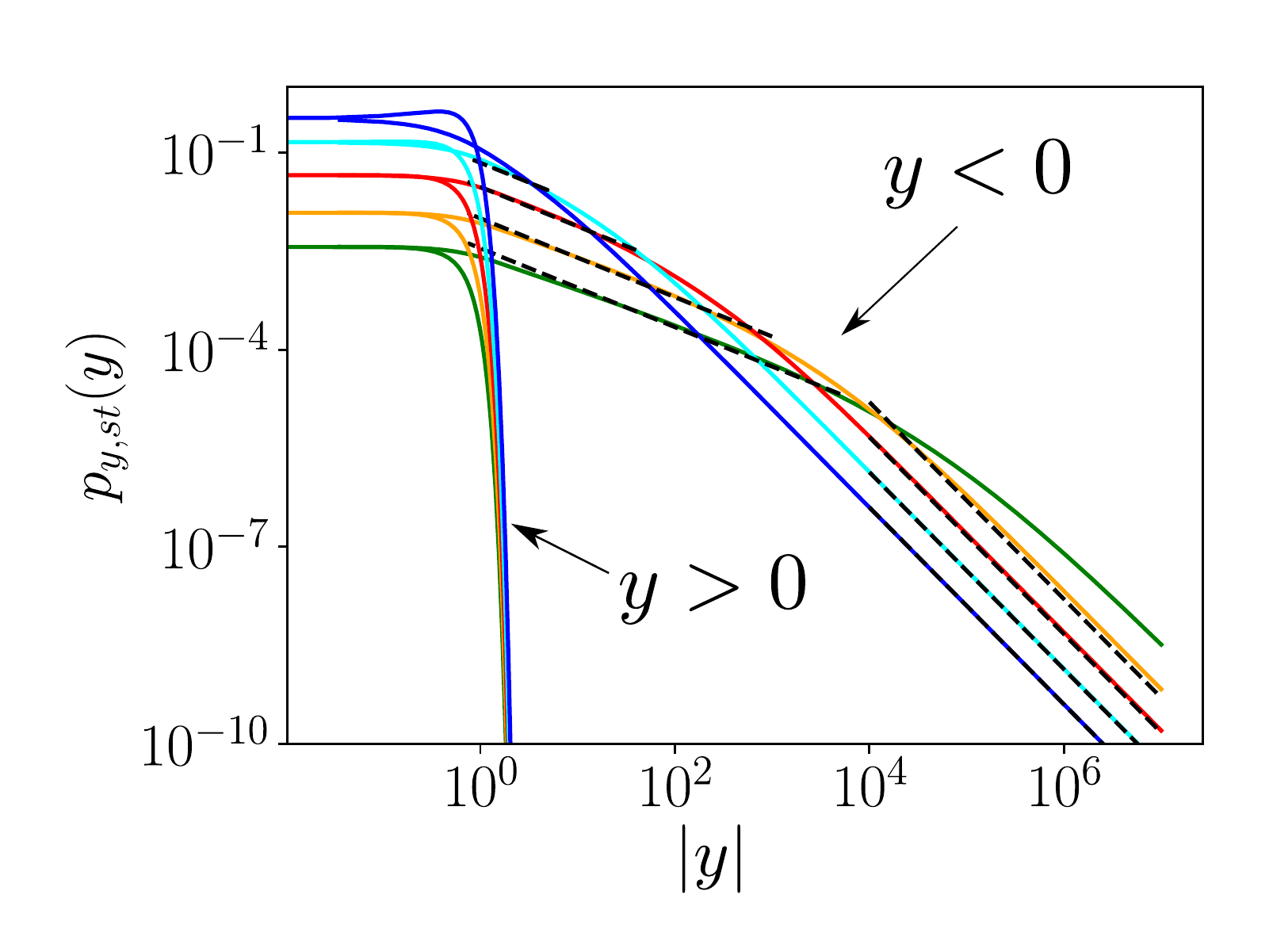}
     \caption{Log-log plot of the numerically obtained stationary PDF $p_{y,st}(y)$ versus $|y|$ for $\alpha=1.5$, $\beta=-1$. Five different values of $\mu$ are shown ($0.01, 0.025,0.085,0.3,1.0$) for $\gamma=1$. The tail at $y\to-\infty$ fits the prediction  (\ref{eq:asymp_y_to_minus_infty}) (dashed lines on the right). Before that limiting scaling is observed at $y\to-\infty$, an intermediate, flatter power-law range occurs at $y<0$, whose exponent is independent of $\mu$, but whose amplitude decreases and whose range increases as $\mu$ decreases. At large positive $y$, there is a faster-than-exponential decay as predicted (compare with figure \ref{fig:pdf_beta0} at $y>0$). }
     \label{fig:pdf_beta-1}
 \end{figure}

Figure \ref{fig:pdf_beta-1} shows the numerically obtained PDF.
It matches the theoretically predicted asymptotics in the $y\to\pm\infty$ limits. In addition, an intermediate, shallower range is observed at intermediate negative $y$ before the predicted asymptotic behavior at $y\to-\infty$ is realized. Figure \ref{fig:pdf_beta-1} suggests that this intermediate range is a power law. A close inspection shows that it is only approximately a power law since it has a finite curvature in the log-log diagram. Notwithstanding this caveat, we propose a simplistic model approximately describing the numerical result 
 \begin{equation}
    p_{y,st}(y) \approx \begin{cases} 0 \hspace{0.8cm}& : y\geq 0 \\  D (-y)^{-\lambda}&: 0>  y \geq y_*\\2C/\mu \text{ }(-y)^{-\alpha} &: y_*> y\end{cases} \label{eq:fit_pdf_beta-1},
\end{equation}
where $C=\Gamma(\pi)\sin(\pi\alpha/2)/\pi$. The portion of the PDF at $y>0$ makes a negligible contribution to its normalisation and any moments of $X$, due to the faster-than-exponential decay at $y>0$. The value of $\lambda$ can only be determined numerically, with relatively large errorbars. Thus fixing $\lambda$ numerically (e.g. $\lambda\approx 0.6$ for  $\alpha=1.5, \beta=-1$ in figure \ref{fig:pdf_beta-1}), there are two unknowns $D$ and $y_*$ which we determine by imposing normalisation of the PDF and the second moment identity (\ref{eq:meanx2_id}). 

\begin{figure}[ht]
    \centering
    \includegraphics[width=8.6cm]{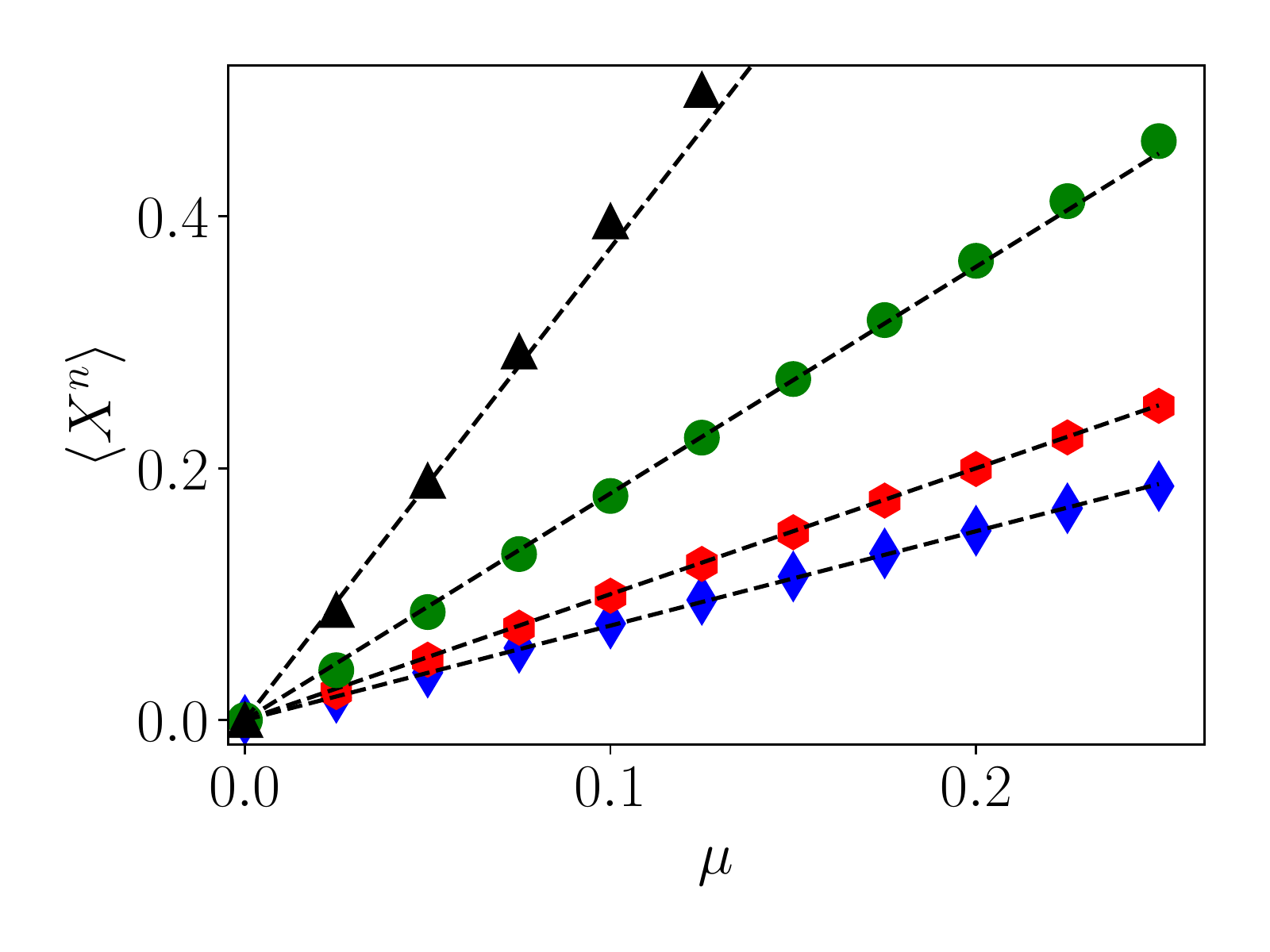}
    \caption{Moments for $\alpha=1.5$, $\beta=-1$ versus $\mu$ at $\gamma=1$ from the numerically computed steady-state PDF. Symbols represent $n=1$ (diamond), $n=2$ (hexagon), $n=3$ (circles), $n=4$ (triangles). The dashed lines show linear scaling.}
    \label{fig:moments_beta_-1}
\end{figure}

Formally,
\begin{align}
    1 = &\frac{D}{1-\lambda} (-y_*)^{1-\lambda} + \frac{2C}{\mu (\alpha-1)} (-y_*)^{1-\alpha}, \label{eq:constr1_pdf_beta-1} \\ 
    \frac{\mu}{\gamma} = & D \int_{y_*}^0 (-y)^{-\lambda} e^{2y} dy + \frac{2C}{\mu}\int_{-\infty}^{y_*} (-y)^{-\alpha}e^{2y} dy. \label{eq:constr2_pdf_beta-1}
\end{align}
Figure \ref{fig:pdf_beta-1} suggests that $y_*\to - \infty$ as $\mu\to 0^+$. In eq. (\ref{eq:constr2_pdf_beta-1}), this implies that the  second integral, from $-\infty$ to $y_*$, is exponentially suppressed for small $\mu$. For large $|y_*|$, the lower limit of the first integral may be replaced by $-\infty$. This leads to 
\begin{align}
    D\approx& \frac{ 2^{1-\lambda}}{\gamma\Gamma(1-\lambda)} \mu, \label{eq:solved_coeff_D_beta-1} \\
    y_* \approx& - \left(\frac{2C}{(\alpha-1)} \right)^{1/(\alpha-1)} \mu^{-1/(\alpha-1)} \label{eq:solved_coeff_ys_beta-1}.
\end{align}
Note that, as expected, $y_* \to -\infty$ as $\mu \to0^+$. The two results (\ref{eq:solved_coeff_D_beta-1}), (\ref{eq:solved_coeff_ys_beta-1}) imply that the moments of $X$ of arbitrary order $n>0$ scale linearly, since large negative $y$ are exponentially suppressed:
\begin{align}
    \langle X^n\rangle = \langle e^{ny} \rangle &\approx D \int_{-\infty}^0 (-y)^{-\lambda} e^{ny} dy \notag \\ & \propto \mu
    \label{eq:linmu}
\end{align}
\NOTE{Note that the critical exponent in the final result is independent of the value of $\lambda$.}
The result of eq. (\ref{eq:linmu}) is confirmed in figure \ref{fig:moments_beta_-1}, where the integer moments up to order four, determined from the numerical solution of the stationary FFPE, are all shown to scale linearly with $\mu$.

\subsubsection{The case $1<\alpha<2$, $|\beta|<1$}
\begin{figure}[ht]
    \centering
    \includegraphics[width=8.6cm]{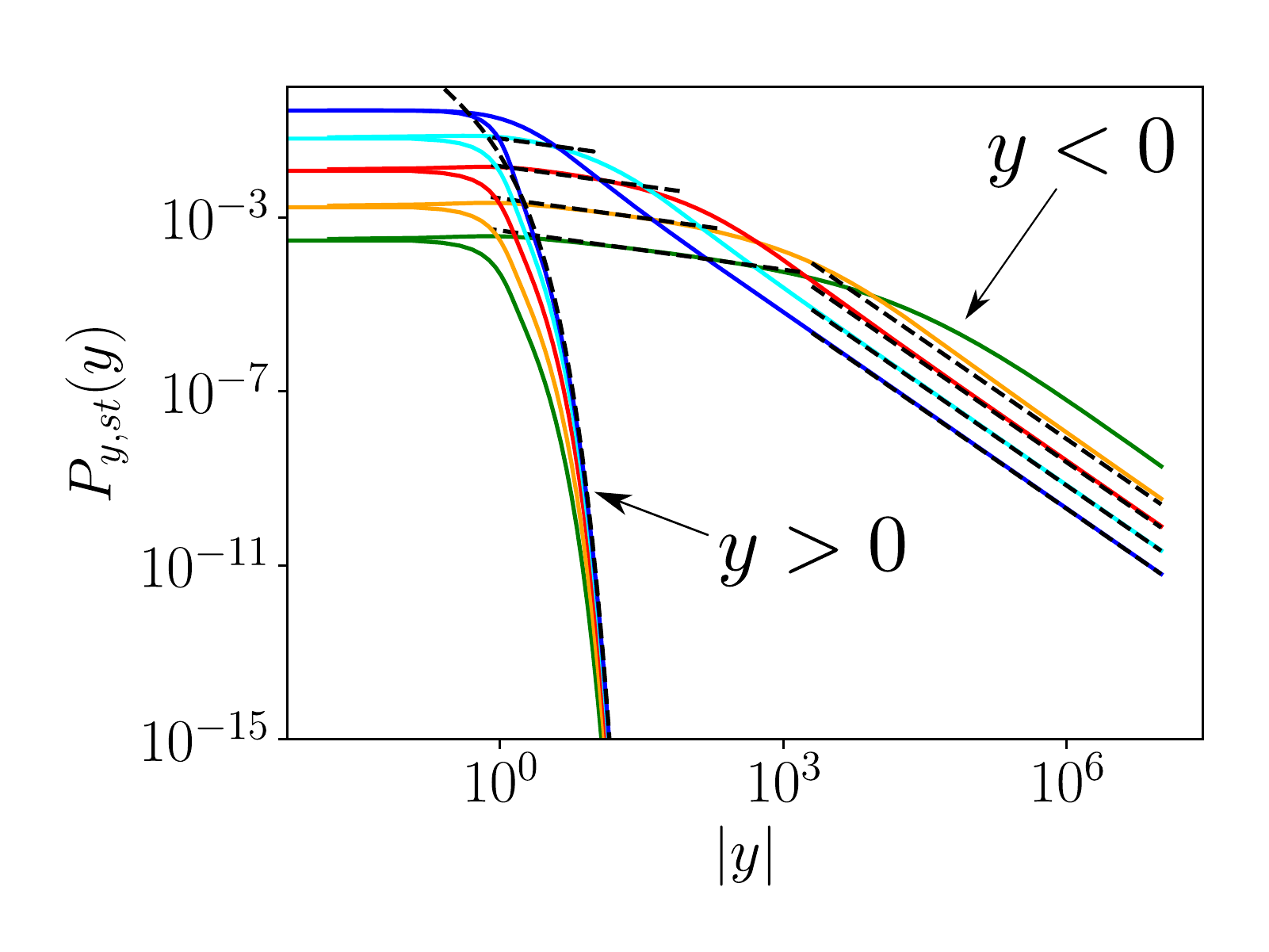}
    \caption{Log-log plot of the numerically obtained stationary PDF $p_{y,st}(y)$ versus $|y|$ for $\alpha=1.5$, $\beta=0$. Five different values of $\mu$ are shown ($0.01, 0.025,0.085,0.3,1.0$) for $\gamma=1$. The tail at $y\to-\infty$ fits the prediction of (\ref{eq:asymp_y_to_minus_infty}) shown in dashed straight lines on the right. At intermediate $y<0$, a flatter power-law range is seen, with exponent independent $\nu$ of $\mu$ ($\nu \approx 0.25$ here), but with amplitude decreasing and width increasing as $\mu$ decreases. The tail at $y\gg 1$  matches the prediction (\ref{eq:asymptotics_x_to_infty_beta_gt_m1}), shown by the curved dashed line.}
    \label{fig:pdf_beta0}
\end{figure}

Figure \ref{fig:pdf_beta0} shows that the PDF, which matches the predicted asymptotics at $y\to \pm \infty$, strongly resembles the case of $\beta=-1$ in that in addition to the asymptotic power-law range, an intermediate, approximately power-law range is seen at negative $y$. Again, close inspection shows that the intermediate range shows small deviations from a power law. However, the most marked difference from the case $\beta=-1$ is that the decay for $|\beta|<1$ is only exponential in $y$ at positive $y$, not faster than exponential as for $\beta=-1$. In particular, the asymptotics at positive $y$ imply a slow, power-law convergence of the second moment since $p_{st}(y)e^{2y}\propto y^{-\alpha}$ at $y\gg 1$. Bearing this in mind, we nonetheless employ the same approximate form for the PDF as for $\beta=-1$,
\begin{equation}
    p_{y,st}(y) \approx \begin{cases} 0 &:\hspace{0.1cm} y\geq 0 \\ E (-y)^{-\nu}&:  0> y\geq y_{**} \\ (1-\beta) C/\mu |y|^{-\alpha} &: y<y_{**}\end{cases},
    \label{eq:fit_pdf_beta0}
\end{equation}
where once more $C=\Gamma(\alpha)\sin(\alpha\pi/2)/\pi$. We may again determine $\nu$ numerically, albeit with significant uncertainty. In figure \ref{fig:pdf_beta0}, where $\alpha=1.5$, $\beta=0$, we observe $\nu \approx 0.25$. As for $\beta=-1$, the portion of the PDF at $y>0$ makes a subdominant contribution to the normalisation and the moments of order $n<2$. To determine $E$, $\nu$, we need two conditions. 
\begin{figure}[ht]
    \centering
    \includegraphics[width=8.6cm]{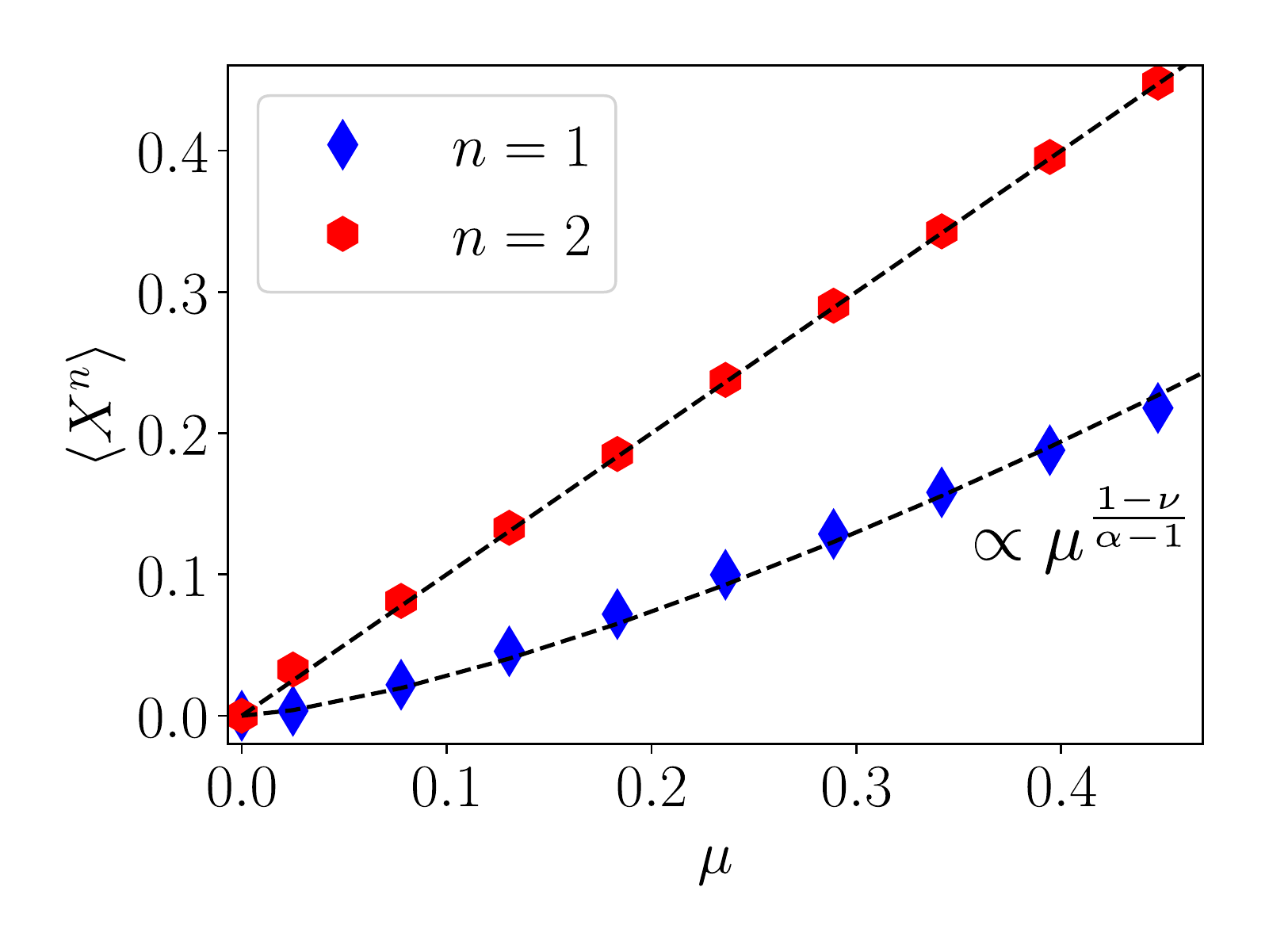}
    \caption{First and second moment of $X$ for $\alpha=1.5$, $\beta=0$ versus $\mu$ at $\gamma=1$. The scaling of the first moment is compatible with the prediction of (\ref{eq:Xmean_vs_mu_beta_lt_1}), shown by the curved dashed line, with $\nu\approx 0.25$ from figure \ref{fig:pdf_beta0}. The second moment is linear in $\mu$, satisfying the identity (\ref{eq:meanx2_id}).}
    \label{fig:moments_beta_0}
\end{figure}

By contrast with the case $\beta=-1$, we impose continuity at $y_{**}$ instead of (\ref{eq:meanx2_id}), normalisation of the PDF. Formally,
\begin{align}
    E (-y_{**})^{-\nu} =& \frac{\left(1-\beta\right)C}{\mu} (-y_{**})^{-\alpha} \label{eq:constr1_pdf_beta0},\\
        1 = &\frac{E}{1-\nu} (-y_{**})^{1-\nu} + \frac{(1-\beta)C}{\mu (\alpha-1)} (-y_{**})^{1-\alpha} \label{eq:constr2_pdf_beta0}.
\end{align}
Solving these two equations gives
\begin{align}
    y_{**} = &\left\lbrace \left[\frac{1}{1-\nu} + \frac{1}{\alpha-1}\right](1-\beta) C \mu \right\rbrace^{-1/(\alpha-1)}\\
    E = & y_{**}^{\nu-\alpha} \frac{(1-\beta)C}{\mu} =  \frac{\lbrace (1-\beta) C \rbrace^{\frac{2\alpha-\nu-1}{\alpha-1}} }{\left[\frac{1}{1-\nu} + \frac{1}{\alpha-1} \right]^{\frac{1}{\alpha-1}}} \mu^{\frac{1-\nu}{\alpha-1}}.
\end{align}
This implies that the first moment exhibits the anomalous scaling
\begin{equation}
    \langle X \rangle = \langle e^y \rangle \approx E \int_{-\infty}^0 e^{y} (-y)^{-\nu} dy \propto  \mu^{\frac{1-\nu}{\alpha-1}}. \label{eq:Xmean_vs_mu_beta_lt_1}
 \end{equation}
Note that, by contrast with the case $\beta=-1$, the critical exponent in (\ref{eq:Xmean_vs_mu_beta_lt_1}) depends on the exponent $\nu$ of the intermediate power-law range at negative $y$, which we have not determined theoretically as a function of $\alpha,\beta$, but only measured numerically. 
The critical scaling of the first moment predicted in eq. (\ref{eq:Xmean_vs_mu_beta_lt_1}) is shown to be consistent with the numerically obtained moments in figure \ref{fig:moments_beta_0} for the case $\alpha=1.5$, $\beta=0$. The prediction (\ref{eq:Xmean_vs_mu_beta_lt_1}) for the critical exponent at $n=1$ was also verified for different values of $\alpha$ at $\beta=0$ (not shown). 

\subsubsection{The case $0<\alpha<1$, $\beta=1$}
In this case the origin is unstable for all $\mu$, and a non-trivial stationary state exists due to the nonlinearity in equation (\ref{eq:langevin}). The point-vortex model presented in the companion paper \cite{vankan2021levy}, where $\alpha=2/3$, $\beta=1$, falls into this parameter range. The asymptotic theoretical results suggest that for $\mu<0$, the PDF can be modeled as
\begin{equation}
    p_{y,st}(y)= \begin{cases} B e^{A_\alpha(\mu)y}\hspace{0.6cm}&: y<y_* \\ \frac{2C}{\gamma} e^{-2y} y^{-\alpha}&: y\geq y_*\end{cases},
\end{equation}
where $C=\sin(\alpha \pi/2)\Gamma(\alpha)/\pi$, $A_\alpha(\mu)$ given by (\ref{eq:A_diff_from_-1}), and the two unknowns $B$ and $y_*$ are in principle determined by continuity at $y_*$ and normalisation. We note that the second moment does not exist because $y^{-\alpha}$ is not integrable at infinity for $\alpha<1$. Moments of order higher than two also diverge. However, $\langle X^n \rangle$ does exist for all $0<n<2$.

For illustration, we consider the special case $\alpha=1/2$, $\gamma=1$ and take the limit $\mu\to 0^-$, where $A_\alpha(\mu)\to\infty$. Clearly then $p_{y,st}(y)\to 0$ at $y<0$. Further, since
\begin{equation}
  2C \int_0^\infty e^{-2y}y^{-1/2} dy = 1,
\end{equation}
taking $y_*\approx 0$ as $\mu\to 0^-$ gives a consistently normalized model of the PDF. For this special case, $\alpha=1/2$, $\gamma=1$, the $n$-th moment of the PDF for $0<n<2$ may be computed to be
\begin{equation}
    \langle X^n \rangle = 2C \int_0^\infty e^{(n-2)y} y^{-\alpha} dy = \sqrt{\frac{2}{2-n}}, \label{eq:first_moment_al0.5}
\end{equation}
for $\mu$ small and negative. Note that the result is independent of $\mu$. For $n=1$, equation (\ref{eq:first_moment_al0.5}) was found to be satisfied  to within a few percent relative error by averaging over sample trajectories (not shown) using the method in appendix \ref{sec:app_exact_sol}. 
\section{Conclusions}
\label{sec:conclusions}
\begin{figure}
    \centering
        \includegraphics[width=8.6cm]{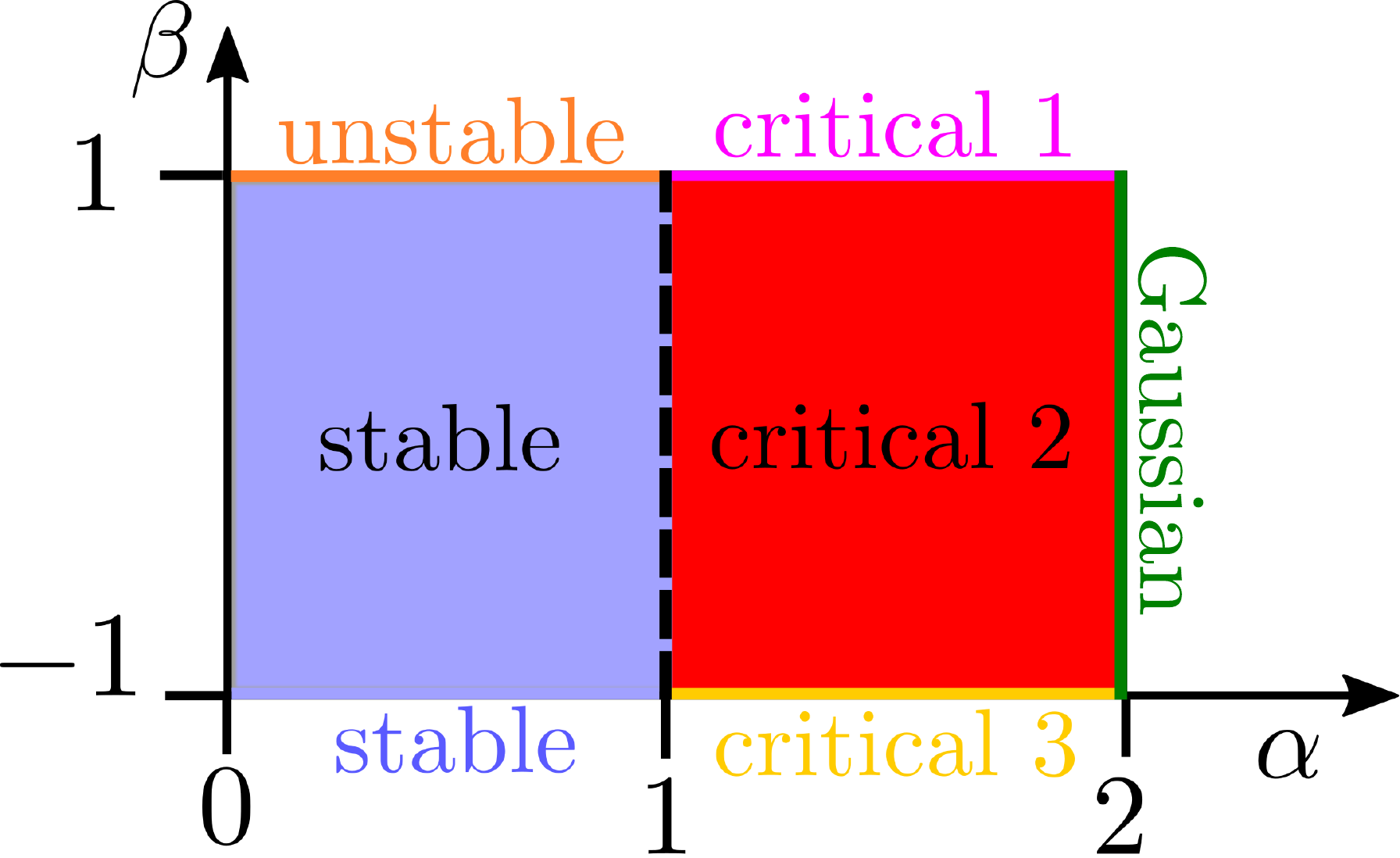}
    \caption{The parameter space of (\ref{eq:langevin}) with white L\'evy noise, $\alpha\in(0,2]$, $\beta\in[-1,1]$. A critical transition occurs at $\mu=0$ for $1<\alpha<2$, and for the Gaussian case $\alpha=2$. For $0<\alpha\leq 1$, the origin is either always stable or always unstable, independently of $\mu$.}
    \label{fig:parameter_space}
\end{figure}
We have studied the stochastic process obeying the Langevin equation (\ref{eq:langevin}) with L\'evy white noise. The theory of on-off intermittency was generalized, from the known case of Gaussian noise, to L\'evy noise by studying the FFPE (\ref{eq:ffpe_stratonovich}) analytically and numerically. First the linear ($\gamma=0$) solution was analysed, which showed leakage of the probability to $x=0$, $x=\infty$ or both, depending on the noise parameters $\alpha$ and $\beta$. Then we computed the nonlinear ($\gamma>0$) stationary solutions of the stationary FFPE, for which the leakage of probability to large $x$ is arrested by the nonlinearity in equation (\ref{eq:langevin}). We showed that for $1<\alpha\leq 2$ the origin is stable at $\mu<0$ and unstable at $\mu>0$. For $0<\alpha<1$, the origin is always stable, or always unstable, for any $\mu$, due to the divergent mean of the noise. In addition to the Gaussian case $\alpha=2$, where the stationary PDF for $\mu>0$ is given by $p_{x,st}(x)=Nx^{-1+\mu}e^{-\frac{\gamma}{2}x^2}$ and all critical exponents are equal to $1$, we identify a total of five qualitatively distinct regimes in the parameter space $\alpha\in(0,2]$, $\beta\in[-1,1]$, illustrated in figure \ref{fig:parameter_space},
\begin{enumerate}
   \item[(i)]  \textit{``Critical 1"} with $1<\alpha<2$, $\beta=1$. For $\mu>0$ and small $x$, the PDF is $p_{x,st}(x) \propto x^{-1+A_\alpha(\mu)} $, with $A_\alpha(\mu)\propto \mu^\frac{1}{\alpha-1}$. This matches the Gaussian small-$x$ result for $\alpha=2$. At $x\gg 1$, the PDF is $p_{x,st}(x) \propto x^{-3} (\log(x))^{-\alpha}$, i.e. $\langle X^n \rangle<\infty$ for $n\leq 2$ but $\langle X^n\rangle =\infty$ for $n>2$. As $\mu\to0^+$, one has $\langle X^n\rangle \propto \mu^{c_n},$ with 
   $c_1=\frac{\alpha}{2(\alpha-1)}$ and $c_2=1$.
   
    \item[(ii)] \textit{``Critical 2"} with $1<\alpha<2$, $|\beta|<1$. The PDF is $p_{x,st}(x) \propto \mu^{-1} x^{-1} (\log(1/x))^{-\alpha}$ at small $x>0$. This is in stark contrast with the Gaussian result; the logarithmic term here is crucial for integrability at $x=0$. At $x\gg 1$, the PDF is $p_{x,st}(x) = C x^{-3} (\log(x))^{-\alpha}$ as in case (i), s.t. only moments of order $n\leq 2$ are finite. At $x<1$, but not too small, there is an intermediate range where approximately $p_{x,st}(x)\propto x^{-1} (\log(x))^{-\nu}$, where the exponent $\nu$ was determined numerically. It remains an open problem to compute it theoretically as a function of $\alpha,\beta$. For small $\mu>0$, we found $\langle X^n\rangle \propto \mu^{c_n},$ with $c_1\approx \frac{1-\nu}{\alpha-1}$ 
    and $c_2=1$.
    
    \item[(iii)]  \textit{``Critical 3"} with $1<\alpha<2$, $\beta=-1$. At small $x$ and $\mu>0$, the PDF is $p_{x,st}(x) \propto \mu^{-1} x^{-1} (\log(1/x))^{-\alpha}$ as for case $(ii)$. For large $x$, the PDF $p_{x,st}(x)$ decays faster than any power of $x$. At $x<1$, but not too small, there is an intermediate range similar to that in (ii) where approximately $p_{x,st}(x)\propto x^{-1} (\log(x))^{-\lambda}$, with a different exponent $\lambda$ which was determined numerically. It remains an open problem to compute $\lambda$ theoretically as a function of $\alpha,\beta$. However, the critical exponents are independent of $\lambda$: for small $\mu>0$, one finds $\langle X^n\rangle \propto \mu^{c_n},$ with $c_n=1$ for all $n>0$.

    \item[(iv)] \textit{``Unstable"} with $0<\alpha< 1$, $\beta=1$. 
    Since the noise is strictly positive and has infinite mean, the origin $x=0$ is always unstable, independently of $\mu$. At small $x$ and for all $\mu<0$, the PDF is $p_{x,st}(x)\propto x^{-1+A_\alpha(\mu)}$. For $\mu> 0$, the PDF vanishes at $x<\sqrt{\mu/\gamma}$. For $\mu<0$ small, in the special case $\alpha=1/2$, $\gamma=1$, the $n$-th moment is shown to be $\langle X^n\rangle =\sqrt{2/(2-n)}$ for $0<n<2$. For $n\geq2$, all moments $\langle X^n\rangle$ diverge.
    
    \item[(v)]  \textit{``Stable"} with $0<\alpha< 1$,  $\beta<1$ or $\alpha=1$ for any $\beta$. The origin is always stable in this case, the stationary PDF is $\delta(x)$ for all $\mu$ as long as $\gamma>0$.
    \end{enumerate}

 In summary, we have shown that instabilities under the influence of multiplicative heavy-tailed noise, modeled as L\'evy white noise, can display anomalous critical exponents differing from those for Gaussian noise, where $c_n=1$ for all $n$. Anomalous critical exponents different from the Gaussian results have been found previously, for instance for instabilities subject to colored noise  \cite{petrelis2012anomalous}. Here, we add the scenario of L\'evy white noise, which leads to several new possibilities of anomalous scaling, as discussed above. 

 \NOTE{Our work serves as a first step in the study of instabilities in the presence of multiplicative L{\'e}vy noise. There are many 
 directions that can be further pursued.
 First of all the values of  the power-law exponents $\lambda,\nu$ 
 in eqs. (\ref{eq:fit_pdf_beta0}), (\ref{eq:fit_pdf_beta-1}) remain unknown leading to only a non-rigorous estimate of the scaling exponents of the different moments with $\mu$.
 Furthermore, the behavior of the system under
 truncated L{\'e}vy noise \cite{mantegna1994stochastic,koponen1995analytic,schinckus2013physicists,chechkin2003bifurcation,chechkin2004levy}, combined L\'evy-Gaussian noise \cite{zan2020stochastic}, a finite-velocity L\'evy walk \cite{xu2020levy}, 
 different nonlinearities \cite{petrelis2006modification}, 
 higher dimensions \cite{graham1982stabilization,mallick2003stability,alexakis2009planar} and its time statistics \cite{heagy1994characterization,hammer1994experimental,rodelsperger1995off,feng1998off,john1999off,vella2003off,huerta2008experimental,aumaitre2007noise,bertin2012off} 
 would also be interesting to understand.
 Finally, since L\'evy statistics are found in many physical systems, we permit ourselves speculate that the anomalous critical exponents predicted here for instabilities in the presence of power-law noise may be observable experimentally.} 
 
\textcolor{black}{We stress that L\'evy noise is a theoretical idealization. From an experimental point of view, one can always compute all the moments of a random signal, firstly since it will be of finite duration $T$, and secondly because on physical grounds, infinite fluctuation amplitudes are unrealistic, such that a large cut-off is required. In a hypothetical experimental observation of L\'evy on-off intermittency, one could repeatedly increase the duration $T$ of the runs and measure the moments from the finite samples of increasing length. As the observation time is increased, the moments which are finite for ideal L\'evy noise will converge as $T$ increases. Those moments that diverge for ideal L\'evy noise will keep growing as $T$ increases. The effect of the noise truncation is to render all moments of the noise increments finite. This implies that the non-generalized central limit theorem applies, implying a convergence to Gaussian statistics at late times. However, it is well known that for large cut-off values, this convergence is "ultra-slow" \cite{mantegna1994stochastic,schinckus2013physicists}. Hence, our predictions for ideal L\'evy noise can be expected to be correct at intermediate observation times, large enough for the tails of the distribution to have been sampled, but short enough to avoid the eventual convergence to Gaussian statistics.}

\section{Acknowledgements}
\textcolor{black}{The authors thank three anonymous referees for their helpful remarks.} The authors also acknowledge helpful discussions with Fran{\c{c}}ois P\'etr\'elis and Stephan Fauve, and thank Olivier B\'enichou for bringing the example of blinking quantum dots to their attention. This work was granted access to the HPC resources of MesoPSL financed by the Region Ile de France and the project Equip@Meso (reference ANR-10-EQPX-29-01) of the programme Investissements d'Avenir supervised by the Agence Nationale pour la Recherche and the HPC resources of GENCI-TGCC \& GENCI-CINES (Project No. A0080511423,  A0090506421). 
This work has also been supported by the Agence nationale de la recherche (ANR DYSTURB project No. ANR-17-CE30-0004). AvK acknowledges support by Studienstiftung des deutschen Volkes.

\begin{appendix}
\section{Solution of the Langevin equation}
\label{sec:app_exact_sol}
Equation (\ref{eq:langevin}) is of the form of a Bernoulli differential equation. Hence, it admits an exact solution, which can be derived by dividing (\ref{eq:langevin}) by $X^3$ and letting $Z(t) = 1/X^2(t)$, such that
\begin{equation}
 \frac{d Z(t)}{dt}  + 2 r(t)Z(t) = 2 \gamma.
\end{equation}
This gives
\begin{equation}
    X(t) = \frac{\mathrm{sign}(x_0)}{\left(\frac{e^{-2\mu t -2 L(t)}}{x_0^2} + 2 \gamma \int_0^t e^{2\mu (t'-t) +2 (L(t')-L(t))}dt' \right)^{\frac{1}{2}}}, \label{eq:exact_soln_langevin}
\end{equation}
which is non-negative if $x_0\geq0$. We denote $dL/dt = f(t)$ with $f(t)$ white Levy noise, i.e. $L(t)$ is a free Levy flight. This solution is also given in \cite{aumaitre2007noise}, where it is stressed that it holds for any type of noise with $L(t)$ being the integral of the noise. By contrast with other nonlinear equations involving multiplicative L\'evy noise, such as \cite{dubkov2008verhulst}, where the analytical solution of the nonlinear Langevin equation gives access to the exact time-dependent PDF, this is impossible here since the result depends the integral of $L(t)$, in addition to $L(t)$. However, the expression is useful for numerical evaluation to generate realisations of the random process. It is advantageous over a direct iterative numerical integration since it does not require smaller time steps at large nonlinearity. Nonetheless, for large values of $L(t')-L(t)$, the exponential in the integrand may produce an overflow error. This can be avoided by choosing integration step $dt$ and the total integration time $t$ not too large. 

\section{Finite-difference numerical solution}
\label{sec:app_num_ffpe}
We recall the stationary space-fractional Fokker-Planck equation in the Stratonovich interpretation in terms of $Y=\log(X)$, which reads
\begin{equation}
    0 = - \partial_y[(\mu-\gamma e^{2y})p_{st}(y)] + \mathcal{D}_y^{\alpha,\beta}p_{st}(y),
\end{equation}
where the fractional derivative is given by
\begin{align}
    \mathcal{D}_y^{\alpha,\beta}p_{st}(y)=& - \frac{[(1+\beta)D_+^\alpha p_{st} + (1-\beta) D_-^\alpha p_{st}]}{2 \cos(\pi \alpha/2)} \\ 
    =& - \frac{[D_+^\alpha + D_-^\alpha +\beta (D_+^\alpha-D_-^\alpha)]p_{st}}{2 \cos(\pi \alpha/2)}.
\end{align}
We consider $1<\alpha<2$, for which Riemann-Liouville derivatives are given by (\ref{eq:Dplus_riemann_derivative}), and (\ref{eq:Dminus_riemann_derivative}). Integrating once in $y$ gives 
\begin{align}
0 =& -(\mu -\gamma e^{2y}) p_{st}(y) \notag \\&+ K_\alpha \frac{d}{dy}\int_{-\infty}^\infty  \frac{p_{st}(z)[1+\beta\mathrm{sign}(y-z)]}{|y-z|^{\alpha-1}} dz,
\end{align}
where $K_\alpha= -(2\cos(\pi\alpha/2)\Gamma(2-\alpha))^{-1}$. To simplify discretization further, it is advantageous to rewrite the term stemming from the fractional derivative in the the Gr\"unwald-Letnikov form, cf. \cite{liu2004numerical}, thereby transferring the $y$-derivative into the integral. This gives
\begin{equation}
    0 = - (\mu - \gamma e^{2y})p_{st}(y) + K_\alpha \int_{-\infty}^\infty \frac{p'_{st}(z)[1+\beta \mathrm{sign}(y-z)]}{|y-z|^{\alpha-1}}dz.
\end{equation}
For discretization we consider a large domain $[y_{min},y_{max}]$, meshed by intervals $[y_{n-1},y_n]$, whose  $N+1$ endpoints are $y_n$, where $p_{st}(y_n)=p_n$, with $n=0,\dots,N$. We prescribe an arbitrary initial condition $p_{-1}>0$. Then, using a backward difference scheme for $f'(z)$ and regularising $|y-z|^{\alpha-1}\to |y-z|^{\alpha-1}+\epsilon$ ($0<\epsilon\ll 1$), we find a matrix equation
\begin{equation}
    b_n = \sum_{m=0}^N L_{n,m} p_m, \label{eq:discretisation_matrix_equation_al_gt_1}
\end{equation}
where $n=1,\dots,N$, 
\begin{equation}
    b_n = -K_\alpha \frac{p_{-1}((1+\beta \mathrm{sign}(y_n-y_0)))}{|y_n-y_0|^{\alpha-1}+\epsilon}
\end{equation}
and \begin{align}
    L_{n,m} =&  -\mu + \gamma e^{2y_0}  + K_\alpha  \frac{(1+\beta\mathrm{sign}(y_0-y_m)))}{|y_n-y_m|^{\alpha-1}+\epsilon} \\ +& K_\alpha \theta(N-1-m) \frac{(1+\beta \mathrm{sign}(y_n-y_{m+1}))}{|y_n-y_{m+1}|^{\alpha-1}+\epsilon},
\end{align}
where $\theta(x)$ designates the Heaviside function. Finally, the steady density $p_m,$ $m=1,\dots,N$ is obtained by inverting the matrix $L_{n,m}$ and normalising the result. For $\beta<1$ we chose a non-uniform grid, composed of a logarithmically spaced grid at $y<-O(100)$, combined with a uniform grid in the region $O(100)>y>-O(100)$. The total grid size was $N=24000$. For $\beta=1$, a uniform grid was used with $N=16000$ (the PDF does not extend to $y<0$ as far). Choosing $\epsilon$ on the order of the smallest grid resolution to the appropriate power $\alpha-1$ gives results consistent with exact theoretical predictions, as described in the text.

\end{appendix}
\bibliography{lit}

\begin{thebibliography}{115}%
\makeatletter
\providecommand \@ifxundefined [1]{%
 \@ifx{#1\undefined}
}%
\providecommand \@ifnum [1]{%
 \ifnum #1\expandafter \@firstoftwo
 \else \expandafter \@secondoftwo
 \fi
}%
\providecommand \@ifx [1]{%
 \ifx #1\expandafter \@firstoftwo
 \else \expandafter \@secondoftwo
 \fi
}%
\providecommand \natexlab [1]{#1}%
\providecommand \enquote  [1]{``#1''}%
\providecommand \bibnamefont  [1]{#1}%
\providecommand \bibfnamefont [1]{#1}%
\providecommand \citenamefont [1]{#1}%
\providecommand \href@noop [0]{\@secondoftwo}%
\providecommand \href [0]{\begingroup \@sanitize@url \@href}%
\providecommand \@href[1]{\@@startlink{#1}\@@href}%
\providecommand \@@href[1]{\endgroup#1\@@endlink}%
\providecommand \@sanitize@url [0]{\catcode `\\12\catcode `\$12\catcode
  `\&12\catcode `\#12\catcode `\^12\catcode `\_12\catcode `\%12\relax}%
\providecommand \@@startlink[1]{}%
\providecommand \@@endlink[0]{}%
\providecommand \url  [0]{\begingroup\@sanitize@url \@url }%
\providecommand \@url [1]{\endgroup\@href {#1}{\urlprefix }}%
\providecommand \urlprefix  [0]{URL }%
\providecommand \Eprint [0]{\href }%
\providecommand \doibase [0]{https://doi.org/}%
\providecommand \selectlanguage [0]{\@gobble}%
\providecommand \bibinfo  [0]{\@secondoftwo}%
\providecommand \bibfield  [0]{\@secondoftwo}%
\providecommand \translation [1]{[#1]}%
\providecommand \BibitemOpen [0]{}%
\providecommand \bibitemStop [0]{}%
\providecommand \bibitemNoStop [0]{.\EOS\space}%
\providecommand \EOS [0]{\spacefactor3000\relax}%
\providecommand \BibitemShut  [1]{\csname bibitem#1\endcsname}%
\let\auto@bib@innerbib\@empty
\bibitem [{\citenamefont {Fujisaka}\ and\ \citenamefont
  {Yamada}(1985)}]{fujisaka1985new}%
  \BibitemOpen
  \bibfield  {author} {\bibinfo {author} {\bibfnamefont {H.}~\bibnamefont
  {Fujisaka}}\ and\ \bibinfo {author} {\bibfnamefont {T.}~\bibnamefont
  {Yamada}},\ }\bibfield  {title} {\bibinfo {title} {A new intermittency in
  coupled dynamical systems},\ }\href@noop {} {\bibfield  {journal} {\bibinfo
  {journal} {Progress of theoretical physics}\ }\textbf {\bibinfo {volume}
  {74}},\ \bibinfo {pages} {918} (\bibinfo {year} {1985})}\BibitemShut
  {NoStop}%
\bibitem [{\citenamefont {Platt}\ \emph {et~al.}(1993)\citenamefont {Platt},
  \citenamefont {Spiegel},\ and\ \citenamefont {Tresser}}]{platt1993off}%
  \BibitemOpen
  \bibfield  {author} {\bibinfo {author} {\bibfnamefont {N.}~\bibnamefont
  {Platt}}, \bibinfo {author} {\bibfnamefont {E.}~\bibnamefont {Spiegel}},\
  and\ \bibinfo {author} {\bibfnamefont {C.}~\bibnamefont {Tresser}},\
  }\bibfield  {title} {\bibinfo {title} {On-off intermittency: A mechanism for
  bursting},\ }\href@noop {} {\bibfield  {journal} {\bibinfo  {journal}
  {Physical Review Letters}\ }\textbf {\bibinfo {volume} {70}},\ \bibinfo
  {pages} {279} (\bibinfo {year} {1993})}\BibitemShut {NoStop}%
\bibitem [{\citenamefont {Ott}\ and\ \citenamefont
  {Sommerer}(1994)}]{ott1994blowout}%
  \BibitemOpen
  \bibfield  {author} {\bibinfo {author} {\bibfnamefont {E.}~\bibnamefont
  {Ott}}\ and\ \bibinfo {author} {\bibfnamefont {J.~C.}\ \bibnamefont
  {Sommerer}},\ }\bibfield  {title} {\bibinfo {title} {Blowout bifurcations:
  the occurrence of riddled basins and on-off intermittency},\ }\href@noop {}
  {\bibfield  {journal} {\bibinfo  {journal} {Physics Letters A}\ }\textbf
  {\bibinfo {volume} {188}},\ \bibinfo {pages} {39} (\bibinfo {year}
  {1994})}\BibitemShut {NoStop}%
\bibitem [{\citenamefont {Heagy}\ \emph {et~al.}(1994)\citenamefont {Heagy},
  \citenamefont {Platt},\ and\ \citenamefont
  {Hammel}}]{heagy1994characterization}%
  \BibitemOpen
  \bibfield  {author} {\bibinfo {author} {\bibfnamefont {J.}~\bibnamefont
  {Heagy}}, \bibinfo {author} {\bibfnamefont {N.}~\bibnamefont {Platt}},\ and\
  \bibinfo {author} {\bibfnamefont {S.}~\bibnamefont {Hammel}},\ }\bibfield
  {title} {\bibinfo {title} {Characterization of on-off intermittency},\
  }\href@noop {} {\bibfield  {journal} {\bibinfo  {journal} {Physical Review
  E}\ }\textbf {\bibinfo {volume} {49}},\ \bibinfo {pages} {1140} (\bibinfo
  {year} {1994})}\BibitemShut {NoStop}%
\bibitem [{\citenamefont {Hammer}\ \emph {et~al.}(1994)\citenamefont {Hammer},
  \citenamefont {Platt}, \citenamefont {Hammel}, \citenamefont {Heagy},\ and\
  \citenamefont {Lee}}]{hammer1994experimental}%
  \BibitemOpen
  \bibfield  {author} {\bibinfo {author} {\bibfnamefont {P.~W.}\ \bibnamefont
  {Hammer}}, \bibinfo {author} {\bibfnamefont {N.}~\bibnamefont {Platt}},
  \bibinfo {author} {\bibfnamefont {S.~M.}\ \bibnamefont {Hammel}}, \bibinfo
  {author} {\bibfnamefont {J.~F.}\ \bibnamefont {Heagy}},\ and\ \bibinfo
  {author} {\bibfnamefont {B.~D.}\ \bibnamefont {Lee}},\ }\bibfield  {title}
  {\bibinfo {title} {Experimental observation of on-off intermittency},\
  }\href@noop {} {\bibfield  {journal} {\bibinfo  {journal} {Physical review
  letters}\ }\textbf {\bibinfo {volume} {73}},\ \bibinfo {pages} {1095}
  (\bibinfo {year} {1994})}\BibitemShut {NoStop}%
\bibitem [{\citenamefont {R{\"o}delsperger}\ \emph {et~al.}(1995)\citenamefont
  {R{\"o}delsperger}, \citenamefont {{\v{C}}enys},\ and\ \citenamefont
  {Benner}}]{rodelsperger1995off}%
  \BibitemOpen
  \bibfield  {author} {\bibinfo {author} {\bibfnamefont {F.}~\bibnamefont
  {R{\"o}delsperger}}, \bibinfo {author} {\bibfnamefont {A.}~\bibnamefont
  {{\v{C}}enys}},\ and\ \bibinfo {author} {\bibfnamefont {H.}~\bibnamefont
  {Benner}},\ }\bibfield  {title} {\bibinfo {title} {On-off intermittency in
  spin-wave instabilities},\ }\href@noop {} {\bibfield  {journal} {\bibinfo
  {journal} {Physical review letters}\ }\textbf {\bibinfo {volume} {75}},\
  \bibinfo {pages} {2594} (\bibinfo {year} {1995})}\BibitemShut {NoStop}%
\bibitem [{\citenamefont {John}\ \emph {et~al.}(1999)\citenamefont {John},
  \citenamefont {Stannarius},\ and\ \citenamefont {Behn}}]{john1999off}%
  \BibitemOpen
  \bibfield  {author} {\bibinfo {author} {\bibfnamefont {T.}~\bibnamefont
  {John}}, \bibinfo {author} {\bibfnamefont {R.}~\bibnamefont {Stannarius}},\
  and\ \bibinfo {author} {\bibfnamefont {U.}~\bibnamefont {Behn}},\ }\bibfield
  {title} {\bibinfo {title} {On-off intermittency in stochastically driven
  electrohydrodynamic convection in nematics},\ }\href@noop {} {\bibfield
  {journal} {\bibinfo  {journal} {Physical review letters}\ }\textbf {\bibinfo
  {volume} {83}},\ \bibinfo {pages} {749} (\bibinfo {year} {1999})}\BibitemShut
  {NoStop}%
\bibitem [{\citenamefont {Vella}\ \emph {et~al.}(2003)\citenamefont {Vella},
  \citenamefont {Setaro}, \citenamefont {Piccirillo},\ and\ \citenamefont
  {Santamato}}]{vella2003off}%
  \BibitemOpen
  \bibfield  {author} {\bibinfo {author} {\bibfnamefont {A.}~\bibnamefont
  {Vella}}, \bibinfo {author} {\bibfnamefont {A.}~\bibnamefont {Setaro}},
  \bibinfo {author} {\bibfnamefont {B.}~\bibnamefont {Piccirillo}},\ and\
  \bibinfo {author} {\bibfnamefont {E.}~\bibnamefont {Santamato}},\ }\bibfield
  {title} {\bibinfo {title} {On-off intermittency in chaotic rotation induced
  in liquid crystals by competition between spin and orbital angular momentum
  of light},\ }\href@noop {} {\bibfield  {journal} {\bibinfo  {journal}
  {Physical Review E}\ }\textbf {\bibinfo {volume} {67}},\ \bibinfo {pages}
  {051704} (\bibinfo {year} {2003})}\BibitemShut {NoStop}%
\bibitem [{\citenamefont {Feng}\ \emph {et~al.}(1998)\citenamefont {Feng},
  \citenamefont {Yu}, \citenamefont {Xie},\ and\ \citenamefont
  {Ding}}]{feng1998off}%
  \BibitemOpen
  \bibfield  {author} {\bibinfo {author} {\bibfnamefont {D.}~\bibnamefont
  {Feng}}, \bibinfo {author} {\bibfnamefont {C.}~\bibnamefont {Yu}}, \bibinfo
  {author} {\bibfnamefont {J.}~\bibnamefont {Xie}},\ and\ \bibinfo {author}
  {\bibfnamefont {W.}~\bibnamefont {Ding}},\ }\bibfield  {title} {\bibinfo
  {title} {On-off intermittencies in gas discharge plasma},\ }\href@noop {}
  {\bibfield  {journal} {\bibinfo  {journal} {Physical Review E}\ }\textbf
  {\bibinfo {volume} {58}},\ \bibinfo {pages} {3678} (\bibinfo {year}
  {1998})}\BibitemShut {NoStop}%
\bibitem [{\citenamefont {Huerta-Cuellar}\ \emph {et~al.}(2008)\citenamefont
  {Huerta-Cuellar}, \citenamefont {Pisarchik},\ and\ \citenamefont
  {Barmenkov}}]{huerta2008experimental}%
  \BibitemOpen
  \bibfield  {author} {\bibinfo {author} {\bibfnamefont {G.}~\bibnamefont
  {Huerta-Cuellar}}, \bibinfo {author} {\bibfnamefont {A.~N.}\ \bibnamefont
  {Pisarchik}},\ and\ \bibinfo {author} {\bibfnamefont {Y.~O.}\ \bibnamefont
  {Barmenkov}},\ }\bibfield  {title} {\bibinfo {title} {Experimental
  characterization of hopping dynamics in a multistable fiber laser},\
  }\href@noop {} {\bibfield  {journal} {\bibinfo  {journal} {Physical Review
  E}\ }\textbf {\bibinfo {volume} {78}},\ \bibinfo {pages} {035202} (\bibinfo
  {year} {2008})}\BibitemShut {NoStop}%
\bibitem [{\citenamefont {Benavides}\ \emph {et~al.}(2020)\citenamefont
  {Benavides}, \citenamefont {Deal}, \citenamefont {Perron}, \citenamefont
  {Venditti}, \citenamefont {Zhang},\ and\ \citenamefont
  {Kamrin}}]{benavides2020multiplicative}%
  \BibitemOpen
  \bibfield  {author} {\bibinfo {author} {\bibfnamefont {S.}~\bibnamefont
  {Benavides}}, \bibinfo {author} {\bibfnamefont {E.}~\bibnamefont {Deal}},
  \bibinfo {author} {\bibfnamefont {J.}~\bibnamefont {Perron}}, \bibinfo
  {author} {\bibfnamefont {J.}~\bibnamefont {Venditti}}, \bibinfo {author}
  {\bibfnamefont {Q.}~\bibnamefont {Zhang}},\ and\ \bibinfo {author}
  {\bibfnamefont {K.}~\bibnamefont {Kamrin}},\ }\href@noop {} {\emph {\bibinfo
  {title} {Multiplicative noise and intermittency in bedload sediment
  transport}}}\ (\bibinfo {year} {2020})\BibitemShut {NoStop}%
\bibitem [{\citenamefont {Cabrera}\ and\ \citenamefont
  {Milton}(2002)}]{cabrera2002off}%
  \BibitemOpen
  \bibfield  {author} {\bibinfo {author} {\bibfnamefont {J.~L.}\ \bibnamefont
  {Cabrera}}\ and\ \bibinfo {author} {\bibfnamefont {J.~G.}\ \bibnamefont
  {Milton}},\ }\bibfield  {title} {\bibinfo {title} {On-off intermittency in a
  human balancing task},\ }\href@noop {} {\bibfield  {journal} {\bibinfo
  {journal} {Physical Review Letters}\ }\textbf {\bibinfo {volume} {89}},\
  \bibinfo {pages} {158702} (\bibinfo {year} {2002})}\BibitemShut {NoStop}%
\bibitem [{\citenamefont {Cabrera}\ and\ \citenamefont
  {Milton}(2004{\natexlab{a}})}]{cabrera2004stick}%
  \BibitemOpen
  \bibfield  {author} {\bibinfo {author} {\bibfnamefont {J.~L.}\ \bibnamefont
  {Cabrera}}\ and\ \bibinfo {author} {\bibfnamefont {J.~G.}\ \bibnamefont
  {Milton}},\ }\bibfield  {title} {\bibinfo {title} {Stick balancing: On-off
  intermittency and survival times},\ }\href@noop {} {\bibfield  {journal}
  {\bibinfo  {journal} {Nonlinear Studies}\ }\textbf {\bibinfo {volume} {11}},\
  \bibinfo {pages} {305} (\bibinfo {year} {2004}{\natexlab{a}})}\BibitemShut
  {NoStop}%
\bibitem [{\citenamefont {Margolin}\ \emph {et~al.}(2005)\citenamefont
  {Margolin}, \citenamefont {Protasenko}, \citenamefont {Kuno},\ and\
  \citenamefont {Barkai}}]{margolin2005power}%
  \BibitemOpen
  \bibfield  {author} {\bibinfo {author} {\bibfnamefont {G.}~\bibnamefont
  {Margolin}}, \bibinfo {author} {\bibfnamefont {V.}~\bibnamefont
  {Protasenko}}, \bibinfo {author} {\bibfnamefont {M.}~\bibnamefont {Kuno}},\
  and\ \bibinfo {author} {\bibfnamefont {E.}~\bibnamefont {Barkai}},\
  }\bibfield  {title} {\bibinfo {title} {Power law blinking quantum dots:
  Stochastic and physical models},\ }\href@noop {} {\bibfield  {journal}
  {\bibinfo  {journal} {arXiv preprint cond-mat/0506512}\ } (\bibinfo {year}
  {2005})}\BibitemShut {NoStop}%
\bibitem [{\citenamefont {Frantsuzov}\ \emph {et~al.}(2008)\citenamefont
  {Frantsuzov}, \citenamefont {Kuno}, \citenamefont {Janko},\ and\
  \citenamefont {Marcus}}]{frantsuzov2008universal}%
  \BibitemOpen
  \bibfield  {author} {\bibinfo {author} {\bibfnamefont {P.}~\bibnamefont
  {Frantsuzov}}, \bibinfo {author} {\bibfnamefont {M.}~\bibnamefont {Kuno}},
  \bibinfo {author} {\bibfnamefont {B.}~\bibnamefont {Janko}},\ and\ \bibinfo
  {author} {\bibfnamefont {R.~A.}\ \bibnamefont {Marcus}},\ }\bibfield  {title}
  {\bibinfo {title} {Universal emission intermittency in quantum dots, nanorods
  and nanowires},\ }\href@noop {} {\bibfield  {journal} {\bibinfo  {journal}
  {Nature Physics}\ }\textbf {\bibinfo {volume} {4}},\ \bibinfo {pages} {519}
  (\bibinfo {year} {2008})}\BibitemShut {NoStop}%
\bibitem [{\citenamefont {Benavides}\ and\ \citenamefont
  {Alexakis}(2017)}]{benavides2017critical}%
  \BibitemOpen
  \bibfield  {author} {\bibinfo {author} {\bibfnamefont {S.~J.}\ \bibnamefont
  {Benavides}}\ and\ \bibinfo {author} {\bibfnamefont {A.}~\bibnamefont
  {Alexakis}},\ }\bibfield  {title} {\bibinfo {title} {Critical transitions in
  thin layer turbulence},\ }\href@noop {} {\bibfield  {journal} {\bibinfo
  {journal} {Journal of Fluid Mechanics}\ }\textbf {\bibinfo {volume} {822}},\
  \bibinfo {pages} {364} (\bibinfo {year} {2017})}\BibitemShut {NoStop}%
\bibitem [{\citenamefont {van Kan}\ and\ \citenamefont
  {Alexakis}(2019)}]{van2019condensates}%
  \BibitemOpen
  \bibfield  {author} {\bibinfo {author} {\bibfnamefont {A.}~\bibnamefont {van
  Kan}}\ and\ \bibinfo {author} {\bibfnamefont {A.}~\bibnamefont {Alexakis}},\
  }\bibfield  {title} {\bibinfo {title} {Condensates in thin-layer
  turbulence},\ }\href@noop {} {\bibfield  {journal} {\bibinfo  {journal}
  {Journal of Fluid Mechanics}\ }\textbf {\bibinfo {volume} {864}},\ \bibinfo
  {pages} {490} (\bibinfo {year} {2019})}\BibitemShut {NoStop}%
\bibitem [{\citenamefont {Sweet}\ \emph {et~al.}(2001)\citenamefont {Sweet},
  \citenamefont {Ott}, \citenamefont {Finn}, \citenamefont {Antonsen~Jr},\ and\
  \citenamefont {Lathrop}}]{sweet2001blowout}%
  \BibitemOpen
  \bibfield  {author} {\bibinfo {author} {\bibfnamefont {D.}~\bibnamefont
  {Sweet}}, \bibinfo {author} {\bibfnamefont {E.}~\bibnamefont {Ott}}, \bibinfo
  {author} {\bibfnamefont {J.~M.}\ \bibnamefont {Finn}}, \bibinfo {author}
  {\bibfnamefont {T.~M.}\ \bibnamefont {Antonsen~Jr}},\ and\ \bibinfo {author}
  {\bibfnamefont {D.~P.}\ \bibnamefont {Lathrop}},\ }\bibfield  {title}
  {\bibinfo {title} {Blowout bifurcations and the onset of magnetic activity in
  turbulent dynamos},\ }\href@noop {} {\bibfield  {journal} {\bibinfo
  {journal} {Physical Review E}\ }\textbf {\bibinfo {volume} {63}},\ \bibinfo
  {pages} {066211} (\bibinfo {year} {2001})}\BibitemShut {NoStop}%
\bibitem [{\citenamefont {Alexakis}\ and\ \citenamefont
  {Ponty}(2008)}]{alexakis2008effect}%
  \BibitemOpen
  \bibfield  {author} {\bibinfo {author} {\bibfnamefont {A.}~\bibnamefont
  {Alexakis}}\ and\ \bibinfo {author} {\bibfnamefont {Y.}~\bibnamefont
  {Ponty}},\ }\bibfield  {title} {\bibinfo {title} {Effect of the lorentz force
  on on-off dynamo intermittency},\ }\href@noop {} {\bibfield  {journal}
  {\bibinfo  {journal} {Physical Review E}\ }\textbf {\bibinfo {volume} {77}},\
  \bibinfo {pages} {056308} (\bibinfo {year} {2008})}\BibitemShut {NoStop}%
\bibitem [{\citenamefont {Raynaud}\ and\ \citenamefont
  {Dormy}(2013)}]{raynaud2013intermittency}%
  \BibitemOpen
  \bibfield  {author} {\bibinfo {author} {\bibfnamefont {R.}~\bibnamefont
  {Raynaud}}\ and\ \bibinfo {author} {\bibfnamefont {E.}~\bibnamefont
  {Dormy}},\ }\bibfield  {title} {\bibinfo {title} {Intermittency in spherical
  couette dynamos},\ }\href@noop {} {\bibfield  {journal} {\bibinfo  {journal}
  {Physical Review E}\ }\textbf {\bibinfo {volume} {87}},\ \bibinfo {pages}
  {033011} (\bibinfo {year} {2013})}\BibitemShut {NoStop}%
\bibitem [{\citenamefont {Strogatz}(2018)}]{strogatz2018nonlinear}%
  \BibitemOpen
  \bibfield  {author} {\bibinfo {author} {\bibfnamefont {S.~H.}\ \bibnamefont
  {Strogatz}},\ }\href@noop {} {\emph {\bibinfo {title} {Nonlinear dynamics and
  chaos with student solutions manual: With applications to physics, biology,
  chemistry, and engineering}}}\ (\bibinfo  {publisher} {CRC press},\ \bibinfo
  {year} {2018})\BibitemShut {NoStop}%
\bibitem [{\citenamefont {Stratonovich}(1966)}]{stratonovich1966new}%
  \BibitemOpen
  \bibfield  {author} {\bibinfo {author} {\bibfnamefont {R.}~\bibnamefont
  {Stratonovich}},\ }\bibfield  {title} {\bibinfo {title} {A new representation
  for stochastic integrals and equations},\ }\href@noop {} {\bibfield
  {journal} {\bibinfo  {journal} {SIAM Journal on Control}\ }\textbf {\bibinfo
  {volume} {4}},\ \bibinfo {pages} {362} (\bibinfo {year} {1966})}\BibitemShut
  {NoStop}%
\bibitem [{\citenamefont {Auma{\^\i}tre}\ \emph {et~al.}(2007)\citenamefont
  {Auma{\^\i}tre}, \citenamefont {Mallick},\ and\ \citenamefont
  {P{\'e}tr{\'e}lis}}]{aumaitre2007noise}%
  \BibitemOpen
  \bibfield  {author} {\bibinfo {author} {\bibfnamefont {S.}~\bibnamefont
  {Auma{\^\i}tre}}, \bibinfo {author} {\bibfnamefont {K.}~\bibnamefont
  {Mallick}},\ and\ \bibinfo {author} {\bibfnamefont {F.}~\bibnamefont
  {P{\'e}tr{\'e}lis}},\ }\bibfield  {title} {\bibinfo {title} {Noise-induced
  bifurcations, multiscaling and on--off intermittency},\ }\href@noop {}
  {\bibfield  {journal} {\bibinfo  {journal} {Journal of Statistical Mechanics:
  Theory and Experiment}\ }\textbf {\bibinfo {volume} {2007}},\ \bibinfo
  {pages} {P07016} (\bibinfo {year} {2007})}\BibitemShut {NoStop}%
\bibitem [{\citenamefont {Schenzle}\ and\ \citenamefont
  {Brand}(1979)}]{schenzle1979multiplicative}%
  \BibitemOpen
  \bibfield  {author} {\bibinfo {author} {\bibfnamefont {A.}~\bibnamefont
  {Schenzle}}\ and\ \bibinfo {author} {\bibfnamefont {H.}~\bibnamefont
  {Brand}},\ }\bibfield  {title} {\bibinfo {title} {Multiplicative stochastic
  processes in statistical physics},\ }\href@noop {} {\bibfield  {journal}
  {\bibinfo  {journal} {Physical Review A}\ }\textbf {\bibinfo {volume} {20}},\
  \bibinfo {pages} {1628} (\bibinfo {year} {1979})}\BibitemShut {NoStop}%
\bibitem [{\citenamefont {Kadanoff}\ \emph {et~al.}(1967)\citenamefont
  {Kadanoff}, \citenamefont {G{\"o}tze}, \citenamefont {Hamblen}, \citenamefont
  {Hecht}, \citenamefont {Lewis}, \citenamefont {Palciauskas}, \citenamefont
  {Rayl}, \citenamefont {Swift}, \citenamefont {Aspnes},\ and\ \citenamefont
  {Kane}}]{kadanoff1967static}%
  \BibitemOpen
  \bibfield  {author} {\bibinfo {author} {\bibfnamefont {L.~P.}\ \bibnamefont
  {Kadanoff}}, \bibinfo {author} {\bibfnamefont {W.}~\bibnamefont {G{\"o}tze}},
  \bibinfo {author} {\bibfnamefont {D.}~\bibnamefont {Hamblen}}, \bibinfo
  {author} {\bibfnamefont {R.}~\bibnamefont {Hecht}}, \bibinfo {author}
  {\bibfnamefont {E.}~\bibnamefont {Lewis}}, \bibinfo {author} {\bibfnamefont
  {V.~V.}\ \bibnamefont {Palciauskas}}, \bibinfo {author} {\bibfnamefont
  {M.}~\bibnamefont {Rayl}}, \bibinfo {author} {\bibfnamefont {J.}~\bibnamefont
  {Swift}}, \bibinfo {author} {\bibfnamefont {D.}~\bibnamefont {Aspnes}},\ and\
  \bibinfo {author} {\bibfnamefont {J.}~\bibnamefont {Kane}},\ }\bibfield
  {title} {\bibinfo {title} {Static phenomena near critical points: theory and
  experiment},\ }\href@noop {} {\bibfield  {journal} {\bibinfo  {journal}
  {Reviews of Modern Physics}\ }\textbf {\bibinfo {volume} {39}},\ \bibinfo
  {pages} {395} (\bibinfo {year} {1967})}\BibitemShut {NoStop}%
\bibitem [{\citenamefont {Goldenfeld}(2018)}]{goldenfeld2018lectures}%
  \BibitemOpen
  \bibfield  {author} {\bibinfo {author} {\bibfnamefont {N.}~\bibnamefont
  {Goldenfeld}},\ }\href@noop {} {\emph {\bibinfo {title} {Lectures on phase
  transitions and the renormalization group}}}\ (\bibinfo  {publisher} {CRC
  Press},\ \bibinfo {year} {2018})\BibitemShut {NoStop}%
\bibitem [{\citenamefont {Eyink}\ and\ \citenamefont
  {Goldenfeld}(1994)}]{eyink1994analogies}%
  \BibitemOpen
  \bibfield  {author} {\bibinfo {author} {\bibfnamefont {G.}~\bibnamefont
  {Eyink}}\ and\ \bibinfo {author} {\bibfnamefont {N.}~\bibnamefont
  {Goldenfeld}},\ }\bibfield  {title} {\bibinfo {title} {Analogies between
  scaling in turbulence, field theory, and critical phenomena},\ }\href@noop {}
  {\bibfield  {journal} {\bibinfo  {journal} {Physical Review E}\ }\textbf
  {\bibinfo {volume} {50}},\ \bibinfo {pages} {4679} (\bibinfo {year}
  {1994})}\BibitemShut {NoStop}%
\bibitem [{\citenamefont {Goldenfeld}\ and\ \citenamefont
  {Shih}(2017)}]{goldenfeld2017turbulence}%
  \BibitemOpen
  \bibfield  {author} {\bibinfo {author} {\bibfnamefont {N.}~\bibnamefont
  {Goldenfeld}}\ and\ \bibinfo {author} {\bibfnamefont {H.-Y.}\ \bibnamefont
  {Shih}},\ }\bibfield  {title} {\bibinfo {title} {Turbulence as a problem in
  non-equilibrium statistical mechanics},\ }\href@noop {} {\bibfield  {journal}
  {\bibinfo  {journal} {Journal of Statistical Physics}\ }\textbf {\bibinfo
  {volume} {167}},\ \bibinfo {pages} {575} (\bibinfo {year}
  {2017})}\BibitemShut {NoStop}%
\bibitem [{\citenamefont {Shlesinger}\ \emph {et~al.}(1995)\citenamefont
  {Shlesinger}, \citenamefont {Zaslavsky},\ and\ \citenamefont
  {Frisch}}]{shlesinger1995levy}%
  \BibitemOpen
  \bibfield  {author} {\bibinfo {author} {\bibfnamefont {M.~F.}\ \bibnamefont
  {Shlesinger}}, \bibinfo {author} {\bibfnamefont {G.~M.}\ \bibnamefont
  {Zaslavsky}},\ and\ \bibinfo {author} {\bibfnamefont {U.}~\bibnamefont
  {Frisch}},\ }\href@noop {} {\emph {\bibinfo {title} {L{\'e}vy flights and
  related topics in physics}}}\ (\bibinfo  {publisher} {Springer},\ \bibinfo
  {year} {1995})\BibitemShut {NoStop}%
\bibitem [{\citenamefont {Chechkin}\ \emph {et~al.}(2008)\citenamefont
  {Chechkin}, \citenamefont {Metzler}, \citenamefont {Klafter}, \citenamefont
  {Gonchar} \emph {et~al.}}]{chechkin2008introduction}%
  \BibitemOpen
  \bibfield  {author} {\bibinfo {author} {\bibfnamefont {A.~V.}\ \bibnamefont
  {Chechkin}}, \bibinfo {author} {\bibfnamefont {R.}~\bibnamefont {Metzler}},
  \bibinfo {author} {\bibfnamefont {J.}~\bibnamefont {Klafter}}, \bibinfo
  {author} {\bibfnamefont {V.~Y.}\ \bibnamefont {Gonchar}}, \emph {et~al.},\
  }\bibfield  {title} {\bibinfo {title} {Introduction to the theory of l{\'e}vy
  flights},\ }\href@noop {} {\bibfield  {journal} {\bibinfo  {journal}
  {Anomalous Transport}\ ,\ \bibinfo {pages} {129}} (\bibinfo {year}
  {2008})}\BibitemShut {NoStop}%
\bibitem [{\citenamefont {Feller}(2008)}]{feller2008introduction}%
  \BibitemOpen
  \bibfield  {author} {\bibinfo {author} {\bibfnamefont {W.}~\bibnamefont
  {Feller}},\ }\href@noop {} {\emph {\bibinfo {title} {An introduction to
  probability theory and its applications, vol 2}}}\ (\bibinfo  {publisher}
  {John Wiley \& Sons},\ \bibinfo {year} {2008})\BibitemShut {NoStop}%
\bibitem [{\citenamefont {Gnedenko}\ \emph {et~al.}(1954)\citenamefont
  {Gnedenko}, \citenamefont {Kolmogorov}, \citenamefont {Gnedenko},\ and\
  \citenamefont {Kolmogorov}}]{gnedenko1954limit}%
  \BibitemOpen
  \bibfield  {author} {\bibinfo {author} {\bibfnamefont {B.}~\bibnamefont
  {Gnedenko}}, \bibinfo {author} {\bibfnamefont {A.}~\bibnamefont
  {Kolmogorov}}, \bibinfo {author} {\bibfnamefont {B.}~\bibnamefont
  {Gnedenko}},\ and\ \bibinfo {author} {\bibfnamefont {A.}~\bibnamefont
  {Kolmogorov}},\ }\bibfield  {title} {\bibinfo {title} {Limit distributions
  for sums of independent},\ }\href@noop {} {\bibfield  {journal} {\bibinfo
  {journal} {Am. J. Math}\ }\textbf {\bibinfo {volume} {105}} (\bibinfo {year}
  {1954})}\BibitemShut {NoStop}%
\bibitem [{\citenamefont {Uchaikin}\ and\ \citenamefont
  {Zolotarev}(2011)}]{uchaikin2011chance}%
  \BibitemOpen
  \bibfield  {author} {\bibinfo {author} {\bibfnamefont {V.~V.}\ \bibnamefont
  {Uchaikin}}\ and\ \bibinfo {author} {\bibfnamefont {V.~M.}\ \bibnamefont
  {Zolotarev}},\ }\href@noop {} {\emph {\bibinfo {title} {Chance and stability:
  stable distributions and their applications}}}\ (\bibinfo  {publisher}
  {Walter de Gruyter},\ \bibinfo {year} {2011})\BibitemShut {NoStop}%
\bibitem [{\citenamefont {Dybiec}\ and\ \citenamefont
  {Gudowska-Nowak}(2004)}]{dybiec2004resonant}%
  \BibitemOpen
  \bibfield  {author} {\bibinfo {author} {\bibfnamefont {B.}~\bibnamefont
  {Dybiec}}\ and\ \bibinfo {author} {\bibfnamefont {E.}~\bibnamefont
  {Gudowska-Nowak}},\ }\bibfield  {title} {\bibinfo {title} {Resonant
  activation in the presence of nonequilibrated baths},\ }\href@noop {}
  {\bibfield  {journal} {\bibinfo  {journal} {Physical Review E}\ }\textbf
  {\bibinfo {volume} {69}},\ \bibinfo {pages} {016105} (\bibinfo {year}
  {2004})}\BibitemShut {NoStop}%
\bibitem [{\citenamefont {Mandelbrot}(1983)}]{mandelbrot1983fractal}%
  \BibitemOpen
  \bibfield  {author} {\bibinfo {author} {\bibfnamefont {B.~B.}\ \bibnamefont
  {Mandelbrot}},\ }\href@noop {} {\emph {\bibinfo {title} {The fractal geometry
  of nature}}},\ Vol.\ \bibinfo {volume} {173}\ (\bibinfo  {publisher} {WH
  freeman New York},\ \bibinfo {year} {1983})\BibitemShut {NoStop}%
\bibitem [{\citenamefont {Shlesinger}\ \emph {et~al.}(1987)\citenamefont
  {Shlesinger}, \citenamefont {West},\ and\ \citenamefont
  {Klafter}}]{shlesinger1987levy}%
  \BibitemOpen
  \bibfield  {author} {\bibinfo {author} {\bibfnamefont {M.}~\bibnamefont
  {Shlesinger}}, \bibinfo {author} {\bibfnamefont {B.}~\bibnamefont {West}},\
  and\ \bibinfo {author} {\bibfnamefont {J.}~\bibnamefont {Klafter}},\
  }\bibfield  {title} {\bibinfo {title} {L{\'e}vy dynamics of enhanced
  diffusion: Application to turbulence},\ }\href@noop {} {\bibfield  {journal}
  {\bibinfo  {journal} {Physical Review Letters}\ }\textbf {\bibinfo {volume}
  {58}},\ \bibinfo {pages} {1100} (\bibinfo {year} {1987})}\BibitemShut
  {NoStop}%
\bibitem [{\citenamefont {Solomon}\ \emph {et~al.}(1993)\citenamefont
  {Solomon}, \citenamefont {Weeks},\ and\ \citenamefont
  {Swinney}}]{solomon1993observation}%
  \BibitemOpen
  \bibfield  {author} {\bibinfo {author} {\bibfnamefont {T.}~\bibnamefont
  {Solomon}}, \bibinfo {author} {\bibfnamefont {E.~R.}\ \bibnamefont {Weeks}},\
  and\ \bibinfo {author} {\bibfnamefont {H.~L.}\ \bibnamefont {Swinney}},\
  }\bibfield  {title} {\bibinfo {title} {Observation of anomalous diffusion and
  l{\'e}vy flights in a two-dimensional rotating flow},\ }\href@noop {}
  {\bibfield  {journal} {\bibinfo  {journal} {Physical Review Letters}\
  }\textbf {\bibinfo {volume} {71}},\ \bibinfo {pages} {3975} (\bibinfo {year}
  {1993})}\BibitemShut {NoStop}%
\bibitem [{\citenamefont {Metzler}\ and\ \citenamefont
  {Klafter}(2000)}]{metzler2000random}%
  \BibitemOpen
  \bibfield  {author} {\bibinfo {author} {\bibfnamefont {R.}~\bibnamefont
  {Metzler}}\ and\ \bibinfo {author} {\bibfnamefont {J.}~\bibnamefont
  {Klafter}},\ }\bibfield  {title} {\bibinfo {title} {The random walk's guide
  to anomalous diffusion: a fractional dynamics approach},\ }\href@noop {}
  {\bibfield  {journal} {\bibinfo  {journal} {Physics reports}\ }\textbf
  {\bibinfo {volume} {339}},\ \bibinfo {pages} {1} (\bibinfo {year}
  {2000})}\BibitemShut {NoStop}%
\bibitem [{\citenamefont {Dubkov}\ \emph {et~al.}(2008)\citenamefont {Dubkov},
  \citenamefont {Spagnolo},\ and\ \citenamefont {Uchaikin}}]{dubkov2008levy}%
  \BibitemOpen
  \bibfield  {author} {\bibinfo {author} {\bibfnamefont {A.~A.}\ \bibnamefont
  {Dubkov}}, \bibinfo {author} {\bibfnamefont {B.}~\bibnamefont {Spagnolo}},\
  and\ \bibinfo {author} {\bibfnamefont {V.~V.}\ \bibnamefont {Uchaikin}},\
  }\bibfield  {title} {\bibinfo {title} {L{\'e}vy flight superdiffusion: an
  introduction},\ }\href@noop {} {\bibfield  {journal} {\bibinfo  {journal}
  {International Journal of Bifurcation and Chaos}\ }\textbf {\bibinfo {volume}
  {18}},\ \bibinfo {pages} {2649} (\bibinfo {year} {2008})}\BibitemShut
  {NoStop}%
\bibitem [{\citenamefont {Dubrulle}\ and\ \citenamefont
  {Laval}(1998)}]{dubrulle1998truncated}%
  \BibitemOpen
  \bibfield  {author} {\bibinfo {author} {\bibfnamefont {B.}~\bibnamefont
  {Dubrulle}}\ and\ \bibinfo {author} {\bibfnamefont {J.-P.}\ \bibnamefont
  {Laval}},\ }\bibfield  {title} {\bibinfo {title} {Truncated l{\'e}vy laws and
  2d turbulence},\ }\href@noop {} {\bibfield  {journal} {\bibinfo  {journal}
  {The European Physical Journal B-Condensed Matter and Complex Systems}\
  }\textbf {\bibinfo {volume} {4}},\ \bibinfo {pages} {143} (\bibinfo {year}
  {1998})}\BibitemShut {NoStop}%
\bibitem [{\citenamefont {del Castillo-Negrete}\ \emph
  {et~al.}(2005)\citenamefont {del Castillo-Negrete}, \citenamefont
  {Carreras},\ and\ \citenamefont {Lynch}}]{del2005nondiffusive}%
  \BibitemOpen
  \bibfield  {author} {\bibinfo {author} {\bibfnamefont {D.}~\bibnamefont {del
  Castillo-Negrete}}, \bibinfo {author} {\bibfnamefont {B.}~\bibnamefont
  {Carreras}},\ and\ \bibinfo {author} {\bibfnamefont {V.}~\bibnamefont
  {Lynch}},\ }\bibfield  {title} {\bibinfo {title} {Nondiffusive transport in
  plasma turbulence: a fractional diffusion approach},\ }\href@noop {}
  {\bibfield  {journal} {\bibinfo  {journal} {Physical review letters}\
  }\textbf {\bibinfo {volume} {94}},\ \bibinfo {pages} {065003} (\bibinfo
  {year} {2005})}\BibitemShut {NoStop}%
\bibitem [{\citenamefont {Schinckus}(2013)}]{schinckus2013physicists}%
  \BibitemOpen
  \bibfield  {author} {\bibinfo {author} {\bibfnamefont {C.}~\bibnamefont
  {Schinckus}},\ }\bibfield  {title} {\bibinfo {title} {How physicists made
  stable l{\'e}vy processes physically plausible},\ }\href@noop {} {\bibfield
  {journal} {\bibinfo  {journal} {Brazilian Journal of Physics}\ }\textbf
  {\bibinfo {volume} {43}},\ \bibinfo {pages} {281} (\bibinfo {year}
  {2013})}\BibitemShut {NoStop}%
\bibitem [{\citenamefont
  {Ditlevsen}(1999{\natexlab{a}})}]{ditlevsen1999anomalous}%
  \BibitemOpen
  \bibfield  {author} {\bibinfo {author} {\bibfnamefont {P.~D.}\ \bibnamefont
  {Ditlevsen}},\ }\bibfield  {title} {\bibinfo {title} {Anomalous jumping in a
  double-well potential},\ }\href@noop {} {\bibfield  {journal} {\bibinfo
  {journal} {Physical Review E}\ }\textbf {\bibinfo {volume} {60}},\ \bibinfo
  {pages} {172} (\bibinfo {year} {1999}{\natexlab{a}})}\BibitemShut {NoStop}%
\bibitem [{\citenamefont
  {Ditlevsen}(1999{\natexlab{b}})}]{ditlevsen1999observation}%
  \BibitemOpen
  \bibfield  {author} {\bibinfo {author} {\bibfnamefont {P.~D.}\ \bibnamefont
  {Ditlevsen}},\ }\bibfield  {title} {\bibinfo {title} {Observation of
  $\alpha$-stable noise induced millennial climate changes from an ice-core
  record},\ }\href@noop {} {\bibfield  {journal} {\bibinfo  {journal}
  {Geophysical Research Letters}\ }\textbf {\bibinfo {volume} {26}},\ \bibinfo
  {pages} {1441} (\bibinfo {year} {1999}{\natexlab{b}})}\BibitemShut {NoStop}%
\bibitem [{\citenamefont {Viswanathan}\ \emph {et~al.}(1996)\citenamefont
  {Viswanathan}, \citenamefont {Afanasyev}, \citenamefont {Buldyrev},
  \citenamefont {Murphy}, \citenamefont {Prince},\ and\ \citenamefont
  {Stanley}}]{viswanathan1996levy}%
  \BibitemOpen
  \bibfield  {author} {\bibinfo {author} {\bibfnamefont {G.~M.}\ \bibnamefont
  {Viswanathan}}, \bibinfo {author} {\bibfnamefont {V.}~\bibnamefont
  {Afanasyev}}, \bibinfo {author} {\bibfnamefont {S.}~\bibnamefont {Buldyrev}},
  \bibinfo {author} {\bibfnamefont {E.}~\bibnamefont {Murphy}}, \bibinfo
  {author} {\bibfnamefont {P.}~\bibnamefont {Prince}},\ and\ \bibinfo {author}
  {\bibfnamefont {H.~E.}\ \bibnamefont {Stanley}},\ }\bibfield  {title}
  {\bibinfo {title} {L{\'e}vy flight search patterns of wandering
  albatrosses},\ }\href@noop {} {\bibfield  {journal} {\bibinfo  {journal}
  {Nature}\ }\textbf {\bibinfo {volume} {381}},\ \bibinfo {pages} {413}
  (\bibinfo {year} {1996})}\BibitemShut {NoStop}%
\bibitem [{\citenamefont {Sims}\ \emph {et~al.}(2008)\citenamefont {Sims},
  \citenamefont {Southall}, \citenamefont {Humphries}, \citenamefont {Hays},
  \citenamefont {Bradshaw}, \citenamefont {Pitchford}, \citenamefont {James},
  \citenamefont {Ahmed}, \citenamefont {Brierley}, \citenamefont {Hindell}
  \emph {et~al.}}]{sims2008scaling}%
  \BibitemOpen
  \bibfield  {author} {\bibinfo {author} {\bibfnamefont {D.~W.}\ \bibnamefont
  {Sims}}, \bibinfo {author} {\bibfnamefont {E.~J.}\ \bibnamefont {Southall}},
  \bibinfo {author} {\bibfnamefont {N.~E.}\ \bibnamefont {Humphries}}, \bibinfo
  {author} {\bibfnamefont {G.~C.}\ \bibnamefont {Hays}}, \bibinfo {author}
  {\bibfnamefont {C.~J.}\ \bibnamefont {Bradshaw}}, \bibinfo {author}
  {\bibfnamefont {J.~W.}\ \bibnamefont {Pitchford}}, \bibinfo {author}
  {\bibfnamefont {A.}~\bibnamefont {James}}, \bibinfo {author} {\bibfnamefont
  {M.~Z.}\ \bibnamefont {Ahmed}}, \bibinfo {author} {\bibfnamefont {A.~S.}\
  \bibnamefont {Brierley}}, \bibinfo {author} {\bibfnamefont {M.~A.}\
  \bibnamefont {Hindell}}, \emph {et~al.},\ }\bibfield  {title} {\bibinfo
  {title} {Scaling laws of marine predator search behaviour},\ }\href@noop {}
  {\bibfield  {journal} {\bibinfo  {journal} {Nature}\ }\textbf {\bibinfo
  {volume} {451}},\ \bibinfo {pages} {1098} (\bibinfo {year}
  {2008})}\BibitemShut {NoStop}%
\bibitem [{\citenamefont {Rhee}\ \emph {et~al.}(2011)\citenamefont {Rhee},
  \citenamefont {Shin}, \citenamefont {Hong}, \citenamefont {Lee},
  \citenamefont {Kim},\ and\ \citenamefont {Chong}}]{rhee2011levy}%
  \BibitemOpen
  \bibfield  {author} {\bibinfo {author} {\bibfnamefont {I.}~\bibnamefont
  {Rhee}}, \bibinfo {author} {\bibfnamefont {M.}~\bibnamefont {Shin}}, \bibinfo
  {author} {\bibfnamefont {S.}~\bibnamefont {Hong}}, \bibinfo {author}
  {\bibfnamefont {K.}~\bibnamefont {Lee}}, \bibinfo {author} {\bibfnamefont
  {S.~J.}\ \bibnamefont {Kim}},\ and\ \bibinfo {author} {\bibfnamefont
  {S.}~\bibnamefont {Chong}},\ }\bibfield  {title} {\bibinfo {title} {On the
  levy-walk nature of human mobility},\ }\href@noop {} {\bibfield  {journal}
  {\bibinfo  {journal} {IEEE/ACM transactions on networking}\ }\textbf
  {\bibinfo {volume} {19}},\ \bibinfo {pages} {630} (\bibinfo {year}
  {2011})}\BibitemShut {NoStop}%
\bibitem [{\citenamefont {Gonzalez}\ \emph {et~al.}(2008)\citenamefont
  {Gonzalez}, \citenamefont {Hidalgo},\ and\ \citenamefont
  {Barabasi}}]{gonzalez2008understanding}%
  \BibitemOpen
  \bibfield  {author} {\bibinfo {author} {\bibfnamefont {M.~C.}\ \bibnamefont
  {Gonzalez}}, \bibinfo {author} {\bibfnamefont {C.~A.}\ \bibnamefont
  {Hidalgo}},\ and\ \bibinfo {author} {\bibfnamefont {A.-L.}\ \bibnamefont
  {Barabasi}},\ }\bibfield  {title} {\bibinfo {title} {Understanding individual
  human mobility patterns},\ }\href@noop {} {\bibfield  {journal} {\bibinfo
  {journal} {nature}\ }\textbf {\bibinfo {volume} {453}},\ \bibinfo {pages}
  {779} (\bibinfo {year} {2008})}\BibitemShut {NoStop}%
\bibitem [{\citenamefont {Edwards}\ \emph {et~al.}(2007)\citenamefont
  {Edwards}, \citenamefont {Phillips}, \citenamefont {Watkins}, \citenamefont
  {Freeman}, \citenamefont {Murphy}, \citenamefont {Afanasyev}, \citenamefont
  {Buldyrev}, \citenamefont {da~Luz}, \citenamefont {Raposo}, \citenamefont
  {Stanley} \emph {et~al.}}]{edwards2007revisiting}%
  \BibitemOpen
  \bibfield  {author} {\bibinfo {author} {\bibfnamefont {A.~M.}\ \bibnamefont
  {Edwards}}, \bibinfo {author} {\bibfnamefont {R.~A.}\ \bibnamefont
  {Phillips}}, \bibinfo {author} {\bibfnamefont {N.~W.}\ \bibnamefont
  {Watkins}}, \bibinfo {author} {\bibfnamefont {M.~P.}\ \bibnamefont
  {Freeman}}, \bibinfo {author} {\bibfnamefont {E.~J.}\ \bibnamefont {Murphy}},
  \bibinfo {author} {\bibfnamefont {V.}~\bibnamefont {Afanasyev}}, \bibinfo
  {author} {\bibfnamefont {S.~V.}\ \bibnamefont {Buldyrev}}, \bibinfo {author}
  {\bibfnamefont {M.~G.}\ \bibnamefont {da~Luz}}, \bibinfo {author}
  {\bibfnamefont {E.~P.}\ \bibnamefont {Raposo}}, \bibinfo {author}
  {\bibfnamefont {H.~E.}\ \bibnamefont {Stanley}}, \emph {et~al.},\ }\bibfield
  {title} {\bibinfo {title} {Revisiting l{\'e}vy flight search patterns of
  wandering albatrosses, bumblebees and deer},\ }\href@noop {} {\bibfield
  {journal} {\bibinfo  {journal} {Nature}\ }\textbf {\bibinfo {volume} {449}},\
  \bibinfo {pages} {1044} (\bibinfo {year} {2007})}\BibitemShut {NoStop}%
\bibitem [{\citenamefont {Gross}\ \emph {et~al.}(2020)\citenamefont {Gross},
  \citenamefont {Zheng}, \citenamefont {Liu}, \citenamefont {Chen},
  \citenamefont {Sela}, \citenamefont {Li}, \citenamefont {Li},\ and\
  \citenamefont {Havlin}}]{gross2020spatio}%
  \BibitemOpen
  \bibfield  {author} {\bibinfo {author} {\bibfnamefont {B.}~\bibnamefont
  {Gross}}, \bibinfo {author} {\bibfnamefont {Z.}~\bibnamefont {Zheng}},
  \bibinfo {author} {\bibfnamefont {S.}~\bibnamefont {Liu}}, \bibinfo {author}
  {\bibfnamefont {X.}~\bibnamefont {Chen}}, \bibinfo {author} {\bibfnamefont
  {A.}~\bibnamefont {Sela}}, \bibinfo {author} {\bibfnamefont {J.}~\bibnamefont
  {Li}}, \bibinfo {author} {\bibfnamefont {D.}~\bibnamefont {Li}},\ and\
  \bibinfo {author} {\bibfnamefont {S.}~\bibnamefont {Havlin}},\ }\bibfield
  {title} {\bibinfo {title} {Spatio-temporal propagation of covid-19
  pandemics},\ }\href@noop {} {\bibfield  {journal} {\bibinfo  {journal} {EPL
  (Europhysics Letters)}\ }\textbf {\bibinfo {volume} {131}},\ \bibinfo {pages}
  {58003} (\bibinfo {year} {2020})}\BibitemShut {NoStop}%
\bibitem [{\citenamefont {Cabrera}\ and\ \citenamefont
  {Milton}(2004{\natexlab{b}})}]{cabrera2004human}%
  \BibitemOpen
  \bibfield  {author} {\bibinfo {author} {\bibfnamefont {J.~L.}\ \bibnamefont
  {Cabrera}}\ and\ \bibinfo {author} {\bibfnamefont {J.~G.}\ \bibnamefont
  {Milton}},\ }\bibfield  {title} {\bibinfo {title} {Human stick balancing:
  tuning l{\'e}vy flights to improve balance control},\ }\href@noop {}
  {\bibfield  {journal} {\bibinfo  {journal} {Chaos: An Interdisciplinary
  Journal of Nonlinear Science}\ }\textbf {\bibinfo {volume} {14}},\ \bibinfo
  {pages} {691} (\bibinfo {year} {2004}{\natexlab{b}})}\BibitemShut {NoStop}%
\bibitem [{\citenamefont {Metzler}\ and\ \citenamefont
  {Klafter}(2004)}]{metzler2004restaurant}%
  \BibitemOpen
  \bibfield  {author} {\bibinfo {author} {\bibfnamefont {R.}~\bibnamefont
  {Metzler}}\ and\ \bibinfo {author} {\bibfnamefont {J.}~\bibnamefont
  {Klafter}},\ }\bibfield  {title} {\bibinfo {title} {The restaurant at the end
  of the random walk: recent developments in the description of anomalous
  transport by fractional dynamics},\ }\href@noop {} {\bibfield  {journal}
  {\bibinfo  {journal} {Journal of Physics A: Mathematical and General}\
  }\textbf {\bibinfo {volume} {37}},\ \bibinfo {pages} {R161} (\bibinfo {year}
  {2004})}\BibitemShut {NoStop}%
\bibitem [{\citenamefont {Applebaum}(2004)}]{applebaum2004levy}%
  \BibitemOpen
  \bibfield  {author} {\bibinfo {author} {\bibfnamefont {D.}~\bibnamefont
  {Applebaum}},\ }\bibfield  {title} {\bibinfo {title} {L{\'e}vy processes-from
  probability to finance and quantum groups},\ }\href@noop {} {\bibfield
  {journal} {\bibinfo  {journal} {Notices of the AMS}\ }\textbf {\bibinfo
  {volume} {51}},\ \bibinfo {pages} {1336} (\bibinfo {year}
  {2004})}\BibitemShut {NoStop}%
\bibitem [{\citenamefont {Roberts}\ \emph {et~al.}(2015)\citenamefont
  {Roberts}, \citenamefont {Boonstra},\ and\ \citenamefont
  {Breakspear}}]{roberts2015heavy}%
  \BibitemOpen
  \bibfield  {author} {\bibinfo {author} {\bibfnamefont {J.~A.}\ \bibnamefont
  {Roberts}}, \bibinfo {author} {\bibfnamefont {T.~W.}\ \bibnamefont
  {Boonstra}},\ and\ \bibinfo {author} {\bibfnamefont {M.}~\bibnamefont
  {Breakspear}},\ }\bibfield  {title} {\bibinfo {title} {The heavy tail of the
  human brain},\ }\href@noop {} {\bibfield  {journal} {\bibinfo  {journal}
  {Current opinion in neurobiology}\ }\textbf {\bibinfo {volume} {31}},\
  \bibinfo {pages} {164} (\bibinfo {year} {2015})}\BibitemShut {NoStop}%
\bibitem [{\citenamefont {Shlesinger}\ and\ \citenamefont
  {Klafter}(1986)}]{shlesinger1986levy}%
  \BibitemOpen
  \bibfield  {author} {\bibinfo {author} {\bibfnamefont {M.~F.}\ \bibnamefont
  {Shlesinger}}\ and\ \bibinfo {author} {\bibfnamefont {J.}~\bibnamefont
  {Klafter}},\ }\bibfield  {title} {\bibinfo {title} {L{\'e}vy walks versus
  l{\'e}vy flights},\ }in\ \href@noop {} {\emph {\bibinfo {booktitle} {On
  growth and form}}}\ (\bibinfo  {publisher} {Springer},\ \bibinfo {year}
  {1986})\ pp.\ \bibinfo {pages} {279--283}\BibitemShut {NoStop}%
\bibitem [{\citenamefont {Zaburdaev}\ \emph {et~al.}(2015)\citenamefont
  {Zaburdaev}, \citenamefont {Denisov},\ and\ \citenamefont
  {Klafter}}]{zaburdaev2015levy}%
  \BibitemOpen
  \bibfield  {author} {\bibinfo {author} {\bibfnamefont {V.}~\bibnamefont
  {Zaburdaev}}, \bibinfo {author} {\bibfnamefont {S.}~\bibnamefont {Denisov}},\
  and\ \bibinfo {author} {\bibfnamefont {J.}~\bibnamefont {Klafter}},\
  }\bibfield  {title} {\bibinfo {title} {L{\'e}vy walks},\ }\href@noop {}
  {\bibfield  {journal} {\bibinfo  {journal} {Reviews of Modern Physics}\
  }\textbf {\bibinfo {volume} {87}},\ \bibinfo {pages} {483} (\bibinfo {year}
  {2015})}\BibitemShut {NoStop}%
\bibitem [{\citenamefont {Jung}\ \emph {et~al.}(2002)\citenamefont {Jung},
  \citenamefont {Barkai},\ and\ \citenamefont {Silbey}}]{jung2002lineshape}%
  \BibitemOpen
  \bibfield  {author} {\bibinfo {author} {\bibfnamefont {Y.}~\bibnamefont
  {Jung}}, \bibinfo {author} {\bibfnamefont {E.}~\bibnamefont {Barkai}},\ and\
  \bibinfo {author} {\bibfnamefont {R.~J.}\ \bibnamefont {Silbey}},\ }\bibfield
   {title} {\bibinfo {title} {Lineshape theory and photon counting statistics
  for blinking quantum dots: a l{\'e}vy walk process},\ }\href@noop {}
  {\bibfield  {journal} {\bibinfo  {journal} {Chemical Physics}\ }\textbf
  {\bibinfo {volume} {284}},\ \bibinfo {pages} {181} (\bibinfo {year}
  {2002})}\BibitemShut {NoStop}%
\bibitem [{\citenamefont {Margolin}\ and\ \citenamefont
  {Barkai}(2005)}]{margolin2005nonergodicity}%
  \BibitemOpen
  \bibfield  {author} {\bibinfo {author} {\bibfnamefont {G.}~\bibnamefont
  {Margolin}}\ and\ \bibinfo {author} {\bibfnamefont {E.}~\bibnamefont
  {Barkai}},\ }\bibfield  {title} {\bibinfo {title} {Nonergodicity of blinking
  nanocrystals and other l{\'e}vy-walk processes},\ }\href@noop {} {\bibfield
  {journal} {\bibinfo  {journal} {Physical review letters}\ }\textbf {\bibinfo
  {volume} {94}},\ \bibinfo {pages} {080601} (\bibinfo {year}
  {2005})}\BibitemShut {NoStop}%
\bibitem [{\citenamefont {van Kan}\ \emph {et~al.}(2020)\citenamefont {van
  Kan}, \citenamefont {Alexakis},\ and\ \citenamefont
  {Brachet}}]{vankan2021levy}%
  \BibitemOpen
  \bibfield  {author} {\bibinfo {author} {\bibfnamefont {A.}~\bibnamefont {van
  Kan}}, \bibinfo {author} {\bibfnamefont {A.}~\bibnamefont {Alexakis}},\ and\
  \bibinfo {author} {\bibfnamefont {M.~E.}\ \bibnamefont {Brachet}},\
  }\bibfield  {title} {\bibinfo {title} {Intermittency of three-dimensional
  perturbations in a point-vortex model},\ }\href@noop {} {\bibfield  {journal}
  {\bibinfo  {journal} {submitted to Physical Review E}\ } (\bibinfo {year}
  {2020})}\BibitemShut {NoStop}%
\bibitem [{\citenamefont {Seshasayanan}\ and\ \citenamefont
  {Gallet}(2020)}]{seshasayanan2020onset}%
  \BibitemOpen
  \bibfield  {author} {\bibinfo {author} {\bibfnamefont {K.}~\bibnamefont
  {Seshasayanan}}\ and\ \bibinfo {author} {\bibfnamefont {B.}~\bibnamefont
  {Gallet}},\ }\bibfield  {title} {\bibinfo {title} {Onset of
  three-dimensionality in rapidly rotating turbulent flows},\ }\href@noop {}
  {\bibfield  {journal} {\bibinfo  {journal} {Journal of Fluid Mechanics}\
  }\textbf {\bibinfo {volume} {901}},\ \bibinfo {pages} {R5} (\bibinfo {year}
  {2020})}\BibitemShut {NoStop}%
\bibitem [{\citenamefont {Jespersen}\ \emph {et~al.}(1999)\citenamefont
  {Jespersen}, \citenamefont {Metzler},\ and\ \citenamefont
  {Fogedby}}]{jespersen1999levy}%
  \BibitemOpen
  \bibfield  {author} {\bibinfo {author} {\bibfnamefont {S.}~\bibnamefont
  {Jespersen}}, \bibinfo {author} {\bibfnamefont {R.}~\bibnamefont {Metzler}},\
  and\ \bibinfo {author} {\bibfnamefont {H.~C.}\ \bibnamefont {Fogedby}},\
  }\bibfield  {title} {\bibinfo {title} {L{\'e}vy flights in external force
  fields: Langevin and fractional fokker-planck equations and their
  solutions},\ }\href@noop {} {\bibfield  {journal} {\bibinfo  {journal}
  {Physical Review E}\ }\textbf {\bibinfo {volume} {59}},\ \bibinfo {pages}
  {2736} (\bibinfo {year} {1999})}\BibitemShut {NoStop}%
\bibitem [{\citenamefont {Chechkin}\ \emph {et~al.}(2002)\citenamefont
  {Chechkin}, \citenamefont {Gonchar}, \citenamefont {Klafter}, \citenamefont
  {Metzler},\ and\ \citenamefont {Tanatarov}}]{chechkin2002stationary}%
  \BibitemOpen
  \bibfield  {author} {\bibinfo {author} {\bibfnamefont {A.}~\bibnamefont
  {Chechkin}}, \bibinfo {author} {\bibfnamefont {V.}~\bibnamefont {Gonchar}},
  \bibinfo {author} {\bibfnamefont {J.}~\bibnamefont {Klafter}}, \bibinfo
  {author} {\bibfnamefont {R.}~\bibnamefont {Metzler}},\ and\ \bibinfo {author}
  {\bibfnamefont {L.}~\bibnamefont {Tanatarov}},\ }\bibfield  {title} {\bibinfo
  {title} {Stationary states of non-linear oscillators driven by l{\'e}vy
  noise},\ }\href@noop {} {\bibfield  {journal} {\bibinfo  {journal} {Chemical
  Physics}\ }\textbf {\bibinfo {volume} {284}},\ \bibinfo {pages} {233}
  (\bibinfo {year} {2002})}\BibitemShut {NoStop}%
\bibitem [{\citenamefont {Chechkin}\ \emph {et~al.}(2003)\citenamefont
  {Chechkin}, \citenamefont {Klafter}, \citenamefont {Gonchar}, \citenamefont
  {Metzler},\ and\ \citenamefont {Tanatarov}}]{chechkin2003bifurcation}%
  \BibitemOpen
  \bibfield  {author} {\bibinfo {author} {\bibfnamefont {A.~V.}\ \bibnamefont
  {Chechkin}}, \bibinfo {author} {\bibfnamefont {J.}~\bibnamefont {Klafter}},
  \bibinfo {author} {\bibfnamefont {V.~Y.}\ \bibnamefont {Gonchar}}, \bibinfo
  {author} {\bibfnamefont {R.}~\bibnamefont {Metzler}},\ and\ \bibinfo {author}
  {\bibfnamefont {L.~V.}\ \bibnamefont {Tanatarov}},\ }\bibfield  {title}
  {\bibinfo {title} {Bifurcation, bimodality, and finite variance in confined
  l{\'e}vy flights},\ }\href@noop {} {\bibfield  {journal} {\bibinfo  {journal}
  {Physical review E}\ }\textbf {\bibinfo {volume} {67}},\ \bibinfo {pages}
  {010102} (\bibinfo {year} {2003})}\BibitemShut {NoStop}%
\bibitem [{\citenamefont {Chechkin}\ \emph {et~al.}(2004)\citenamefont
  {Chechkin}, \citenamefont {Gonchar}, \citenamefont {Klafter}, \citenamefont
  {Metzler},\ and\ \citenamefont {Tanatarov}}]{chechkin2004levy}%
  \BibitemOpen
  \bibfield  {author} {\bibinfo {author} {\bibfnamefont {A.~V.}\ \bibnamefont
  {Chechkin}}, \bibinfo {author} {\bibfnamefont {V.~Y.}\ \bibnamefont
  {Gonchar}}, \bibinfo {author} {\bibfnamefont {J.}~\bibnamefont {Klafter}},
  \bibinfo {author} {\bibfnamefont {R.}~\bibnamefont {Metzler}},\ and\ \bibinfo
  {author} {\bibfnamefont {L.~V.}\ \bibnamefont {Tanatarov}},\ }\bibfield
  {title} {\bibinfo {title} {L{\'e}vy flights in a steep potential well},\
  }\href@noop {} {\bibfield  {journal} {\bibinfo  {journal} {Journal of
  Statistical Physics}\ }\textbf {\bibinfo {volume} {115}},\ \bibinfo {pages}
  {1505} (\bibinfo {year} {2004})}\BibitemShut {NoStop}%
\bibitem [{\citenamefont {Dybiec}\ \emph
  {et~al.}(2007{\natexlab{a}})\citenamefont {Dybiec}, \citenamefont
  {Gudowska-Nowak},\ and\ \citenamefont {Sokolov}}]{dybiec2007stationary}%
  \BibitemOpen
  \bibfield  {author} {\bibinfo {author} {\bibfnamefont {B.}~\bibnamefont
  {Dybiec}}, \bibinfo {author} {\bibfnamefont {E.}~\bibnamefont
  {Gudowska-Nowak}},\ and\ \bibinfo {author} {\bibfnamefont {I.}~\bibnamefont
  {Sokolov}},\ }\bibfield  {title} {\bibinfo {title} {Stationary states in
  langevin dynamics under asymmetric l{\'e}vy noises},\ }\href@noop {}
  {\bibfield  {journal} {\bibinfo  {journal} {Physical Review E}\ }\textbf
  {\bibinfo {volume} {76}},\ \bibinfo {pages} {041122} (\bibinfo {year}
  {2007}{\natexlab{a}})}\BibitemShut {NoStop}%
\bibitem [{\citenamefont {Denisov}\ \emph {et~al.}(2008)\citenamefont
  {Denisov}, \citenamefont {Horsthemke},\ and\ \citenamefont
  {H{\"a}nggi}}]{denisov2008steady}%
  \BibitemOpen
  \bibfield  {author} {\bibinfo {author} {\bibfnamefont {S.}~\bibnamefont
  {Denisov}}, \bibinfo {author} {\bibfnamefont {W.}~\bibnamefont
  {Horsthemke}},\ and\ \bibinfo {author} {\bibfnamefont {P.}~\bibnamefont
  {H{\"a}nggi}},\ }\bibfield  {title} {\bibinfo {title} {Steady-state l{\'e}vy
  flights in a confined domain},\ }\href@noop {} {\bibfield  {journal}
  {\bibinfo  {journal} {Physical Review E}\ }\textbf {\bibinfo {volume} {77}},\
  \bibinfo {pages} {061112} (\bibinfo {year} {2008})}\BibitemShut {NoStop}%
\bibitem [{\citenamefont {Dybiec}\ \emph {et~al.}(2010)\citenamefont {Dybiec},
  \citenamefont {Sokolov},\ and\ \citenamefont
  {Chechkin}}]{dybiec2010stationary}%
  \BibitemOpen
  \bibfield  {author} {\bibinfo {author} {\bibfnamefont {B.}~\bibnamefont
  {Dybiec}}, \bibinfo {author} {\bibfnamefont {I.~M.}\ \bibnamefont
  {Sokolov}},\ and\ \bibinfo {author} {\bibfnamefont {A.~V.}\ \bibnamefont
  {Chechkin}},\ }\bibfield  {title} {\bibinfo {title} {Stationary states in
  single-well potentials under symmetric l{\'e}vy noises},\ }\href@noop {}
  {\bibfield  {journal} {\bibinfo  {journal} {Journal of Statistical Mechanics:
  Theory and Experiment}\ }\textbf {\bibinfo {volume} {2010}},\ \bibinfo
  {pages} {P07008} (\bibinfo {year} {2010})}\BibitemShut {NoStop}%
\bibitem [{\citenamefont {Padash}\ \emph {et~al.}(2019)\citenamefont {Padash},
  \citenamefont {Chechkin}, \citenamefont {Dybiec}, \citenamefont
  {Pavlyukevich}, \citenamefont {Shokri},\ and\ \citenamefont
  {Metzler}}]{padash2019first}%
  \BibitemOpen
  \bibfield  {author} {\bibinfo {author} {\bibfnamefont {A.}~\bibnamefont
  {Padash}}, \bibinfo {author} {\bibfnamefont {A.~V.}\ \bibnamefont
  {Chechkin}}, \bibinfo {author} {\bibfnamefont {B.}~\bibnamefont {Dybiec}},
  \bibinfo {author} {\bibfnamefont {I.}~\bibnamefont {Pavlyukevich}}, \bibinfo
  {author} {\bibfnamefont {B.}~\bibnamefont {Shokri}},\ and\ \bibinfo {author}
  {\bibfnamefont {R.}~\bibnamefont {Metzler}},\ }\bibfield  {title} {\bibinfo
  {title} {First-passage properties of asymmetric l{\'e}vy flights},\
  }\href@noop {} {\bibfield  {journal} {\bibinfo  {journal} {Journal of Physics
  A: Mathematical and Theoretical}\ }\textbf {\bibinfo {volume} {52}},\
  \bibinfo {pages} {454004} (\bibinfo {year} {2019})}\BibitemShut {NoStop}%
\bibitem [{\citenamefont
  {Srokowski}(2009{\natexlab{a}})}]{srokowski2009fractional}%
  \BibitemOpen
  \bibfield  {author} {\bibinfo {author} {\bibfnamefont {T.}~\bibnamefont
  {Srokowski}},\ }\bibfield  {title} {\bibinfo {title} {Fractional
  fokker-planck equation for l{\'e}vy flights in nonhomogeneous environments},\
  }\href@noop {} {\bibfield  {journal} {\bibinfo  {journal} {Physical Review
  E}\ }\textbf {\bibinfo {volume} {79}},\ \bibinfo {pages} {040104} (\bibinfo
  {year} {2009}{\natexlab{a}})}\BibitemShut {NoStop}%
\bibitem [{\citenamefont
  {Srokowski}(2009{\natexlab{b}})}]{srokowski2009multiplicative}%
  \BibitemOpen
  \bibfield  {author} {\bibinfo {author} {\bibfnamefont {T.}~\bibnamefont
  {Srokowski}},\ }\bibfield  {title} {\bibinfo {title} {Multiplicative l{\'e}vy
  processes: It{\^o} versus stratonovich interpretation},\ }\href@noop {}
  {\bibfield  {journal} {\bibinfo  {journal} {Physical Review E}\ }\textbf
  {\bibinfo {volume} {80}},\ \bibinfo {pages} {051113} (\bibinfo {year}
  {2009}{\natexlab{b}})}\BibitemShut {NoStop}%
\bibitem [{\citenamefont {La~Cognata}\ \emph {et~al.}(2010)\citenamefont
  {La~Cognata}, \citenamefont {Valenti}, \citenamefont {Dubkov},\ and\
  \citenamefont {Spagnolo}}]{la2010dynamics}%
  \BibitemOpen
  \bibfield  {author} {\bibinfo {author} {\bibfnamefont {A.}~\bibnamefont
  {La~Cognata}}, \bibinfo {author} {\bibfnamefont {D.}~\bibnamefont {Valenti}},
  \bibinfo {author} {\bibfnamefont {A.}~\bibnamefont {Dubkov}},\ and\ \bibinfo
  {author} {\bibfnamefont {B.}~\bibnamefont {Spagnolo}},\ }\bibfield  {title}
  {\bibinfo {title} {Dynamics of two competing species in the presence of
  l{\'e}vy noise sources},\ }\href@noop {} {\bibfield  {journal} {\bibinfo
  {journal} {Physical Review E}\ }\textbf {\bibinfo {volume} {82}},\ \bibinfo
  {pages} {011121} (\bibinfo {year} {2010})}\BibitemShut {NoStop}%
\bibitem [{\citenamefont {Srokowski}(2010)}]{srokowski2010nonlinear}%
  \BibitemOpen
  \bibfield  {author} {\bibinfo {author} {\bibfnamefont {T.}~\bibnamefont
  {Srokowski}},\ }\bibfield  {title} {\bibinfo {title} {Nonlinear stochastic
  equations with multiplicative l{\'e}vy noise},\ }\href@noop {} {\bibfield
  {journal} {\bibinfo  {journal} {Physical Review E}\ }\textbf {\bibinfo
  {volume} {81}},\ \bibinfo {pages} {051110} (\bibinfo {year}
  {2010})}\BibitemShut {NoStop}%
\bibitem [{\citenamefont {Srokowski}(2012)}]{srokowski2012multiplicative}%
  \BibitemOpen
  \bibfield  {author} {\bibinfo {author} {\bibfnamefont {T.}~\bibnamefont
  {Srokowski}},\ }\bibfield  {title} {\bibinfo {title} {Multiplicative l{\'e}vy
  noise in bistable systems},\ }\href@noop {} {\bibfield  {journal} {\bibinfo
  {journal} {The European Physical Journal B}\ }\textbf {\bibinfo {volume}
  {85}},\ \bibinfo {pages} {65} (\bibinfo {year} {2012})}\BibitemShut {NoStop}%
\bibitem [{\citenamefont {Chechkin}\ \emph {et~al.}(2005)\citenamefont
  {Chechkin}, \citenamefont {Gonchar}, \citenamefont {Klafter},\ and\
  \citenamefont {Metzler}}]{chechkin2005barrier}%
  \BibitemOpen
  \bibfield  {author} {\bibinfo {author} {\bibfnamefont {A.~V.}\ \bibnamefont
  {Chechkin}}, \bibinfo {author} {\bibfnamefont {V.~Y.}\ \bibnamefont
  {Gonchar}}, \bibinfo {author} {\bibfnamefont {J.}~\bibnamefont {Klafter}},\
  and\ \bibinfo {author} {\bibfnamefont {R.}~\bibnamefont {Metzler}},\
  }\bibfield  {title} {\bibinfo {title} {Barrier crossing of a l{\'e}vy
  flight},\ }\href@noop {} {\bibfield  {journal} {\bibinfo  {journal} {EPL
  (Europhysics Letters)}\ }\textbf {\bibinfo {volume} {72}},\ \bibinfo {pages}
  {348} (\bibinfo {year} {2005})}\BibitemShut {NoStop}%
\bibitem [{\citenamefont {Dybiec}\ \emph
  {et~al.}(2007{\natexlab{b}})\citenamefont {Dybiec}, \citenamefont
  {Gudowska-Nowak},\ and\ \citenamefont {H{\"a}nggi}}]{dybiec2007escape}%
  \BibitemOpen
  \bibfield  {author} {\bibinfo {author} {\bibfnamefont {B.}~\bibnamefont
  {Dybiec}}, \bibinfo {author} {\bibfnamefont {E.}~\bibnamefont
  {Gudowska-Nowak}},\ and\ \bibinfo {author} {\bibfnamefont {P.}~\bibnamefont
  {H{\"a}nggi}},\ }\bibfield  {title} {\bibinfo {title} {Escape driven by
  $\alpha$-stable white noises},\ }\href@noop {} {\bibfield  {journal}
  {\bibinfo  {journal} {Physical Review E}\ }\textbf {\bibinfo {volume} {75}},\
  \bibinfo {pages} {021109} (\bibinfo {year} {2007}{\natexlab{b}})}\BibitemShut
  {NoStop}%
\bibitem [{\citenamefont {Chechkin}\ \emph {et~al.}(2007)\citenamefont
  {Chechkin}, \citenamefont {Sliusarenko}, \citenamefont {Metzler},\ and\
  \citenamefont {Klafter}}]{chechkin2007barrier}%
  \BibitemOpen
  \bibfield  {author} {\bibinfo {author} {\bibfnamefont {A.~V.}\ \bibnamefont
  {Chechkin}}, \bibinfo {author} {\bibfnamefont {O.~Y.}\ \bibnamefont
  {Sliusarenko}}, \bibinfo {author} {\bibfnamefont {R.}~\bibnamefont
  {Metzler}},\ and\ \bibinfo {author} {\bibfnamefont {J.}~\bibnamefont
  {Klafter}},\ }\bibfield  {title} {\bibinfo {title} {Barrier crossing driven
  by l{\'e}vy noise: Universality and the role of noise intensity},\
  }\href@noop {} {\bibfield  {journal} {\bibinfo  {journal} {Physical Review
  E}\ }\textbf {\bibinfo {volume} {75}},\ \bibinfo {pages} {041101} (\bibinfo
  {year} {2007})}\BibitemShut {NoStop}%
\bibitem [{\citenamefont {Koren}\ \emph {et~al.}(2007)\citenamefont {Koren},
  \citenamefont {Lomholt}, \citenamefont {Chechkin}, \citenamefont {Klafter},\
  and\ \citenamefont {Metzler}}]{koren2007leapover}%
  \BibitemOpen
  \bibfield  {author} {\bibinfo {author} {\bibfnamefont {T.}~\bibnamefont
  {Koren}}, \bibinfo {author} {\bibfnamefont {M.~A.}\ \bibnamefont {Lomholt}},
  \bibinfo {author} {\bibfnamefont {A.~V.}\ \bibnamefont {Chechkin}}, \bibinfo
  {author} {\bibfnamefont {J.}~\bibnamefont {Klafter}},\ and\ \bibinfo {author}
  {\bibfnamefont {R.}~\bibnamefont {Metzler}},\ }\bibfield  {title} {\bibinfo
  {title} {Leapover lengths and first passage time statistics for l{\'e}vy
  flights},\ }\href@noop {} {\bibfield  {journal} {\bibinfo  {journal}
  {Physical review letters}\ }\textbf {\bibinfo {volume} {99}},\ \bibinfo
  {pages} {160602} (\bibinfo {year} {2007})}\BibitemShut {NoStop}%
\bibitem [{\citenamefont {Capa{\l}a}\ \emph {et~al.}(2020)\citenamefont
  {Capa{\l}a}, \citenamefont {Padash}, \citenamefont {Chechkin}, \citenamefont
  {Shokri}, \citenamefont {Metzler},\ and\ \citenamefont
  {Dybiec}}]{capala2020levy}%
  \BibitemOpen
  \bibfield  {author} {\bibinfo {author} {\bibfnamefont {K.}~\bibnamefont
  {Capa{\l}a}}, \bibinfo {author} {\bibfnamefont {A.}~\bibnamefont {Padash}},
  \bibinfo {author} {\bibfnamefont {A.~V.}\ \bibnamefont {Chechkin}}, \bibinfo
  {author} {\bibfnamefont {B.}~\bibnamefont {Shokri}}, \bibinfo {author}
  {\bibfnamefont {R.}~\bibnamefont {Metzler}},\ and\ \bibinfo {author}
  {\bibfnamefont {B.}~\bibnamefont {Dybiec}},\ }\bibfield  {title} {\bibinfo
  {title} {L{\'e}vy noise-driven escape from arctangent potential wells},\
  }\href@noop {} {\bibfield  {journal} {\bibinfo  {journal} {Chaos: An
  Interdisciplinary Journal of Nonlinear Science}\ }\textbf {\bibinfo {volume}
  {30}},\ \bibinfo {pages} {123103} (\bibinfo {year} {2020})}\BibitemShut
  {NoStop}%
\bibitem [{\citenamefont {Zeng}\ \emph {et~al.}(2007)\citenamefont {Zeng},
  \citenamefont {Bao},\ and\ \citenamefont {Xu}}]{zeng2007effects}%
  \BibitemOpen
  \bibfield  {author} {\bibinfo {author} {\bibfnamefont {L.}~\bibnamefont
  {Zeng}}, \bibinfo {author} {\bibfnamefont {R.}~\bibnamefont {Bao}},\ and\
  \bibinfo {author} {\bibfnamefont {B.}~\bibnamefont {Xu}},\ }\bibfield
  {title} {\bibinfo {title} {Effects of l{\'e}vy noise in aperiodic stochastic
  resonance},\ }\href@noop {} {\bibfield  {journal} {\bibinfo  {journal}
  {Journal of physics A: Mathematical and Theoretical}\ }\textbf {\bibinfo
  {volume} {40}},\ \bibinfo {pages} {7175} (\bibinfo {year}
  {2007})}\BibitemShut {NoStop}%
\bibitem [{\citenamefont {Dybiec}(2009)}]{dybiec2009levy}%
  \BibitemOpen
  \bibfield  {author} {\bibinfo {author} {\bibfnamefont {B.}~\bibnamefont
  {Dybiec}},\ }\bibfield  {title} {\bibinfo {title} {L{\'e}vy noises: Double
  stochastic resonance in a single-well potential},\ }\href@noop {} {\bibfield
  {journal} {\bibinfo  {journal} {Physical Review E}\ }\textbf {\bibinfo
  {volume} {80}},\ \bibinfo {pages} {041111} (\bibinfo {year}
  {2009})}\BibitemShut {NoStop}%
\bibitem [{\citenamefont {Dybiec}\ and\ \citenamefont
  {Gudowska-Nowak}(2009)}]{dybiec2009levy2}%
  \BibitemOpen
  \bibfield  {author} {\bibinfo {author} {\bibfnamefont {B.}~\bibnamefont
  {Dybiec}}\ and\ \bibinfo {author} {\bibfnamefont {E.}~\bibnamefont
  {Gudowska-Nowak}},\ }\bibfield  {title} {\bibinfo {title} {L{\'e}vy stable
  noise-induced transitions: stochastic resonance, resonant activation and
  dynamic hysteresis},\ }\href@noop {} {\bibfield  {journal} {\bibinfo
  {journal} {Journal of Statistical Mechanics: Theory and Experiment}\ }\textbf
  {\bibinfo {volume} {2009}},\ \bibinfo {pages} {P05004} (\bibinfo {year}
  {2009})}\BibitemShut {NoStop}%
\bibitem [{\citenamefont {Xu}\ \emph {et~al.}(2013)\citenamefont {Xu},
  \citenamefont {Li}, \citenamefont {Feng}, \citenamefont {Zhang},
  \citenamefont {Xu},\ and\ \citenamefont {Duan}}]{xu2013levy}%
  \BibitemOpen
  \bibfield  {author} {\bibinfo {author} {\bibfnamefont {Y.}~\bibnamefont
  {Xu}}, \bibinfo {author} {\bibfnamefont {J.}~\bibnamefont {Li}}, \bibinfo
  {author} {\bibfnamefont {J.}~\bibnamefont {Feng}}, \bibinfo {author}
  {\bibfnamefont {H.}~\bibnamefont {Zhang}}, \bibinfo {author} {\bibfnamefont
  {W.}~\bibnamefont {Xu}},\ and\ \bibinfo {author} {\bibfnamefont
  {J.}~\bibnamefont {Duan}},\ }\bibfield  {title} {\bibinfo {title} {L{\'e}vy
  noise-induced stochastic resonance in a bistable system},\ }\href@noop {}
  {\bibfield  {journal} {\bibinfo  {journal} {The European Physical Journal B}\
  }\textbf {\bibinfo {volume} {86}},\ \bibinfo {pages} {198} (\bibinfo {year}
  {2013})}\BibitemShut {NoStop}%
\bibitem [{\citenamefont {Yamapi}\ \emph {et~al.}(2019)\citenamefont {Yamapi},
  \citenamefont {Yonkeu}, \citenamefont {Filatrella},\ and\ \citenamefont
  {Kurths}}]{yamapi2019levy}%
  \BibitemOpen
  \bibfield  {author} {\bibinfo {author} {\bibfnamefont {R.}~\bibnamefont
  {Yamapi}}, \bibinfo {author} {\bibfnamefont {R.~M.}\ \bibnamefont {Yonkeu}},
  \bibinfo {author} {\bibfnamefont {G.}~\bibnamefont {Filatrella}},\ and\
  \bibinfo {author} {\bibfnamefont {J.}~\bibnamefont {Kurths}},\ }\bibfield
  {title} {\bibinfo {title} {L{\'e}vy noise induced transitions and enhanced
  stability in a birhythmic van der pol system},\ }\href@noop {} {\bibfield
  {journal} {\bibinfo  {journal} {The European Physical Journal B}\ }\textbf
  {\bibinfo {volume} {92}},\ \bibinfo {pages} {152} (\bibinfo {year}
  {2019})}\BibitemShut {NoStop}%
\bibitem [{\citenamefont {Sokolov}\ \emph {et~al.}(2011)\citenamefont
  {Sokolov}, \citenamefont {Ebeling},\ and\ \citenamefont
  {Dybiec}}]{sokolov2011harmonic}%
  \BibitemOpen
  \bibfield  {author} {\bibinfo {author} {\bibfnamefont {I.~M.}\ \bibnamefont
  {Sokolov}}, \bibinfo {author} {\bibfnamefont {W.}~\bibnamefont {Ebeling}},\
  and\ \bibinfo {author} {\bibfnamefont {B.}~\bibnamefont {Dybiec}},\
  }\bibfield  {title} {\bibinfo {title} {Harmonic oscillator under l{\'e}vy
  noise: Unexpected properties in the phase space},\ }\href@noop {} {\bibfield
  {journal} {\bibinfo  {journal} {Physical Review E}\ }\textbf {\bibinfo
  {volume} {83}},\ \bibinfo {pages} {041118} (\bibinfo {year}
  {2011})}\BibitemShut {NoStop}%
\bibitem [{\citenamefont {Tanaka}(2020)}]{tanaka2020low}%
  \BibitemOpen
  \bibfield  {author} {\bibinfo {author} {\bibfnamefont {T.}~\bibnamefont
  {Tanaka}},\ }\bibfield  {title} {\bibinfo {title} {Low-dimensional dynamics
  of phase oscillators driven by cauchy noise},\ }\href@noop {} {\bibfield
  {journal} {\bibinfo  {journal} {Physical Review E}\ }\textbf {\bibinfo
  {volume} {102}},\ \bibinfo {pages} {042220} (\bibinfo {year}
  {2020})}\BibitemShut {NoStop}%
\bibitem [{\citenamefont {Dubkov}\ and\ \citenamefont
  {Spagnolo}(2008)}]{dubkov2008verhulst}%
  \BibitemOpen
  \bibfield  {author} {\bibinfo {author} {\bibfnamefont {A.}~\bibnamefont
  {Dubkov}}\ and\ \bibinfo {author} {\bibfnamefont {B.}~\bibnamefont
  {Spagnolo}},\ }\bibfield  {title} {\bibinfo {title} {Verhulst model with
  l{\'e}vy white noise excitation},\ }\href@noop {} {\bibfield  {journal}
  {\bibinfo  {journal} {The European Physical Journal B}\ }\textbf {\bibinfo
  {volume} {65}},\ \bibinfo {pages} {361} (\bibinfo {year} {2008})}\BibitemShut
  {NoStop}%
\bibitem [{\citenamefont {Dybiec}\ \emph {et~al.}(2008)\citenamefont {Dybiec},
  \citenamefont {Gudowska-Nowak},\ and\ \citenamefont
  {Sokolov}}]{dybiec2008transport}%
  \BibitemOpen
  \bibfield  {author} {\bibinfo {author} {\bibfnamefont {B.}~\bibnamefont
  {Dybiec}}, \bibinfo {author} {\bibfnamefont {E.}~\bibnamefont
  {Gudowska-Nowak}},\ and\ \bibinfo {author} {\bibfnamefont {I.}~\bibnamefont
  {Sokolov}},\ }\bibfield  {title} {\bibinfo {title} {Transport in a l{\'e}vy
  ratchet: Group velocity and distribution spread},\ }\href@noop {} {\bibfield
  {journal} {\bibinfo  {journal} {Physical Review E}\ }\textbf {\bibinfo
  {volume} {78}},\ \bibinfo {pages} {011117} (\bibinfo {year}
  {2008})}\BibitemShut {NoStop}%
\bibitem [{\citenamefont {Guarcello}\ \emph {et~al.}(2013)\citenamefont
  {Guarcello}, \citenamefont {Valenti}, \citenamefont {Augello},\ and\
  \citenamefont {Spagnolo}}]{guarcello2013role}%
  \BibitemOpen
  \bibfield  {author} {\bibinfo {author} {\bibfnamefont {C.}~\bibnamefont
  {Guarcello}}, \bibinfo {author} {\bibfnamefont {D.}~\bibnamefont {Valenti}},
  \bibinfo {author} {\bibfnamefont {G.}~\bibnamefont {Augello}},\ and\ \bibinfo
  {author} {\bibfnamefont {B.}~\bibnamefont {Spagnolo}},\ }\bibfield  {title}
  {\bibinfo {title} {The role of non-gaussian sources in the transient dynamics
  of long josephson junctions.},\ }\href@noop {} {\bibfield  {journal}
  {\bibinfo  {journal} {Acta Physica Polonica B}\ }\textbf {\bibinfo {volume}
  {44}} (\bibinfo {year} {2013})}\BibitemShut {NoStop}%
\bibitem [{\citenamefont {Guarcello}\ \emph {et~al.}(2020)\citenamefont
  {Guarcello}, \citenamefont {Filatrella}, \citenamefont {Spagnolo},
  \citenamefont {Pierro},\ and\ \citenamefont
  {Valenti}}]{guarcello2020voltage}%
  \BibitemOpen
  \bibfield  {author} {\bibinfo {author} {\bibfnamefont {C.}~\bibnamefont
  {Guarcello}}, \bibinfo {author} {\bibfnamefont {G.}~\bibnamefont
  {Filatrella}}, \bibinfo {author} {\bibfnamefont {B.}~\bibnamefont
  {Spagnolo}}, \bibinfo {author} {\bibfnamefont {V.}~\bibnamefont {Pierro}},\
  and\ \bibinfo {author} {\bibfnamefont {D.}~\bibnamefont {Valenti}},\
  }\bibfield  {title} {\bibinfo {title} {Voltage drop across josephson
  junctions for l{\'e}vy noise detection},\ }\href@noop {} {\bibfield
  {journal} {\bibinfo  {journal} {Physical Review Research}\ }\textbf {\bibinfo
  {volume} {2}},\ \bibinfo {pages} {043332} (\bibinfo {year}
  {2020})}\BibitemShut {NoStop}%
\bibitem [{\citenamefont {Guarcello}\ \emph {et~al.}(2017)\citenamefont
  {Guarcello}, \citenamefont {Valenti}, \citenamefont {Spagnolo}, \citenamefont
  {Pierro},\ and\ \citenamefont {Filatrella}}]{guarcello2017anomalous}%
  \BibitemOpen
  \bibfield  {author} {\bibinfo {author} {\bibfnamefont {C.}~\bibnamefont
  {Guarcello}}, \bibinfo {author} {\bibfnamefont {D.}~\bibnamefont {Valenti}},
  \bibinfo {author} {\bibfnamefont {B.}~\bibnamefont {Spagnolo}}, \bibinfo
  {author} {\bibfnamefont {V.}~\bibnamefont {Pierro}},\ and\ \bibinfo {author}
  {\bibfnamefont {G.}~\bibnamefont {Filatrella}},\ }\bibfield  {title}
  {\bibinfo {title} {Anomalous transport effects on switching currents of
  graphene-based josephson junctions},\ }\href@noop {} {\bibfield  {journal}
  {\bibinfo  {journal} {Nanotechnology}\ }\textbf {\bibinfo {volume} {28}},\
  \bibinfo {pages} {134001} (\bibinfo {year} {2017})}\BibitemShut {NoStop}%
\bibitem [{\citenamefont {Valenti}\ \emph {et~al.}(2014)\citenamefont
  {Valenti}, \citenamefont {Guarcello},\ and\ \citenamefont
  {Spagnolo}}]{valenti2014switching}%
  \BibitemOpen
  \bibfield  {author} {\bibinfo {author} {\bibfnamefont {D.}~\bibnamefont
  {Valenti}}, \bibinfo {author} {\bibfnamefont {C.}~\bibnamefont {Guarcello}},\
  and\ \bibinfo {author} {\bibfnamefont {B.}~\bibnamefont {Spagnolo}},\
  }\bibfield  {title} {\bibinfo {title} {Switching times in long-overlap
  josephson junctions subject to thermal fluctuations and non-gaussian noise
  sources},\ }\href@noop {} {\bibfield  {journal} {\bibinfo  {journal}
  {Physical Review B}\ }\textbf {\bibinfo {volume} {89}},\ \bibinfo {pages}
  {214510} (\bibinfo {year} {2014})}\BibitemShut {NoStop}%
\bibitem [{\citenamefont {Guarcello}\ \emph {et~al.}(2016)\citenamefont
  {Guarcello}, \citenamefont {Valenti}, \citenamefont {Carollo},\ and\
  \citenamefont {Spagnolo}}]{guarcello2016effects}%
  \BibitemOpen
  \bibfield  {author} {\bibinfo {author} {\bibfnamefont {C.}~\bibnamefont
  {Guarcello}}, \bibinfo {author} {\bibfnamefont {D.}~\bibnamefont {Valenti}},
  \bibinfo {author} {\bibfnamefont {A.}~\bibnamefont {Carollo}},\ and\ \bibinfo
  {author} {\bibfnamefont {B.}~\bibnamefont {Spagnolo}},\ }\bibfield  {title}
  {\bibinfo {title} {Effects of l{\'e}vy noise on the dynamics of sine-gordon
  solitons in long josephson junctions},\ }\href@noop {} {\bibfield  {journal}
  {\bibinfo  {journal} {Journal of Statistical Mechanics: Theory and
  Experiment}\ }\textbf {\bibinfo {volume} {2016}},\ \bibinfo {pages} {054012}
  (\bibinfo {year} {2016})}\BibitemShut {NoStop}%
\bibitem [{\citenamefont {Guarcello}\ \emph {et~al.}(2019)\citenamefont
  {Guarcello}, \citenamefont {Valenti}, \citenamefont {Spagnolo}, \citenamefont
  {Pierro},\ and\ \citenamefont {Filatrella}}]{guarcello2019josephson}%
  \BibitemOpen
  \bibfield  {author} {\bibinfo {author} {\bibfnamefont {C.}~\bibnamefont
  {Guarcello}}, \bibinfo {author} {\bibfnamefont {D.}~\bibnamefont {Valenti}},
  \bibinfo {author} {\bibfnamefont {B.}~\bibnamefont {Spagnolo}}, \bibinfo
  {author} {\bibfnamefont {V.}~\bibnamefont {Pierro}},\ and\ \bibinfo {author}
  {\bibfnamefont {G.}~\bibnamefont {Filatrella}},\ }\bibfield  {title}
  {\bibinfo {title} {Josephson-based threshold detector for
  l{\'e}vy-distributed current fluctuations},\ }\href@noop {} {\bibfield
  {journal} {\bibinfo  {journal} {Physical Review Applied}\ }\textbf {\bibinfo
  {volume} {11}},\ \bibinfo {pages} {044078} (\bibinfo {year}
  {2019})}\BibitemShut {NoStop}%
\bibitem [{\citenamefont {Ding}\ and\ \citenamefont
  {Yang}(1995)}]{ding1995distribution}%
  \BibitemOpen
  \bibfield  {author} {\bibinfo {author} {\bibfnamefont {M.}~\bibnamefont
  {Ding}}\ and\ \bibinfo {author} {\bibfnamefont {W.}~\bibnamefont {Yang}},\
  }\bibfield  {title} {\bibinfo {title} {Distribution of the first return time
  in fractional brownian motion and its application to the study of on-off
  intermittency},\ }\href@noop {} {\bibfield  {journal} {\bibinfo  {journal}
  {Physical Review E}\ }\textbf {\bibinfo {volume} {52}},\ \bibinfo {pages}
  {207} (\bibinfo {year} {1995})}\BibitemShut {NoStop}%
\bibitem [{\citenamefont {Auma{\^\i}tre}\ \emph {et~al.}(2005)\citenamefont
  {Auma{\^\i}tre}, \citenamefont {P{\'e}tr{\'e}lis},\ and\ \citenamefont
  {Mallick}}]{aumaitre2005low}%
  \BibitemOpen
  \bibfield  {author} {\bibinfo {author} {\bibfnamefont {S.}~\bibnamefont
  {Auma{\^\i}tre}}, \bibinfo {author} {\bibfnamefont {F.}~\bibnamefont
  {P{\'e}tr{\'e}lis}},\ and\ \bibinfo {author} {\bibfnamefont {K.}~\bibnamefont
  {Mallick}},\ }\bibfield  {title} {\bibinfo {title} {Low-frequency noise
  controls on-off intermittency of bifurcating systems},\ }\href@noop {}
  {\bibfield  {journal} {\bibinfo  {journal} {Physical review letters}\
  }\textbf {\bibinfo {volume} {95}},\ \bibinfo {pages} {064101} (\bibinfo
  {year} {2005})}\BibitemShut {NoStop}%
\bibitem [{\citenamefont {Auma{\^\i}tre}\ \emph {et~al.}(2006)\citenamefont
  {Auma{\^\i}tre}, \citenamefont {Mallick},\ and\ \citenamefont
  {P{\'e}tr{\'e}lis}}]{aumaitre2006effects}%
  \BibitemOpen
  \bibfield  {author} {\bibinfo {author} {\bibfnamefont {S.}~\bibnamefont
  {Auma{\^\i}tre}}, \bibinfo {author} {\bibfnamefont {K.}~\bibnamefont
  {Mallick}},\ and\ \bibinfo {author} {\bibfnamefont {F.}~\bibnamefont
  {P{\'e}tr{\'e}lis}},\ }\bibfield  {title} {\bibinfo {title} {Effects of the
  low frequencies of noise on on--off intermittency},\ }\href@noop {}
  {\bibfield  {journal} {\bibinfo  {journal} {Journal of statistical physics}\
  }\textbf {\bibinfo {volume} {123}},\ \bibinfo {pages} {909} (\bibinfo {year}
  {2006})}\BibitemShut {NoStop}%
\bibitem [{\citenamefont {Alexakis}\ and\ \citenamefont
  {P{\'e}tr{\'e}lis}(2012)}]{alexakis2012critical}%
  \BibitemOpen
  \bibfield  {author} {\bibinfo {author} {\bibfnamefont {A.}~\bibnamefont
  {Alexakis}}\ and\ \bibinfo {author} {\bibfnamefont {F.}~\bibnamefont
  {P{\'e}tr{\'e}lis}},\ }\bibfield  {title} {\bibinfo {title} {Critical
  exponents in zero dimensions},\ }\href@noop {} {\bibfield  {journal}
  {\bibinfo  {journal} {Journal of Statistical Physics}\ }\textbf {\bibinfo
  {volume} {149}},\ \bibinfo {pages} {738} (\bibinfo {year}
  {2012})}\BibitemShut {NoStop}%
\bibitem [{\citenamefont {P{\'e}tr{\'e}lis}\ and\ \citenamefont
  {Alexakis}(2012)}]{petrelis2012anomalous}%
  \BibitemOpen
  \bibfield  {author} {\bibinfo {author} {\bibfnamefont {F.}~\bibnamefont
  {P{\'e}tr{\'e}lis}}\ and\ \bibinfo {author} {\bibfnamefont {A.}~\bibnamefont
  {Alexakis}},\ }\bibfield  {title} {\bibinfo {title} {Anomalous exponents at
  the onset of an instability},\ }\href@noop {} {\bibfield  {journal} {\bibinfo
   {journal} {Physical Review Letters}\ }\textbf {\bibinfo {volume} {108}},\
  \bibinfo {pages} {014501} (\bibinfo {year} {2012})}\BibitemShut {NoStop}%
\bibitem [{\citenamefont {Alexakis}\ and\ \citenamefont
  {P{\'e}tr{\'e}lis}(2009)}]{alexakis2009planar}%
  \BibitemOpen
  \bibfield  {author} {\bibinfo {author} {\bibfnamefont {A.}~\bibnamefont
  {Alexakis}}\ and\ \bibinfo {author} {\bibfnamefont {F.}~\bibnamefont
  {P{\'e}tr{\'e}lis}},\ }\bibfield  {title} {\bibinfo {title} {Planar
  bifurcation subject to multiplicative noise: Role of symmetry},\ }\href@noop
  {} {\bibfield  {journal} {\bibinfo  {journal} {Physical Review E}\ }\textbf
  {\bibinfo {volume} {80}},\ \bibinfo {pages} {041134} (\bibinfo {year}
  {2009})}\BibitemShut {NoStop}%
\bibitem [{\citenamefont {Chambers}\ \emph {et~al.}(1976)\citenamefont
  {Chambers}, \citenamefont {Mallows},\ and\ \citenamefont
  {Stuck}}]{chambers1976method}%
  \BibitemOpen
  \bibfield  {author} {\bibinfo {author} {\bibfnamefont {J.~M.}\ \bibnamefont
  {Chambers}}, \bibinfo {author} {\bibfnamefont {C.~L.}\ \bibnamefont
  {Mallows}},\ and\ \bibinfo {author} {\bibfnamefont {B.}~\bibnamefont
  {Stuck}},\ }\bibfield  {title} {\bibinfo {title} {A method for simulating
  stable random variables},\ }\href@noop {} {\bibfield  {journal} {\bibinfo
  {journal} {Journal of the american statistical association}\ }\textbf
  {\bibinfo {volume} {71}},\ \bibinfo {pages} {340} (\bibinfo {year}
  {1976})}\BibitemShut {NoStop}%
\bibitem [{\citenamefont {It{\^o}}(1944)}]{ito1944stochasticintegral}%
  \BibitemOpen
  \bibfield  {author} {\bibinfo {author} {\bibfnamefont {K.}~\bibnamefont
  {It{\^o}}},\ }\bibfield  {title} {\bibinfo {title} {Stochastic integral},\
  }\href@noop {} {\bibfield  {journal} {\bibinfo  {journal} {Proceedings of the
  Imperial Academy}\ }\textbf {\bibinfo {volume} {20}},\ \bibinfo {pages} {519}
  (\bibinfo {year} {1944})}\BibitemShut {NoStop}%
\bibitem [{\citenamefont {Denisov}\ \emph {et~al.}(2009)\citenamefont
  {Denisov}, \citenamefont {Horsthemke},\ and\ \citenamefont
  {H{\"a}nggi}}]{denisov2009generalized}%
  \BibitemOpen
  \bibfield  {author} {\bibinfo {author} {\bibfnamefont {S.~I.}\ \bibnamefont
  {Denisov}}, \bibinfo {author} {\bibfnamefont {W.}~\bibnamefont
  {Horsthemke}},\ and\ \bibinfo {author} {\bibfnamefont {P.}~\bibnamefont
  {H{\"a}nggi}},\ }\bibfield  {title} {\bibinfo {title} {Generalized
  fokker-planck equation: Derivation and exact solutions},\ }\href@noop {}
  {\bibfield  {journal} {\bibinfo  {journal} {The European Physical Journal B}\
  }\textbf {\bibinfo {volume} {68}},\ \bibinfo {pages} {567} (\bibinfo {year}
  {2009})}\BibitemShut {NoStop}%
\bibitem [{\citenamefont {Mainardi}\ \emph {et~al.}()\citenamefont {Mainardi},
  \citenamefont {Luchko}, \citenamefont {Pagnini},\ and\ \citenamefont
  {Gorenflo}}]{Mainardi_thefundamental}%
  \BibitemOpen
  \bibfield  {author} {\bibinfo {author} {\bibfnamefont {F.}~\bibnamefont
  {Mainardi}}, \bibinfo {author} {\bibfnamefont {Y.}~\bibnamefont {Luchko}},
  \bibinfo {author} {\bibfnamefont {G.}~\bibnamefont {Pagnini}},\ and\ \bibinfo
  {author} {\bibfnamefont {D.~T.~R.}\ \bibnamefont {Gorenflo}},\ }\bibfield
  {title} {\bibinfo {title} {The fundamental solution of the space-time
  fractional diffusion equation},\ }\href@noop {} {\bibinfo  {journal} {Fract.
  Calc. Appl. Anal}\ ,\ \bibinfo {pages} {153}}\BibitemShut {NoStop}%
\bibitem [{\citenamefont {Samko}\ \emph {et~al.}(1993)\citenamefont {Samko},
  \citenamefont {Kilbas}, \citenamefont {Marichev} \emph
  {et~al.}}]{samko1993fractional}%
  \BibitemOpen
\bibfield  {journal} {  }\bibfield  {author} {\bibinfo {author} {\bibfnamefont
  {S.~G.}\ \bibnamefont {Samko}}, \bibinfo {author} {\bibfnamefont {A.~A.}\
  \bibnamefont {Kilbas}}, \bibinfo {author} {\bibfnamefont {O.~I.}\
  \bibnamefont {Marichev}}, \emph {et~al.},\ }\href@noop {} {\emph {\bibinfo
  {title} {Fractional integrals and derivatives}}},\ Vol.~\bibinfo {volume}
  {1}\ (\bibinfo  {publisher} {Gordon and Breach Science Publishers, Yverdon
  Yverdon-les-Bains, Switzerland},\ \bibinfo {year} {1993})\BibitemShut
  {NoStop}%
\bibitem [{\citenamefont {Carr}\ and\ \citenamefont
  {Wu}(2003)}]{carr2003finite}%
  \BibitemOpen
  \bibfield  {author} {\bibinfo {author} {\bibfnamefont {P.}~\bibnamefont
  {Carr}}\ and\ \bibinfo {author} {\bibfnamefont {L.}~\bibnamefont {Wu}},\
  }\bibfield  {title} {\bibinfo {title} {The finite moment log stable process
  and option pricing},\ }\href@noop {} {\bibfield  {journal} {\bibinfo
  {journal} {The journal of finance}\ }\textbf {\bibinfo {volume} {58}},\
  \bibinfo {pages} {753} (\bibinfo {year} {2003})}\BibitemShut {NoStop}%
\bibitem [{\citenamefont {Seshasayanan}\ and\ \citenamefont
  {P{\'e}tr{\'e}lis}(2018)}]{seshasayanan2018growth}%
  \BibitemOpen
  \bibfield  {author} {\bibinfo {author} {\bibfnamefont {K.}~\bibnamefont
  {Seshasayanan}}\ and\ \bibinfo {author} {\bibfnamefont {F.}~\bibnamefont
  {P{\'e}tr{\'e}lis}},\ }\bibfield  {title} {\bibinfo {title} {Growth rate
  distribution and intermittency in kinematic turbulent dynamos: Which moment
  predicts the dynamo onset?},\ }\href@noop {} {\bibfield  {journal} {\bibinfo
  {journal} {EPL (Europhysics Letters)}\ }\textbf {\bibinfo {volume} {122}},\
  \bibinfo {pages} {64004} (\bibinfo {year} {2018})}\BibitemShut {NoStop}%
\bibitem [{\citenamefont {Mantegna}\ and\ \citenamefont
  {Stanley}(1994)}]{mantegna1994stochastic}%
  \BibitemOpen
  \bibfield  {author} {\bibinfo {author} {\bibfnamefont {R.~N.}\ \bibnamefont
  {Mantegna}}\ and\ \bibinfo {author} {\bibfnamefont {H.~E.}\ \bibnamefont
  {Stanley}},\ }\bibfield  {title} {\bibinfo {title} {Stochastic process with
  ultraslow convergence to a gaussian: the truncated l{\'e}vy flight},\
  }\href@noop {} {\bibfield  {journal} {\bibinfo  {journal} {Physical Review
  Letters}\ }\textbf {\bibinfo {volume} {73}},\ \bibinfo {pages} {2946}
  (\bibinfo {year} {1994})}\BibitemShut {NoStop}%
\bibitem [{\citenamefont {Koponen}(1995)}]{koponen1995analytic}%
  \BibitemOpen
  \bibfield  {author} {\bibinfo {author} {\bibfnamefont {I.}~\bibnamefont
  {Koponen}},\ }\bibfield  {title} {\bibinfo {title} {Analytic approach to the
  problem of convergence of truncated l{\'e}vy flights towards the gaussian
  stochastic process},\ }\href@noop {} {\bibfield  {journal} {\bibinfo
  {journal} {Physical Review E}\ }\textbf {\bibinfo {volume} {52}},\ \bibinfo
  {pages} {1197} (\bibinfo {year} {1995})}\BibitemShut {NoStop}%
\bibitem [{\citenamefont {Zan}\ \emph {et~al.}(2020)\citenamefont {Zan},
  \citenamefont {Xu}, \citenamefont {Kurths}, \citenamefont {Chechkin},\ and\
  \citenamefont {Metzler}}]{zan2020stochastic}%
  \BibitemOpen
  \bibfield  {author} {\bibinfo {author} {\bibfnamefont {W.}~\bibnamefont
  {Zan}}, \bibinfo {author} {\bibfnamefont {Y.}~\bibnamefont {Xu}}, \bibinfo
  {author} {\bibfnamefont {J.}~\bibnamefont {Kurths}}, \bibinfo {author}
  {\bibfnamefont {A.~V.}\ \bibnamefont {Chechkin}},\ and\ \bibinfo {author}
  {\bibfnamefont {R.}~\bibnamefont {Metzler}},\ }\bibfield  {title} {\bibinfo
  {title} {Stochastic dynamics driven by combined l{\'e}vy--gaussian noise:
  fractional fokker--planck--kolmogorov equation and solution},\ }\href@noop {}
  {\bibfield  {journal} {\bibinfo  {journal} {Journal of Physics A:
  Mathematical and Theoretical}\ }\textbf {\bibinfo {volume} {53}},\ \bibinfo
  {pages} {385001} (\bibinfo {year} {2020})}\BibitemShut {NoStop}%
\bibitem [{\citenamefont {Xu}\ \emph {et~al.}(2020)\citenamefont {Xu},
  \citenamefont {Zhou}, \citenamefont {Metzler},\ and\ \citenamefont
  {Deng}}]{xu2020levy}%
  \BibitemOpen
  \bibfield  {author} {\bibinfo {author} {\bibfnamefont {P.}~\bibnamefont
  {Xu}}, \bibinfo {author} {\bibfnamefont {T.}~\bibnamefont {Zhou}}, \bibinfo
  {author} {\bibfnamefont {R.}~\bibnamefont {Metzler}},\ and\ \bibinfo {author}
  {\bibfnamefont {W.}~\bibnamefont {Deng}},\ }\bibfield  {title} {\bibinfo
  {title} {L{\'e}vy walk dynamics in an external harmonic potential},\
  }\href@noop {} {\bibfield  {journal} {\bibinfo  {journal} {Physical Review
  E}\ }\textbf {\bibinfo {volume} {101}},\ \bibinfo {pages} {062127} (\bibinfo
  {year} {2020})}\BibitemShut {NoStop}%
\bibitem [{\citenamefont {P{\'e}tr{\'e}lis}\ and\ \citenamefont
  {Auma{\^\i}tre}(2006)}]{petrelis2006modification}%
  \BibitemOpen
  \bibfield  {author} {\bibinfo {author} {\bibfnamefont {F.}~\bibnamefont
  {P{\'e}tr{\'e}lis}}\ and\ \bibinfo {author} {\bibfnamefont {S.}~\bibnamefont
  {Auma{\^\i}tre}},\ }\bibfield  {title} {\bibinfo {title} {Modification of
  instability processes by multiplicative noises},\ }\href@noop {} {\bibfield
  {journal} {\bibinfo  {journal} {The European Physical Journal B-Condensed
  Matter and Complex Systems}\ }\textbf {\bibinfo {volume} {51}},\ \bibinfo
  {pages} {357} (\bibinfo {year} {2006})}\BibitemShut {NoStop}%
\bibitem [{\citenamefont {Graham}\ and\ \citenamefont
  {Schenzle}(1982)}]{graham1982stabilization}%
  \BibitemOpen
  \bibfield  {author} {\bibinfo {author} {\bibfnamefont {R.}~\bibnamefont
  {Graham}}\ and\ \bibinfo {author} {\bibfnamefont {A.}~\bibnamefont
  {Schenzle}},\ }\bibfield  {title} {\bibinfo {title} {Stabilization by
  multiplicative noise},\ }\href@noop {} {\bibfield  {journal} {\bibinfo
  {journal} {Physical Review A}\ }\textbf {\bibinfo {volume} {26}},\ \bibinfo
  {pages} {1676} (\bibinfo {year} {1982})}\BibitemShut {NoStop}%
\bibitem [{\citenamefont {Mallick}\ and\ \citenamefont
  {Marcq}(2003)}]{mallick2003stability}%
  \BibitemOpen
  \bibfield  {author} {\bibinfo {author} {\bibfnamefont {K.}~\bibnamefont
  {Mallick}}\ and\ \bibinfo {author} {\bibfnamefont {P.}~\bibnamefont
  {Marcq}},\ }\bibfield  {title} {\bibinfo {title} {Stability analysis of a
  noise-induced hopf bifurcation},\ }\href@noop {} {\bibfield  {journal}
  {\bibinfo  {journal} {The European Physical Journal B-Condensed Matter and
  Complex Systems}\ }\textbf {\bibinfo {volume} {36}},\ \bibinfo {pages} {119}
  (\bibinfo {year} {2003})}\BibitemShut {NoStop}%
\bibitem [{\citenamefont {Bertin}(2012)}]{bertin2012off}%
  \BibitemOpen
  \bibfield  {author} {\bibinfo {author} {\bibfnamefont {E.}~\bibnamefont
  {Bertin}},\ }\bibfield  {title} {\bibinfo {title} {On-off intermittency over
  an extended range of control parameter},\ }\href@noop {} {\bibfield
  {journal} {\bibinfo  {journal} {Physical Review E}\ }\textbf {\bibinfo
  {volume} {85}},\ \bibinfo {pages} {042104} (\bibinfo {year}
  {2012})}\BibitemShut {NoStop}%
\bibitem [{\citenamefont {Liu}\ \emph {et~al.}(2004)\citenamefont {Liu},
  \citenamefont {Anh},\ and\ \citenamefont {Turner}}]{liu2004numerical}%
  \BibitemOpen
  \bibfield  {author} {\bibinfo {author} {\bibfnamefont {F.}~\bibnamefont
  {Liu}}, \bibinfo {author} {\bibfnamefont {V.}~\bibnamefont {Anh}},\ and\
  \bibinfo {author} {\bibfnamefont {I.}~\bibnamefont {Turner}},\ }\bibfield
  {title} {\bibinfo {title} {Numerical solution of the space fractional
  fokker--planck equation},\ }\href@noop {} {\bibfield  {journal} {\bibinfo
  {journal} {Journal of Computational and Applied Mathematics}\ }\textbf
  {\bibinfo {volume} {166}},\ \bibinfo {pages} {209} (\bibinfo {year}
  {2004})}\BibitemShut {NoStop}%
\end{thebibliography}%
\end{document}